\documentclass{article}

\usepackage{graphicx}
\usepackage{multirow}
\usepackage{amssymb} 
\usepackage{amsthm}
\usepackage{amsmath}
\usepackage{fullpage}
\usepackage[longnamesfirst]{natbib}
\usepackage{enumerate}
\usepackage{amsfonts, bm, amsmath, color, natbib, rotating, amsthm, subcaption, lscape, multicol, epstopdf, verbatim, hyperref}  
\usepackage{authblk}
\usepackage{xr}
\bibpunct{(}{)}{;}{a}{,}{,}

\usepackage{setspace}
\theoremstyle{plain} \newtheorem{theorem}{Theorem} \newtheorem{proposition}{Proposition}  

\theoremstyle{definition} \newtheorem{definition}{Definition}   

\theoremstyle{remark}

\begin{document}

\newif\ifblinded
 %\blindedtrue

\title{Dynamic Shrinkage Processes}

\ifblinded
\author{}
\else

\author{Daniel R. Kowal, David S. Matteson, and David Ruppert\thanks{Kowal is Assistant Professor, Department of Statistics, Rice University, Duncan Hall, Houston, TX 77021 (E-mail: \href{mailto:daniel.kowal@rice.edu}{daniel.kowal@rice.edu}). Matteson is Associate Professor, Department of Statistical Science and ILR School, Cornell University, 1196 Comstock Hall, Ithaca, NY 14853 (E-mail: \href{mailto:matteson@cornell.edu}{matteson@cornell.edu}; Webpage: \url{http://www.stat.cornell.edu/\~matteson/}). Ruppert is Andrew Schultz, Jr. Professor of Engineering,  Department of Statistical Science and School of Operations Research and Information Engineering, Cornell University, 1170 Comstock Hall, Ithaca, NY 14853 (E-mail: \href{mailto:dr24@cornell.edu}{dr24@cornell.edu}; Webpage: \url{http://people.orie.cornell.edu/\~davidr/}). Financial support from NSF grant AST-1312903 (Kowal and Ruppert), a Xerox PARC Faculty Research Award (Kowal and Matteson), NSF grant DMS-1455172 (Matteson), and Cornell University Atkinson's Center for a Sustainable Future AVF-2017 (Matteson) is gratefully acknowledged. }}

\fi

\maketitle
\large 

\vspace{-10mm}

\begin{abstract}
We propose a novel class of dynamic shrinkage processes for Bayesian time series and regression analysis. Building upon a global-local framework of prior construction, in which continuous scale mixtures of Gaussian distributions are employed for both desirable shrinkage properties and computational tractability, we model dependence among the local scale parameters.
The resulting processes inherit the desirable shrinkage behavior of popular global-local priors, such as the horseshoe prior, but provide additional localized adaptivity, which is important for modeling time series data or regression functions with local features. 
We construct a computationally efficient Gibbs sampling algorithm based on a P{\'o}lya-Gamma scale mixture representation of the proposed process. 
Using dynamic shrinkage processes, we develop a Bayesian trend filtering model that produces more accurate estimates and tighter posterior credible intervals than competing methods, and apply the model for irregular curve-fitting of minute-by-minute Twitter CPU usage data.
In addition, we develop an adaptive time-varying parameter regression model to assess the efficacy of the Fama-French five-factor asset pricing model with momentum added as a sixth factor. Our dynamic analysis of manufacturing and healthcare industry data shows that with the exception of the market risk, no other risk factors are significant except for brief periods. 
\end{abstract}

\clearpage

\section{Introduction}
The global-local class of prior distributions is a popular and successful mechanism for providing shrinkage and regularization in a broad variety of models and applications. Global-local priors use continuous scale mixtures of Gaussian distributions to produce desirable shrinkage properties, such as (approximate) sparsity or smoothness, often leading to highly competitive and computationally tractable estimation procedures. For example, in the variable selection context, exact sparsity-inducing priors such as the spike-and-slab prior become intractable for even a moderate number of predictors. By comparison, global-local priors that shrink toward sparsity, such as the horseshoe prior \citep{carvalho2010horseshoe}, produce competitive estimators with greater scalability, and are validated by theoretical results, simulation studies, and a variety of applications \citep{carvalho2009handling,datta2013asymptotic,van2014horseshoe}. Unlike non-Bayesian counterparts such as the lasso \citep{tibshirani1996regression}, shrinkage priors also provide adequate uncertainty quantification for parameters of interest \citep{kyung2010penalized,van2014horseshoe}. 

%In general, global-local priors are employed for approximate variable selection, regularization, and smoothing. 

%and produce estimators analogous to a model-averaging procedure \citep{polson2012local}

%The horseshoe prior has been demonstrated to be an effective shrinkage prior in both sparse and non-sparse settings via theoretical results, simulation studies, and a variety of applications \citep{carvalho2009handling,carvalho2010horseshoe,datta2013asymptotic,van2014horseshoe}.

%Motivate the global-local framework:
% Approximate variable selection
% Regularization (e.g., BTF)
% Highly competitive
% Computationally efficient
% Model-averaged coefficients will be nonzero anyway \citep{polson2012local}
% Point null hypotheses are unrealistic (shrink rather than select) \citep{polson2012local}
% Cite some applications of global-local shrinkage priors

The class of global-local scale mixtures of Gaussian distributions (e.g., \citealp{carvalho2010horseshoe,polson2010shrink,polson2012local}) is defined as follows:
\begin{equation} \label{globalLocalObs}
[\omega_t|\tau,\lambda_t]  \stackrel{indep}{\sim}  N(0, \tau^2 \lambda_t^2), \quad t=1,\ldots,T
\end{equation}
where $\tau > 0$ controls the global shrinkage for all $\{\omega_t\}_{t=1}^T$, while  $\lambda_t > 0$ tunes the local shrinkage for a particular $\omega_t$. Such a model is particularly well-suited for sparse data: $\tau$ determines the global level of sparsity for $\{\omega_t\}_{t=1}^T$, while large $\lambda_t$ allows large absolute deviations of $\omega_t$ from its prior mean (zero) and small $\lambda_t$ provides extreme shrinkage to zero. Careful choice of priors for $\lambda_t^2$ and $\tau^2$ provide both the flexibility to accomodate large signals and adequate shrinkage of noise (e.g., \citealp{carvalho2010horseshoe}), so the framework of \eqref{globalLocalObs} is widely applicable. The prior in \eqref{globalLocalObs} is commonly paired with the likelihood $[y_t | \omega_t, \sigma^2] \stackrel{indep}{\sim} N(\omega_t, \sigma^2)$, but we will consider dynamic generalizations. 

Most commonly, the local scale parameters $\{\lambda_t\}$ are assumed to be  \emph{a priori} independent and identically distributed (iid). However, it can be advantageous to forgo the independence assumption. 
In the dynamic setting, in which the observations $y_t$ are time-ordered and $t$ denotes a time index, it is natural to allow the local scale parameter, $\lambda_t$, to depend on the history of the shrinkage process $\{\lambda_s\}_{s <t}$. As a result, the probability of large (or small) deviations of $\omega_t$ from the prior mean (zero), as determined by $\lambda_t$, is informed by the previous shrinkage behavior $\{\lambda_s\}_{s <t}$. Such model-based dependence may improve the ability of the model to adapt dynamically, which is important for time series estimation, forecasting, and inference. %However, the standard global-local prior independence assumption precludes dependence in the shrinkage process. 

%Since large values of $\lambda_t$ permit large absolute deviations of $\omega_t$ from the prior mean (zero),

%we may prefer to relax the independence assumption \eqref{globalLocalindep} and instead model $\{\lambda_t\}_{t=1}^T$ as a dynamic process, so that the probability of large (or small) deviations from the prior mean at time $t$ are informed by the history of the shrinkage process, $\{\lambda_s\}_{s <t}$. Such model-based dependence may improve the ability of the model to adapt dynamically, which is important for time series estimation, forecasting, and inference. %However, the standard global-local prior independence assumption in \eqref{globalLocalindep} precludes such dependence in the shrinkage process. 

We propose to model dependence in the process $\{\lambda_t\}$ using a novel log-scale representation of a broad class of global-local shrinkage priors. By considering $\{\lambda_t\}$ on the log-scale, we gain access to a variety of widely successful models for dependent data, such as (vector) autoregressions, linear regressions, Gaussian processes, and factor models, among others. An important contribution of the manuscript is to provide (i) a framework for incorporating dependent data models into popular shrinkage priors  and (ii) an accompanying Gibbs sampling algorithm, which relies on new parameter expansion techniques for computational efficiency. 

We propose to model dependence of the log-variance process $h_t = \log(\tau^2 \lambda_t^2)$ in \eqref{globalLocalObs} using the general dependent data model 
\begin{equation}\label{dsp}
h_t =  \mu + \psi_t + \eta_t,  \quad \eta_t \stackrel{iid}{\sim} Z(\alpha, \beta, 0, 1)
\end{equation} 
where $\mu = \log(\tau^2)$ corresponds to the global scale parameter, $(\psi_t + \eta_t) = \log (\lambda_t^2)$ corresponds to the local scale parameter, and $Z(\alpha, \beta, \mu_z, \sigma_z)$ denotes the \emph{$Z$-distribution} with density function 
\begin{equation}\label{zdist}
[z] = \big[\sigma B(\alpha, \beta)\big]^{-1}  \big\{ \exp\big[(z-\mu_z)/\sigma_z\big]\big\}^\alpha
 \big\{ 1 + \exp\big[(z-\mu_z)/\sigma_z\big]\big\}^{-(\alpha+\beta)}, z \in \mathbb{R}
\end{equation}
where $B(\cdot, \cdot)$ is the Beta function. In model \eqref{dsp}, the local scale parameter $\lambda_t = \exp[(\psi_t + \eta_t)/2]$ has two components: $\psi_t$, which models dependence (see below), and $\eta_t$, which corresponds to the usual iid (log-) local scale parameter. When $\psi_t = 0$, model \eqref{dsp} reduces to the static setting, and implies an \emph{inverted-Beta} prior for $\lambda_t^2$ (see Section \ref{distSMN} for more details). Notably, the class of priors represented in \eqref{dsp} includes the important shrinkage distributions in Table \ref{ibDist}, 
in each case extended to the dependent data setting.

\begin{table}[h]
\centering
\begin{tabular}{lll}
\hline
$\alpha = \beta = 1/2 $ &Horseshoe Prior & \cite{carvalho2010horseshoe} \\ 
$\alpha = 1/2 , \beta=1$ & Strawderman-Berger Prior & \cite{strawderman1971proper,berger1980robust} \\ 
$\alpha= 1, \beta = c-2, c > 0 $ &Normal-Exponential-Gamma Prior & \cite{griffin2005alternative} \\ 
$\alpha = \beta  \rightarrow 0  $ & (Improper) Normal-Jeffreys' Prior & \cite{figueiredo2003adaptive,bae2004gene} \\ 
\hline
\end{tabular}
\caption{Special cases of the inverted-Beta prior. \label{ibDist}}
\end{table}
The role of $\psi_t$ in \eqref{dsp} is to provide locally adaptive shrinkage by modeling dependence. For dynamic dependence, we propose the \emph{dynamic shrinkage process}
\begin{equation}\label{dhs}
h_{t+1} =  \mu + \phi (h_t - \mu) + \eta_t,  \quad \eta_t \stackrel{iid}{\sim} Z(\alpha, \beta, 0, 1)
%\log \lambda_{t+1}^2 = \phi \log \lambda_t^2 + \eta_t, \quad \eta_t \stackrel{iid}{\sim} Z(\alpha, \beta, 0, 1)
\end{equation} 
which is equivalent to \eqref{dsp} with $\psi_t = \phi(h_{t-1} - \mu)$. Relative to static shrinkage priors, model \eqref{dhs} only adds one additional parameter, $\phi$, and reduces to the static setting when $\phi = 0$. Importantly, the proposed Gibbs sampler for the parameters in \eqref{dhs} is linear in the number of time points, $T$, and therefore scalable. Other examples of \eqref{dsp} include linear regression, $\psi_t = \bm z_t'\bm \alpha$ for a vector of predictors $\bm z_t$, Gaussian processes, and various multivariate models (see \eqref{SVfullMult} in Section \ref{tvp}). We focus on dynamic dependence, but our modeling framework and computational techniques may be extended to incorporate more general dependence among shrinkage parameters.

Despite the apparent complexity of the model, we develop a new Gibbs sampling algorithm via a parameter expansion of model \eqref{dsp}. The MCMC algorithm combines a log-variance sampler \citep{kastner2014ancillarity} and a P{\'o}lya-Gamma sampler \citep{polson2013bayesian} to produce a conditionally Gaussian representation of model \eqref{dsp}, which permits flexibility in specification of $\psi_t$. The resulting model is  easy to implement, computationally efficient, and widely applicable.

\begin{figure}[h]
\centering
\begin{subfigure}[b]{0.49 \textwidth}
\includegraphics[width=\textwidth]{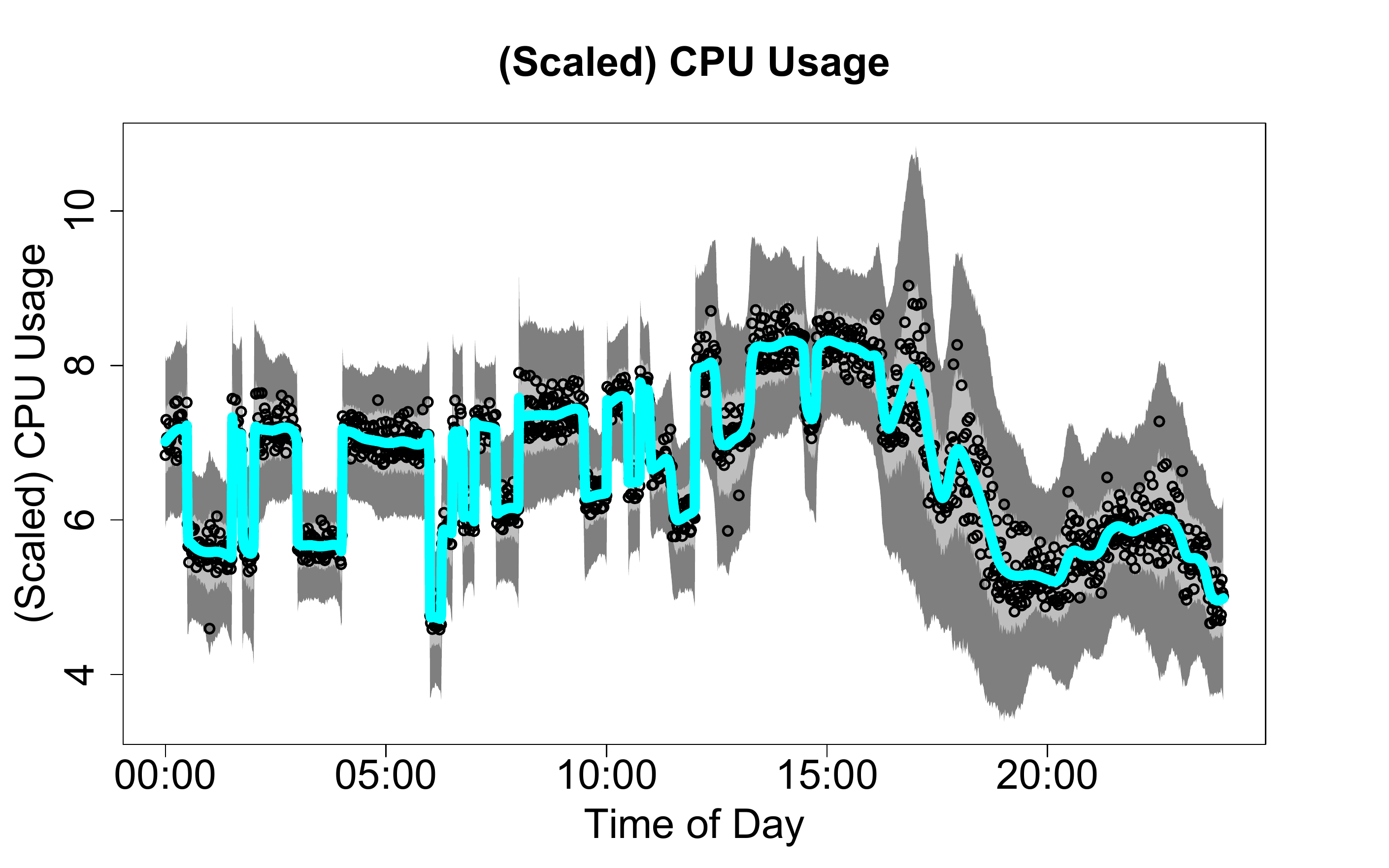}
\caption{}
\label{fig:totalCPUa}
\end{subfigure}
\begin{subfigure}[b]{0.49\textwidth}
\includegraphics[width=\textwidth]{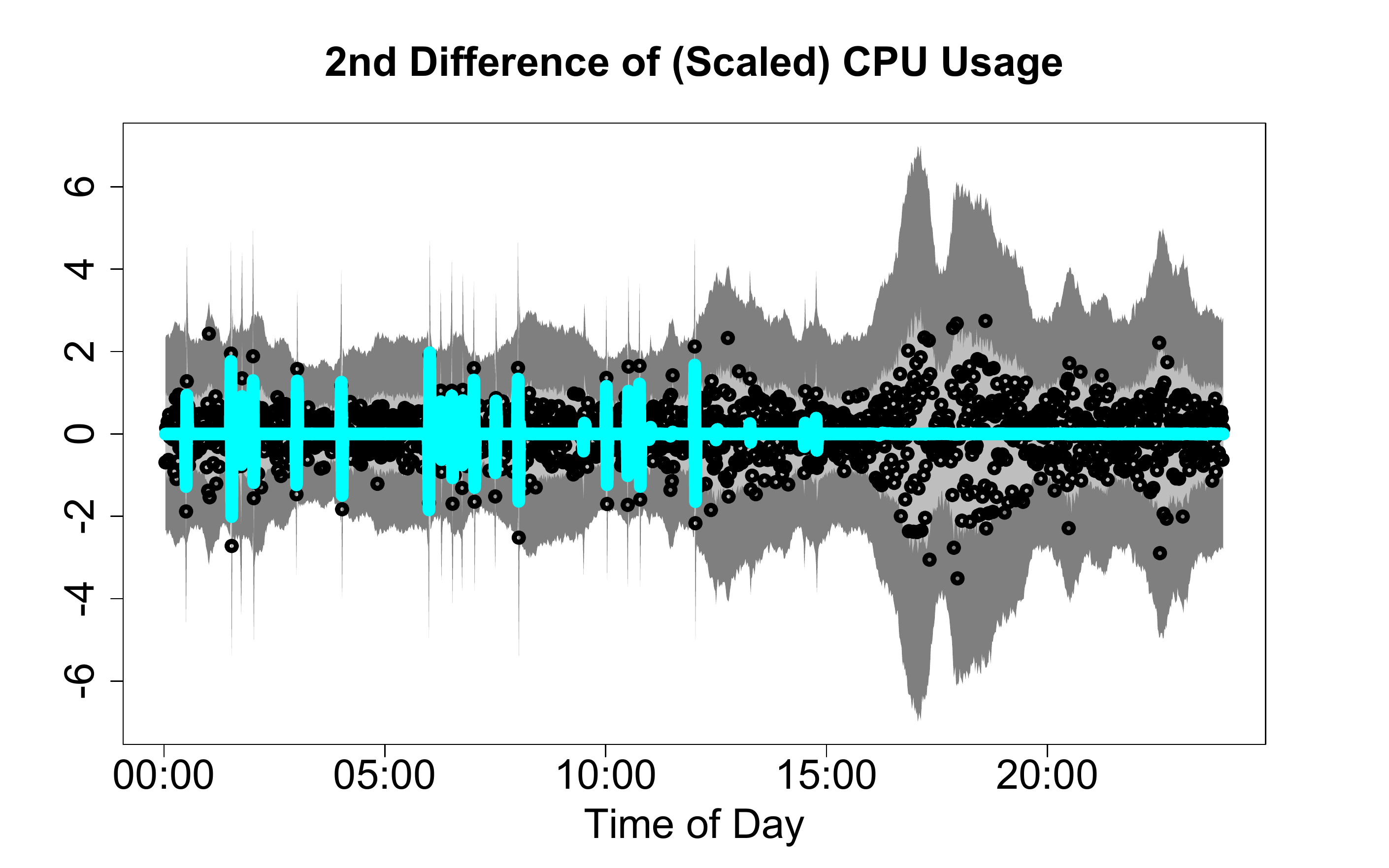}
\caption{}
\label{fig:totalCPUb}
\end{subfigure}
\begin{subfigure}[b]{0.49 \textwidth}
\includegraphics[width=\textwidth]{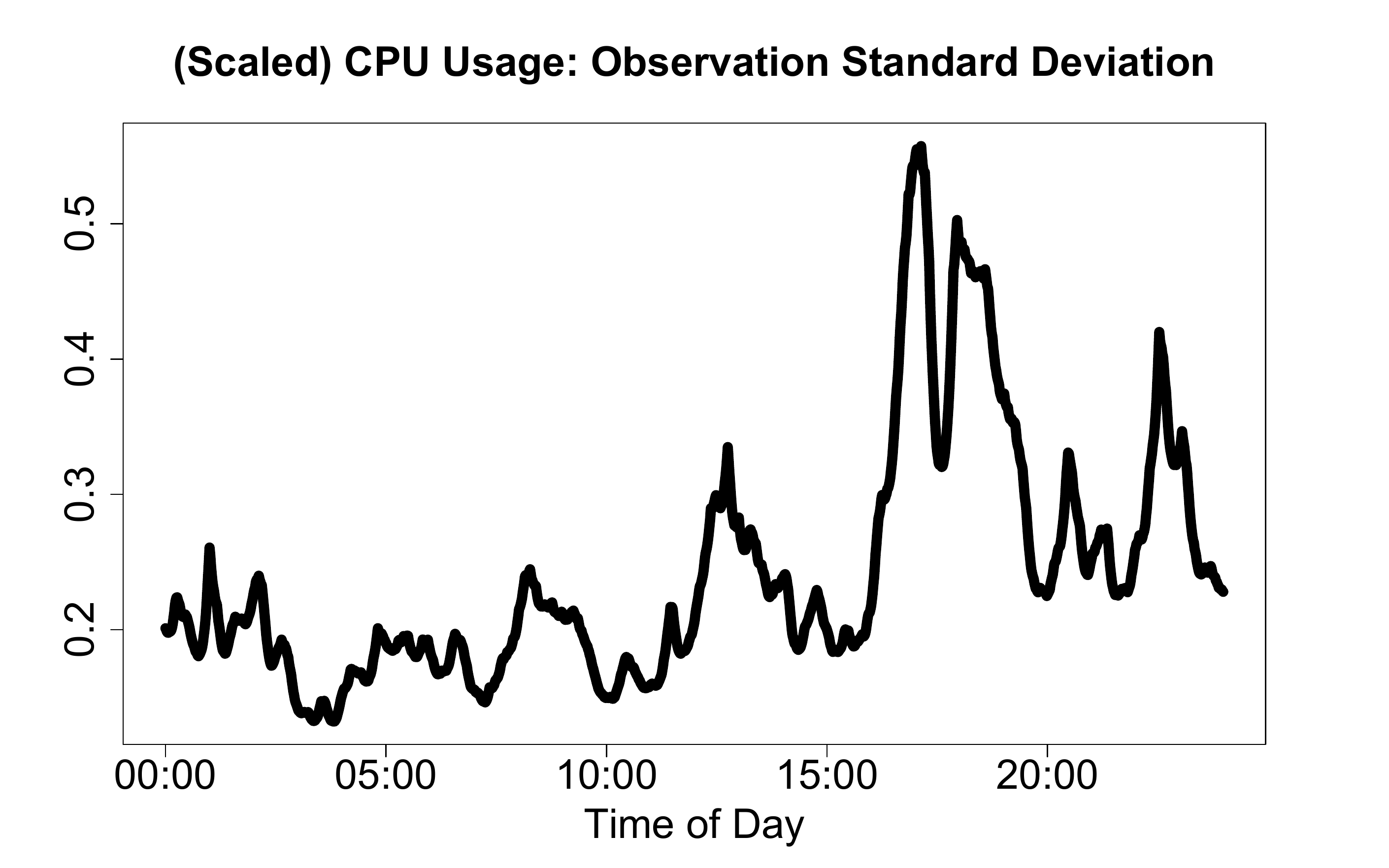}
\caption{}
\label{fig:totalCPUc}
\end{subfigure}
\begin{subfigure}[b]{0.49 \textwidth}
\includegraphics[width=\textwidth]{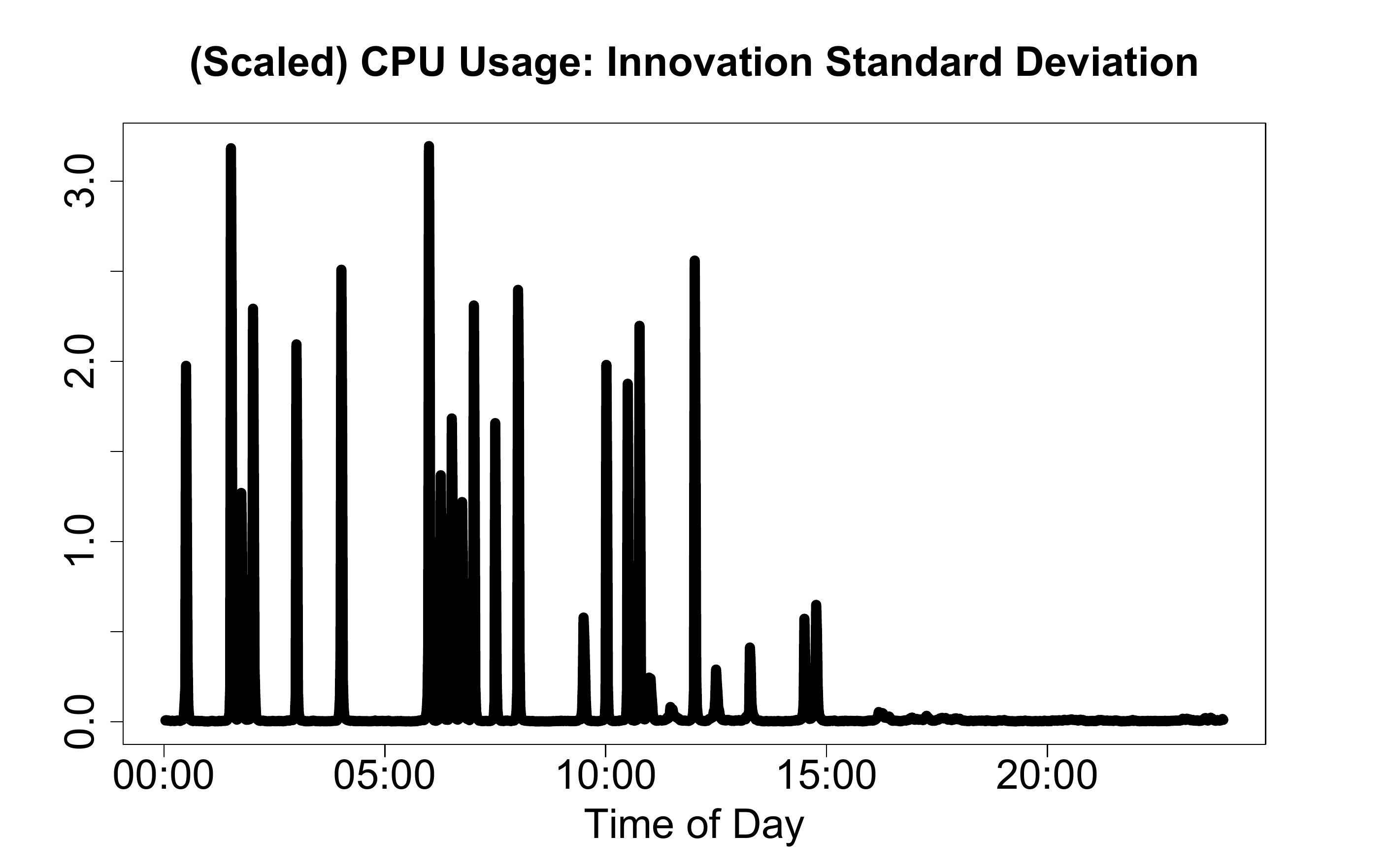}
\caption{}
\label{fig:totalCPUd}
\end{subfigure}
\caption{\small Bayesian trend filtering ($D=2$) with dynamic horseshoe process innovations of minute-by-minute Twitter CPU usage data. (a) Observed data $y_t$ (points), posterior expectation (cyan) of $\beta_t$, and 95\% pointwise highest posterior density (HPD) credible intervals (light gray) and 95\% simultaneous credible bands (dark gray) for the posterior predictive distribution of $y_t$. (b) Second difference of observed data $\Delta^2 y_t$ (points), posterior expectation of $\omega_t = \Delta^2\beta_t$ (cyan), and 95\% pointwise HPD intervals (light gray) and simultaneous credible bands (dark gray) for the posterior predictive distribution of $\Delta^2 y_t$. 
(c) Posterior expectation of time-dependent observation standard deviations, $\sigma_t$. 
(d) Posterior expectation of time-dependent innovation (prior) standard deviations, $\tau\lambda_t$.% = \exp(h_t/2)$.
\label{fig:totalCPU}}
%\caption{\small Bayesian trend filtering ($D=2$) with dynamic horseshoe innovations of minute-by-minute CPU usage data. (a) Observed data $y_t$ (points) and posterior expectation (cyan), 95\% pointwise highest posterior density (HPD) credible intervals (light gray) and 95\% simultaneous credible bands (dark gray) of $\beta_t$. (b) Second difference of observed data $\Delta^2 y_t$ (points) and posterior expectation (cyan), 95\% pointwise HPD credible intervals (light gray)  and 95\% simultaneous credible bands (dark gray) of $\omega_t = \Delta^2\beta_t$. (c) Posterior expectation of time-dependent observation standard deviations, $\sigma_t$. (d) Posterior expectation of time-dependent innovation standard deviations, $\tau\lambda_t = \exp(h_t/2)$.\label{fig:totalCPU}}
\end{figure}

For a motivating example, consider the minute-by-minute Twitter CPU usage data in Figure \ref{fig:totalCPUa} \citep{james2016leveraging}. The data show an overall smooth trend interrupted by irregular jumps throughout the morning and early afternoon, with increased volatility from 16:00-18:00. It is important to identify both abrupt changes as well as slowly-varying intraday trends. To model these features, we combine the likelihood $y_t \stackrel{indep}{\sim} N(\beta_t, \sigma_t^2)$ with a standard SV model for the observation error variance, $\sigma_t^2$, and a \emph{dynamic horseshoe process} as the prior on the second differences of the conditional mean, $\omega_t = \Delta^2\beta_t = \Delta \beta_t - \Delta \beta_{t-1}$, given by \eqref{dhs} with $\alpha = \beta = 1/2$ (see Section \ref{trendCPU} for details). The dynamic horseshoe process  either drives $\omega_t$ to zero, in which case $\beta_t$ is locally linear, or leaves $\omega_t$ effectively unpenalized, in which case large changes in slope are permissible (see Figure \ref{fig:totalCPUb}). The resulting posterior expectation of $\beta_t$ and credible bands for the posterior predictive distribution of $\{y_t\}$ adapt to both irregular jumps and smooth trends (see Figure \ref{fig:totalCPUa}).

For comparison, Figure \ref{fig:totalCPU} provides the posterior expectations of both the observation error standard deviations, $\sigma_t$ (Figure \ref{fig:totalCPUc}) and the prior standard deviations, $[\tau\lambda_t] = \exp(h_t/2)$ (Figure \ref{fig:totalCPUd}). The horseshoe-like shrinkage behavior of $\lambda_t$ is evident: values of $\lambda_t$ are either near zero, corresponding to aggressive shrinkage of $\omega_t = \Delta^2\beta_t$ to zero, or large, corresponding to large absolute changes in the slope of $\beta_t$. Importantly, Figure \ref{fig:totalCPU} also provides motivation for a \emph{dynamic} shrinkage process: there is clear \emph{volatility clustering} of $\{\lambda_t\}$, in which the shrinkage induced by $\lambda_t$ persists for consecutive time points. The volatility clustering reflects---and motivates---the temporally adaptive shrinkage behavior of the dynamic shrinkage process.

Shrinkage priors and variable selection have been used successfully for time series modeling in a broad variety of settings. \cite{belmonte2014hierarchical} propose a Bayesian Lasso prior and \cite{bitto2016achieving} use a normal-gamma prior for shrinkage in dynamic linear models, while \cite{korobilis2013hierarchical} consider several (non-dynamic) scale mixture priors for time series regression. 
In each case, the lack of a local (dynamic) scale parameter implies a time-invariant rate of shrinkage for each variable. \cite{fruhwirth2010stochastic} introduce indicator variables to discern between static and dynamic parameters, but the model cannot shrink adaptively for local time periods. \cite{nakajima2013bayesian} provide a procedure for local thresholding of dynamic coefficients, but the computational challenges of model implementation are significant. 
\cite{chan2012time} propose a class of time-varying dimension models, but due to the computational complexity of the model, only consider inclusion or exclusion of a variable for all times, which produces non-dynamic variable selection and a limited set of models.   \cite{rockovadynamic} develop an optimization approach for dynamic variable selection. However, their method only provides point estimates, and does so under the restrictive assumptions that (i) the regression coefficients follow identical time series models (AR(1) processes) with known parameters and (ii) the observation error variance is known and non-dynamic. These key limitations are not present in our framework.

Perhaps most comparable to the proposed methodology, \cite{kalli2014time} propose a class of priors which exhibit dynamic shrinkage using normal-gamma autoregressive processes. The \cite{kalli2014time} prior is a dynamic extension of the normal-gamma prior of \cite{griffin2010inference}, and provides improvements in forecasting performance relative to non-dynamic shrinkage priors. However, the  \cite{kalli2014time} model requires careful specification of several hyperparameters and hyperpriors, and the computation requires sophisticated adaptive MCMC techniques, which results in lengthy computation times. By comparison, our proposed class of dynamic shrinkage processes is far more general, and includes the dynamic horseshoe process as a special case---which notably does not require tuning of sensitive hyperparameters.
Empirically, for time-varying parameter regression models with dynamic shrinkage, the \cite{kalli2014time} MCMC sampler requires several hours, while our proposed MCMC sampler runs in only a few minutes (see Section \ref{multSims} for details and a comparison of these methods). 

%Furthermore, our proposed MCMC sampling algorithm combines existing samplers for large blocks of parameters, which produces a straightforward yet efficient Gibbs sampler, with computations linear in the number of time points. 

We apply dynamic shrinkage processes to develop a dynamic fundamental factor model for asset pricing. We build upon the five-factor Fama-French model \citep{fama2015five}, which extends the three-factor Fama-French model \citep{fama1993common} for modeling equity returns with common risk factors. We propose a dynamic extension which allows for time-varying factor loadings, possibly with localized or irregular features, and include a sixth factor, momentum \citep{carhart1997persistence}. Despite the popularity of the three-factor Fama-French model, there is not yet consensus regarding the necessity of all five factors in \cite{fama2015five} or the momentum factor. Dynamic shrinkage processes provide a mechanism for addressing this question: within a time-varying parameter regression model, dynamic shrinkage processes provide the necessary flexibility to adapt to rapidly-changing features, while shrinking unnecessary factors to zero. Our dynamic analysis shows that with the exception of the market risk factor, no other risk factors are important except for brief periods.

We introduce the dynamic shrinkage process in Section \ref{sectionDSP} and discuss relevant properties, including the P{\'o}lya-Gamma parameter expansion for efficient computations. In Section \ref{trend}, we apply the prior to develop a more adaptive Bayesian trend filtering model for irregular curve-fitting, and we compare the proposed procedure with state-of-the-art alternatives through simulations and a CPU usage application. We propose in Section \ref{tvp} a time-varying parameter regression model with dynamic shrinkage processes for adaptive regularization and evaluate the model using simulations and an asset pricing example. In Section \ref{algorithm}, we discuss the details of the Gibbs sampling algorithm, and conclude with Section \ref{conclusions}. Proofs are in the Appendix, with additional details in the supplement.

%[Mention  \cite{lang2002function} somewhere?]

\section{Dynamic Shrinkage Processes}\label{sectionDSP}
The proposed dynamic shrinkage process contains three prominent features: (i) a  dependent model for the local scale parameters, $\lambda_t$; (ii) a log-scale representation of a broad class of global-local priors to propagate desirable shrinkage properties to the dynamic setting; and (iii) a Gaussian scale-mixture representation of the implied log-variance innovations to produce an efficient Gibbs sampling algorithm. In this section, we provide the relevant details regarding these features, and explore the properties of the resulting process. 

%MCMC implementations of the SV model commonly represent the likelihood for $h_t$ on the log-scale, $\log \omega_t^2 = h_t + \epsilon_t^*$ with $\epsilon_t^* \stackrel{indep}{\sim}  \log \chi_1^2$, and approximate the $\log \chi_1^2$ distribution for the errors via a known discrete mixture of Gaussian distributions (e.g.,  \citealp{kim1998stochastic,omori2007stochastic,kastner2014ancillarity}). The approximation is quite accurate, and posterior quantities may be reweighted to produce exact (up to MCMC error) estimates. Importantly, the discrete mixture of Gaussians approximation provides a framework for a fast and efficient MCMC sampler: conditional on the mixing component, the model for $\{h_t\}_{t=1}^T$ is a Gaussian dynamic linear model, and therefore $\{h_t\}_{t=1}^T$ may be sampled jointly in $\mathcal{O}(T)$ computations. We provide the details in Section \ref{algorithm}.

\subsection{Log-Scale Representations of Global-Local Priors} \label{distSMN}
In model \eqref{dsp} and   \eqref{dhs}, we propose to model the (dynamic) dependence of the local scale parameters $\lambda_t$ via the log-variance $h_t = \log(\tau^2 \lambda_t^2)$ of the Gaussian prior \eqref{globalLocalObs}. As we demonstrate below, our specification of \eqref{dsp} and \eqref{dhs} produces desirable locally adaptive shrinkage properties. However, such shrinkage behavior is not automatic: we must consider appropriate distributions for $\mu$ and $\eta_t$. To illustrate this point, suppose $\eta_t \stackrel{iid}{\sim}  N(0, \sigma_\eta^2)$ in \eqref{dhs}, which is a common assumption in stochastic volatility (SV) modeling \citep{kim1998stochastic}. For the likelihood $y_t \sim N(\omega_t, 1)$ and the prior \eqref{globalLocalObs}, the posterior expectation of $\omega_t$ is 
$\mathbb{E}[\omega_t | \{y_s\}, \tau] = \left(1 - \mathbb{E}[\kappa_t | \{y_s\}, \tau]\right)y_t
$, where 
\begin{equation}\label{shrinkageParameter}
\kappa_t \equiv \frac{1}{1 + \mbox{Var}\left[\omega_t |\tau, \lambda_t\right]}=  \frac{1}{1+ \tau^2\lambda_t^2}
\end{equation}
is the \emph{shrinkage parameter}. 
As noted by \cite{carvalho2010horseshoe}, $\mathbb{E}[\kappa_t | \{y_s\}, \tau]$ is interpretable as the amount of shrinkage toward zero \emph{a posteriori}: $\kappa_t \approx 0$  yields minimal shrinkage (for signals), while $\kappa_t \approx 1$ yields maximal shrinkage to zero (for noise). For the standard SV model and fixing $\phi=\mu = 0$ for simplicity, $\lambda_t = \exp(\eta_t/2)$ is log-normally distributed, and the shrinkage parameter has density $[\kappa_t] \propto  \frac{1}{\kappa_t(1-\kappa_t)} \exp \Big\{-\frac{1}{2\sigma_\eta^2}\Big[\log\Big(\frac{1-\kappa_t}{\kappa_t}\Big)\Big]^2\Big\}.$
Notably, the density for $\kappa_t$ approaches zero as $\kappa_t \rightarrow 0$ and as $\kappa_t \rightarrow 1$. As a result, direct application of the Gaussian SV model may overshrink true signals and undershrink noise. 

\begin{figure}[h]
\begin{center}
\includegraphics[width=.75\textwidth]{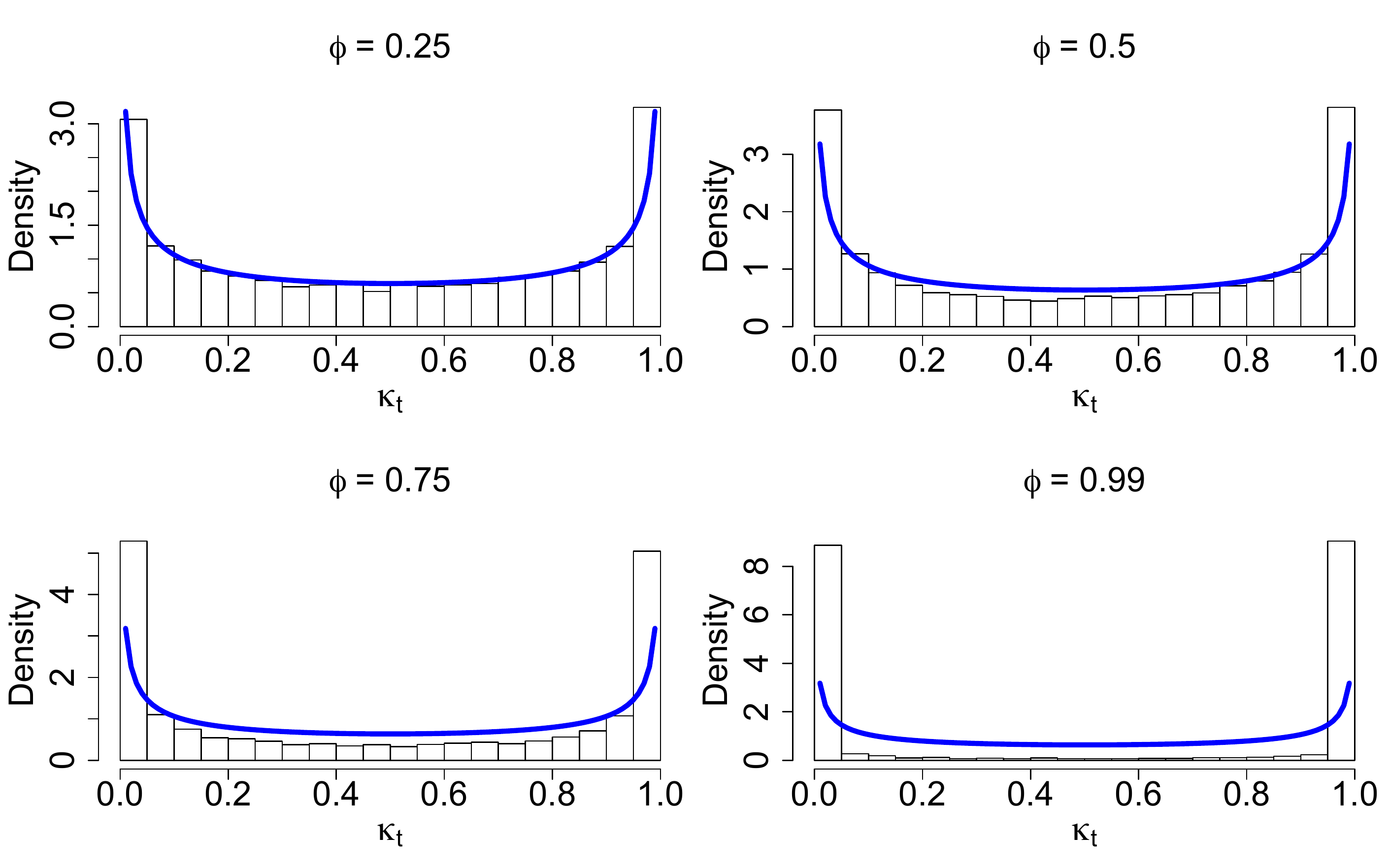}
\caption{Simulation-based estimate of the stationary distribution of $\kappa_t$ for various AR(1) coefficients $\phi$. The blue line indicates the density of $\kappa_t$ in the static ($\phi = 0$) horseshoe, $\left[\kappa\right] \sim \mbox{Beta}\left(1/2, 1/2\right)$. \label{fig:stationaryKappa}}
\end{center}
\end{figure}

By comparison, consider the horseshoe prior of \cite{carvalho2010horseshoe}. The horseshoe prior combines \eqref{globalLocalObs} with $[\lambda_t] \stackrel{iid}{\sim} C^+(0,1),$ where $C^+$ denotes the half-Cauchy distribution. For fixed $\tau = 1$, the half-Cauchy prior on $\lambda_t$ is equivalent to $\kappa_t \stackrel{iid}{\sim} \mbox{Beta}(1/2, 1/2)$, which induces a ``horseshoe" shape for the shrinkage parameter (see Figure \ref{fig:stationaryKappa}). The horseshoe-like behavior is ideal in sparse settings, since the prior density allocates most of its mass near zero (minimal shrinkage of signals) and one (maximal shrinkage of noise). Theoretical results, simulation studies, and a variety of applications confirm the effectiveness of the horseshoe prior \citep{carvalho2009handling,carvalho2010horseshoe,datta2013asymptotic,van2014horseshoe}.
%The horseshoe prior has been demonstrated to be an effective shrinkage prior in both sparse and non-sparse settings via theoretical results, simulation studies, and a variety of applications \citep{carvalho2009handling,carvalho2010horseshoe,datta2013asymptotic,van2014horseshoe}.

To emulate the robustness and sparsity properties of the horseshoe and other shrinkage priors in the dynamic setting, we represent a general class of global-local shrinkage priors on the log-scale.  
%\cite{polson2012local} use the log-scale representation of the horseshoe prior to characterize a broad class of priors, including the horseshoe, in terms of penalty functions and self-similar L{\'e}vy process. Similarly, \cite{polson2012half} suggest that log-scale representations of existing shrinkage priors  may be promising for further generalizations or theoretical arguments. 
As a motivating example, consider the special case of \eqref{globalLocalObs} and \eqref{dhs} with $\phi = 0$: $\omega_t \stackrel{indep}{\sim}  N(0, \tau^2 \lambda_t^2)$ with $\log(\lambda_t^2) = \eta_t$.
This example is illuminating: we equivalently express the (static) horseshoe prior by letting $\eta_t \stackrel{D}{=}\log \lambda_t^2$,  where $\stackrel{D}{=}$ denotes equality in distribution. In particular,  $\lambda_t \sim C^+(0,1)$ implies 
$$
\left[\lambda_t^2\right] \propto \left(\lambda_t^2\right)^{-1/2} \left(1+ \lambda_t^2\right)^{-1}
$$
which implies  
$$
\left[\eta_t \right] = \pi^{-1} \exp(\eta_t/2)\left[1 + \exp(\eta_t)\right]^{-1}
$$
so $\eta_t$ is $Z$-distributed with $\eta_t \sim Z(\alpha = 1/2, \beta = 1/2, \mu_z = 0, \sigma_z=1)$.
%The $Z$-distribution has log-linear tails and is symmetric for $\alpha = \beta$. 
Importantly, $Z$-distributions may be written as mean-variance scale mixtures of Gaussian distributions \citep{barndorff1982normal}, which produces a useful framework for a parameter-expanded Gibbs sampler.

More generally, consider the inverted-Beta prior, denoted $IB(\beta, \alpha)$, for $\lambda^2$ with density $$[\lambda^2] \propto \big(\lambda^2\big)^{\alpha-1}\big(1 + \lambda^2\big)^{-(\alpha+\beta)}, \lambda > 0$$
%\begin{equation}\label{ib}
%[\lambda^2] = \frac{\left(\lambda^2\right)^{\alpha-1}\left(1 + \lambda^2\right)^{-(\alpha+\beta)}}{Be(\beta,\alpha)}
%\end{equation}
(e.g., \citealp{armagan2011generalized,polson2012local,polson2012half}). Special cases of the inverted-Beta distribution are provided in Table \ref{ibDist}.

\iffalse 
include: 
\begin{enumerate}
\item The horseshoe prior: $\alpha = \beta = 1/2$ \citep{carvalho2010horseshoe};
\item The Strawderman-Berger prior \citep{strawderman1971proper,berger1980robust}:  $\beta=1$, $\alpha = 1/2$;
\item The Normal-Exponential-Gamma prior \citep{griffin2005alternative}: $\beta = c-2$, $\alpha= 1$, where $c > 0$ is a hyperparameter; and 
\item The improper Normal-Jeffreys' prior \citep{figueiredo2003adaptive,bae2004gene}: $\beta = \alpha \rightarrow 0$.
\end{enumerate}
\begin{equation*}
\begin{cases}
\alpha = \beta = 1/2 &  \mbox{ Horseshoe} \\
\alpha = 1/2 , \beta=1 & \mbox{ Strawderman-Berger} \\
\alpha= 1, \beta = c-2, c > 0 & \mbox{ Normal-Exponential-Gamma} \\
\alpha = \beta  \rightarrow 0 & \mbox{ (Improper) Normal-Jeffreys' }
\end{cases}
\end{equation*}
\fi

This broad class of priors may be equivalently constructed via the variances $\lambda_t^2$, the shrinkage parameters $\kappa_t$, or the log-variances $\eta_t$.
\begin{proposition}\label{prop:distequiv}
The following distributions are equivalent:
%\begin{equation}\label{distequiv} \lambda^2 \sim IB(\beta, \alpha) \iff \kappa = \frac{1}{1+\lambda^2} \sim \mbox{Beta}(\beta, \alpha)\iff \eta = \log(\lambda^2) = \log(\kappa^{-1} - 1)\sim Z(\alpha, \beta, 0 ,1) \end{equation}
\begin{enumerate}
\item $\lambda^2 \sim IB(\beta, \alpha)$; \vspace{-2mm}
\item $\kappa = 1/\big(1+\lambda^2\big) \sim \mbox{Beta}(\beta, \alpha)$; \vspace{-2mm}
\item $\eta = \log(\lambda^2) = \log(\kappa^{-1} - 1)\sim Z(\alpha, \beta, 0 ,1)$.
\end{enumerate}
\end{proposition}
%The proof of Proposition \ref{prop:distequiv} is in the supplement.  
Note that the ordering of the parameters $\alpha,\beta$ is identical for the inverted-Beta and Beta distributions, but reversed for the $Z$-distribution.

%The implied prior distribution of $\eta \equiv \log \lambda^2$ is $\pi(\eta) \propto \left[\exp(\eta)\right]^\alpha\left[1 + \exp(\eta)\right]^{-(\alpha+\beta)}$, from which it follows that $\eta \sim Z(\alpha, \beta,0,  1).$ This result also appears in \cite{polson2012local}. 

Now consider the dynamic setting in which $\phi \ne 0$. Model \eqref{dhs} implies that the conditional prior variance for $\omega_t$ in \eqref{globalLocalObs} is $\exp(h_t) = \exp(\mu  + \phi (h_{t-1} - \mu) + \eta_t) = \tau^2\lambda_{t-1}^{2\phi} \tilde \lambda_t^2 $, where $\tau^2 = \exp(\mu)$, $\lambda_{t-1}^2 = \exp(h_{t-1} - \mu)$, and $ \tilde \lambda_t^2 = \exp(\eta_t) \stackrel{iid}{\sim} IB(\beta, \alpha)$, as in the non-dynamic setting. This prior generalizes the $IB(\beta, \alpha)$ prior via the local variance term, $\lambda_{t-1}^{2\phi}$, which incorporates information about the shrinkage behavior at the previous time $t-1$ in the prior for $\omega_t$. We formalize the role of this local adjustment term with the following results.

\begin{proposition}\label{prop:distequiv2}
Suppose $\eta \sim Z(\alpha, \beta, \mu_z, 1)$ for $\mu_z \in \mathbb{R}$. Then $\kappa =  1/(1+\exp(\eta)) \sim \mbox{TPB}(\beta, \alpha, \exp(\mu_z))$, where $\kappa \sim \mbox{TPB}(\beta, \alpha, \gamma)$ denote the \emph{three-parameter Beta distribution} with density
$
[\kappa] = [B(\beta, \alpha)]^{-1} \gamma^\beta \kappa^{\beta - 1}(1-\kappa)^{\alpha - 1}\left[1 + (\gamma - 1)\kappa\right]^{-(\alpha + \beta)}, \kappa \in (0,1), \gamma > 0$.
\end{proposition}
%The proof of Proposition \ref{prop:distequiv2} is in the supplement. 
The three-parameter Beta (TPB) distribution  \citep{armagan2011generalized} generalizes the Beta distribution: $\gamma = 1$ produces the $\mbox{Beta}(\beta, \alpha)$ distribution, while $\gamma > 1$ (respectively, $\gamma < 1$) allocates more mass near zero (respectively, one) relative to the $\mbox{Beta}(\beta, \alpha)$ distribution. For dynamic shrinkage processes, the TPB distribution arises as the conditional prior distribution of $\kappa_{t+1}$ given $\{\kappa_s\}_{s \le t}.$
\begin{theorem}\label{theorem:condTPB}
For the dynamic shrinkage process \eqref{dhs}, the conditional prior distribution of the shrinkage parameter $\kappa_{t+1} = 1/\big(1 + \tau^2\lambda_{t+1}^2\big)$ is 
\begin{equation}\label{condTPB}
[\kappa_{t+1} | \{\kappa_s\}_{s \le t}, \phi, \tau] \sim \mbox{TPB}\left(\beta, \alpha, \tau^{2(1-\phi)} \left[\frac{1 - \kappa_t}{\kappa_t}\right]^\phi\right)
\end{equation}
or equivalently, $[\kappa_{t+1} | \{\lambda_s\}_{s \le t}, \phi, \tau] \sim \mbox{TPB}(\beta, \alpha, \tau^2\lambda_t^{2\phi})$. 
\end{theorem}
The proof of Theorem \ref{theorem:condTPB} is in the Appendix. 
Naturally, the previous value of the shrinkage parameter, $\kappa_t$, together with the AR(1) coefficient $\phi$, inform both the magnitude and the direction of the distributional shift of $\kappa_{t+1}$. 
\begin{theorem}\label{theorem:conc}
For the dynamic horseshoe process of \eqref{dhs} with $\alpha=\beta=1/2$ and fixed $\tau = 1$, the conditional prior distribution \eqref{condTPB} satisfies
$
\mathbb{P}\big(\kappa_{t+1} < \varepsilon | \{\kappa_s\}_{s \le t}, \phi\big) \rightarrow 1
$ as $\kappa_t \rightarrow 0$ for any $\varepsilon \in (0,1)$ and fixed $\phi \ne 0$.
\end{theorem}
The proof of Theorem \ref{theorem:conc} is in the Appendix. Importantly, Theorem \ref{theorem:conc} demonstrates that the mass of the conditional prior distribution for $\kappa_{t+1}$ concentrates near zero---corresponding to minimal shrinkage of signals---when $\kappa_t$ is near zero, so the shrinkage behavior at time $t$ informs the (prior) shrinkage behavior at time $t+1$.

We similarly characterize the posterior distribution of $\kappa_{t+1}$ given $\{\kappa_s\}_{s\le t}$ in the following theorem, which extends the results of \cite{datta2013asymptotic} to the dynamic setting.

\begin{theorem}\label{theorem:DattaGhosh}
Under the likelihood $y_t \stackrel{indep}{\sim} N(\omega_t, 1)$, the prior \eqref{globalLocalObs}, and the dynamic horseshoe process \eqref{dhs} with $\alpha=\beta=1/2$ and fixed $\phi \ne 0$, the posterior distribution of $\kappa_{t+1}$ given the history of the shrinkage process $\{\kappa_s\}_{s \le t}$ satisfies the following properties:
\begin{enumerate}[(a)]
\item For any fixed $\varepsilon \in (0,1)$, 
$\mathbb{P}\big(\kappa_{t+1} > 1 - \varepsilon  \big| y_{t+1}, \{\kappa_s\}_{s \le t}, \phi, \tau\big) \rightarrow 1$ as $\gamma_t \rightarrow 0$ uniformly in $y_{t+1}\in\mathbb{R}$, where $\gamma_t = \tau^{2(1-\phi)} \left[(1 - \kappa_t)/\kappa_t\right]^\phi.$
\item For any fixed $\varepsilon \in (0,1)$ and $\gamma_t  <1$, $\mathbb{P}\big(\kappa_{t+1} < \varepsilon  \big| y_{t+1}, \{\kappa_s\}_{s \le t}, \phi, \tau\big) \rightarrow 1$ as $|y_{t+1}| \rightarrow \infty$.  
\end{enumerate}
\end{theorem}
The proof of Theorem \ref{theorem:DattaGhosh} is in the supplementary material, and uses the 
observation that marginally, $[y_{t+1} | \{\kappa_s\}] \stackrel{indep}{\sim} N(0, \kappa_{t+1}^{-1})$, so the posterior distribution of $\kappa_{t+1}$ is 
\begin{align*}
[\kappa_{t+1} | y_{t+1}, \{\kappa_s\}_{s \le t}, \phi, \tau] &\propto \left\{\kappa_{t+1}^{\beta- 1} (1-\kappa_{t+1})^{\alpha - 1} \big[1 + (\gamma_t - 1)\kappa_{t+1}\big]^{-(\alpha + \beta)}\right\} \left\{ \kappa_{t+1}^{1/2}\exp\left(-y_{t+1}^2\kappa_{t+1}/2\right)\right\} \\
&\propto (1-\kappa_{t+1})^{-1/2} \big[1 + (\gamma_t - 1)\kappa_{t+1}\big]^{-1} \exp\left(-y_{t+1}^2\kappa_{t+1}/2\right).
\end{align*}
Theorem \ref{theorem:DattaGhosh}(a) demonstrates that the posterior mass of $[\kappa_{t+1} | \{\kappa_s\}_{s \le t}]$ concentrates near one as $\tau \rightarrow 0$, as in the non-dynamic horseshoe, but also as $\kappa_t \rightarrow 1$. Therefore, the dynamic horseshoe process provides an additional mechanism for shrinkage of noise, besides the global scale parameter $\tau$, via the previous shrinkage parameter $\kappa_t$. Moreover, Theorem \ref{theorem:DattaGhosh}(b) shows that, despite the additional shrinkage capabilities, the posterior mass of $[\kappa_{t+1} | \{\kappa_s\}_{s \le t}]$ concentrates near zero for large absolute signals $|y_{t+1}|$, which indicates robustness of the dynamic horseshoe process to large signals analogous to the static horseshoe prior.

%We also consider the stationary distribution of $\{\kappa_t\}$ implied by model \eqref{dhs}. 
When $|\phi| < 1$, the log-variance process $\{h_t\}$ is stationary, 
%with unconditional mean $\mathbb{E} \left[h_t\right] = \mu$ and variance $\mbox{Var}\left[h_t\right] = \pi^2/\left(1-\phi^2\right)$, 
which implies  $\{\kappa_t\}$ is stationary. In Figure \ref{fig:stationaryKappa}, we plot a simulation-based estimate of the stationary distribution of $\kappa_t$ for various values of $\phi$ under the dynamic horseshoe process. The stationary distribution of $\kappa_t$ is similar to the static horseshoe distribution ($\phi = 0$) for $\phi < 0.5$, while for large values of $\phi$ the distribution becomes more peaked at zero (less shrinkage of $\omega_t$) and one (more shrinkage of $\omega_t$). The result is intuitive: larger $|\phi|$ corresponds to greater persistence in shrinkage behavior, so marginally we expect states of aggressive shrinkage or little shrinkage.

\subsection{Scale Mixtures via  P{\'o}lya-Gamma Processes}
For efficient computations, we develop a parameter expansion of the general model \eqref{dsp} and the dynamic shrinkage process \eqref{dhs} using a conditionally Gaussian representation for $\eta_t$. In doing so, we may incorporate Gaussian models---and accompanying sampling algorithms---for dependent data in \eqref{dsp}. Given a conditionally Gaussian parameter expansion, a Gibbs sampler for \eqref{dsp} proceeds as follows: (i) draw the log-variances $h_t$, for which the conditional prior \eqref{dsp} is  Gaussian, and (ii) draw the parameters in $\mu$ and $\psi_t$, for which the conditional likelihood \eqref{dsp} is Gaussian. For the log-variance sampler, we represent the likelihood for $h_t$ in \eqref{globalLocalObs} on the log-scale and approximate the resulting distribution using a known discrete mixture of Gaussian distributions (see Section \ref{algorithm}).  This approach is popular in SV modeling (e.g.,  \citealp{kim1998stochastic}), which is analogous to the dynamic shrinkage process in \eqref{dhs}. Importantly, the proposed parameter expansion inherits the computational complexity of the samplers for $h_t$ and $\psi_t$: for the dynamic shrinkage processes in \eqref{dhs}, the proposed parameter expansion implies that the log-variance $\{h_t\}_{t=1}^T$ is a Gaussian dynamic linear model, and therefore $\{h_t\}_{t=1}^T$ may be sampled jointly in $\mathcal{O}(T)$ computations. We provide the relevant details in Section \ref{algorithm}.

The proposed algorithm requires a parameter expansion of $ \eta_t \stackrel{iid}{\sim} Z(\alpha, \beta, 0, 1)$ in \eqref{dhs}  as a scale mixture of Gaussian distributions.
The representation of a $Z$-distribution as a mean-variance scale mixtures of Gaussian distributions is due to \cite{barndorff1982normal}. For implementation, we build on the framework of  \cite{polson2013bayesian}, who propose a P{\'o}lya-Gamma scale mixture of Gaussians representation for Bayesian logistic regression.
A \emph{P{\'o}lya-Gamma} random variable $\xi$  with parameters $b>0$ and $c \in \mathbb{R}$, denoted $\xi \sim \mbox{PG}(b,c)$, is an infinite convolution of Gamma random variables:
\begin{equation}\label{defPG}
\xi \stackrel{D}{=} \frac{1}{2\pi^2} \sum_{k=1}^\infty \frac{g_k}{(k - 1/2)^2 - c^2/(4\pi^2)}
\end{equation}
where $g_k \stackrel{iid}{\sim}\mbox{Gamma}(b, 1)$.
\iffalse
The P{\'o}lya-Gamma random variable is defined below.
\begin{definition}[\cite{polson2013bayesian}]
A random variable $\xi$ has a P{\'o}lya-Gamma distribution with parameters $b>0$ and $c \in \mathbb{R}$, denoted $\xi \sim \mbox{PG}(b,c)$, if 
\begin{equation}\label{defPG}
\xi \stackrel{D}{=} \frac{1}{2\pi^2} \sum_{k=1}^\infty \frac{g_k}{(k - 1/2)^2 - c^2/(4\pi^2)}
\end{equation}
where $g_k \stackrel{iid}{\sim}\mbox{Gamma}(b, 1)$ and $\stackrel{D}{=} $ denotes equality in distribution. 
\end{definition}
\fi
Properties of  P{\'o}lya-Gamma random variables may be found in \cite{barndorff1982normal} and \cite{polson2013bayesian}. Our interest in P{\'o}lya-Gamma random variables derives from their role in representing the $Z$-distribution as a mean-variance scale mixture of Gaussians.
\begin{theorem}\label{PGthm}
The random variable $\eta \sim Z(\alpha, \beta, 0 ,1)$, or equivalently $\eta = \log(\lambda^2)$ with $\lambda^2 \sim IB(\beta, \alpha)$, is a mean-variance scale mixture of Gaussian distributions with 
\begin{equation}\label{smnGP}
\begin{cases}
[\eta | \xi] \sim N\left( \xi^{-1}[\alpha-\beta]/2, \xi^{-1}\right) \\
[\xi] \sim \mbox{PG}(\alpha + \beta, 0).
\end{cases}
\end{equation}
Moreover, the conditional distribution of $\xi$ is $[\xi | \eta]\sim \mbox{PG}(\alpha + \beta, \eta)$.
\end{theorem}
The proof of Theorem \ref{PGthm} is in the Appendix. When $\alpha = \beta$, the $Z$-distribution is symmetric, and the conditional expectation in \eqref{smnGP} simplifies to $\mathbb{E}[\eta | \xi] = 0$. \cite{polson2013bayesian} propose a sampling algorithm for P{\'o}lya-Gamma random variables, which is available in the \texttt{R} package \texttt{BayesLogit}, and is extremely efficient when $b = 1$. In our setting, this corresponds to $\alpha + \beta  = 1$, for which the horseshoe prior is the prime example.  Importantly, this representation allows us to construct an efficient sampling algorithm that combines an $\mathcal{O}(T)$ sampling algorithm for the log-volatilities $\{h_t\}_{t=1}^T$ with a P{\'o}lya-Gamma sampler for the mixing parameters.

\section{Bayesian Trend Filtering with Dynamic Shrinkage Processes}\label{trend}
Dynamic shrinkage processes are particularly appropriate for dynamic linear models (DLMs). DLMs combine an observation equation, which relates the observed data to latent state variables, and an evolution equation, which allows the state variables---and therefore the conditional mean of the data---to be dynamic. By construction, DLMs contain many parameters, and therefore may benefit from structured regularization. The proposed dynamic shrinkage processes offer such regularization, and unlike existing methods, do so adaptively.

Consider the following DLM with a $D$th order random walk on the state variable, $\beta_t$:
\begin{equation}\label{locLinTrend}
\begin{cases}
y_t = \beta_t + \epsilon_t, & [\epsilon_t | \sigma_\epsilon] \stackrel{iid}{\sim}  N(0, \sigma_\epsilon^2), \quad t=1,\ldots,T\\
\Delta^D \beta_{t+1} =  \omega_t, &  [\omega_t | \tau, \{\lambda_s\}]  \stackrel{indep}{\sim}   N(0, \tau^2 \lambda_t^2), \quad t = D,\ldots,T
\end{cases}
\end{equation}
and $\beta_{t+1} =\omega_t  \sim  N(0, \tau^2 \lambda_t^2)$ for $ t = 0,\ldots,D-1$, where $\Delta$ is the differencing operator and $D \in \mathbb{Z}^+$ is the degree of differencing. By imposing a shrinkage prior on $\lambda_t$, model \eqref{locLinTrend} may be viewed as a Bayesian adaptation of the \emph{trend filtering} model of \cite{kim2009ell_1} and \cite{tibshirani2014adaptive}: model \eqref{locLinTrend} features a penalty encouraging sparsity of the $D$th order differences of the conditional mean, $\beta_t$.  \cite{faulkner2016locally} provide an implementation based on the (static) horseshoe prior and the Bayesian lasso, and further allow for non-Gaussian likelihoods. We refer to model \eqref{locLinTrend} as a \emph{Bayesian trend filtering} (BTF) model, with various choices available for the distribution of the innovation standard deviations, $(\tau\lambda_t)$. 

We propose a dynamic horseshoe process as the prior for the innovations $\omega_t$ in model \eqref{locLinTrend}. The aggressive shrinkage of the horseshoe prior forces small values of $|\omega_t| = |\Delta^D \beta_{t+1}|$ toward zero, while the robustness of the horseshoe prior permits large values of $|\Delta^D \beta_{t+1}|$. When $D=2$, model \eqref{locLinTrend} will shrink the conditional mean $\beta_t$ toward a piecewise linear function with breakpoints determined adaptively, while allowing large absolute changes in the slopes.  
Further, using the \emph{dynamic} horseshoe process, the shrinkage effects induced by $\lambda_t$ are time-dependent, which provides localized adaptability to regions with rapidly- or slowly-changing features. Following \cite{carvalho2010horseshoe} and \cite{polson2012half}, we assume a half-Cauchy prior for the global scale parameter $\tau \sim C^+(0, \sigma_\epsilon/\sqrt T)$, in which we scale by the observation error variance and the sample size \citep{piironen2016hyperprior}. Using P{\'o}lya-Gamma mixtures, the implied conditional prior on $\mu = \log(\tau^2)$ is $[\mu |\sigma_\epsilon, \xi_\mu] \sim N(\log \sigma_\epsilon^2 - \log T, \xi_\mu^{-1})$ with $\xi_\mu \sim \mbox{PG}(1,0)$. We include the details of the Gibbs sampling algorithm for model \eqref{locLinTrend} in Section \ref{algorithm}, which is notably \emph{linear} in the number of time points, $T$: the full conditional posterior precision matrices for $\bm \beta = (\beta_1,\ldots,\beta_T)'$ and $\bm h = (h_1,\ldots,h_T)'$ are $D$-banded and tridiagonal, respectively, which admit highly efficient $\mathcal{O}(T)$ back-band substitution sampling algorithms  (see the supplement for empirical evidence).

\subsection{Bayesian Trend Filtering: Simulations}\label{trendSims}
To assess the performance of the Bayesian trend filtering (BTF) model \eqref{locLinTrend} with dynamic horseshoe innovations ({\bf BTF-DHS}), we compared the proposed methods to several competitive alternatives using simulated data. We considered the following variations on BTF model \eqref{locLinTrend}:  normal-inverse-Gamma ({\bf BTF-NIG}) innovations via $\tau^{-2} \sim \mbox{Gamma}(0.001, 0.001)$ with $\lambda_t=1$; and (static) horseshoe priors for the innovations ({\bf BTF-HS}) via $\tau, \lambda_t \stackrel{iid}{\sim} C^+(0,1)$.
% implemented via the parameter expansion $\big[\lambda_t^{-2}| \gamma_{\lambda_t}\big] \stackrel{indep}{\sim} \mbox{Gamma}\big(\frac{1}{2}, \gamma_{\lambda_t}\big)$, $\big[\gamma_{\lambda_t} \big]\stackrel{indep}{\sim} \mbox{Gamma}\big(\frac{1}{2}, 1\big)$ and similarly, $\big[\tau^{-2}| \gamma_{\tau}\big] \stackrel{indep}{\sim} \mbox{Gamma}\big(\frac{1}{2}, \gamma_{\tau}\big)$, $\gamma_{\tau} \stackrel{indep}{\sim} \mbox{Gamma}\big(\frac{1}{2}, 1\big)$ \citep{wand2011mean}. 
In addition, we include the (non-Bayesian) trend filtering model of \cite{tibshirani2014adaptive} implemented using the \texttt{R} package \texttt{genlasso} \citep{genlasso}, for which the regularization tuning parameter is chosen using cross-validation ({\bf Trend Filtering}). For all trend filtering models, we select $D=2$, but the relative performance is similar for $D=1$. Among non-trend filtering models, we include a smoothing spline estimator implemented via \texttt{smooth.spline()} in \texttt{R} ({\bf Smoothing Spline}); the wavelet-based estimator of \cite{abramovich1998wavelet} ({\bf BayesThresh}) implemented in the \texttt{wavethresh} package \citep{wavethresh}; and the nested Gaussian Process ({\bf nGP}) model of \cite{zhu2013locally}, which relies on a state space model framework for efficient computations, comparable to---but empirically less efficient than---the BTF model \eqref{locLinTrend}.

%For further comparisons, we include competitive alternatives to (Bayesian) trend filtering. A useful baseline is a smoothing spline, implemented via \texttt{smooth.spline()} in \texttt{R}, which is well known to be effective for fitting smooth curves, but is less effective for non-smooth curves ({\bf Smoothing Spline}). A more effective alternative for non-smooth curves is the wavelet-based estimator of \cite{abramovich1998wavelet} ({\bf BayesThresh}) implemented in the \texttt{wavethresh} package \citep{wavethresh}. Finally, we include the nested Gaussian Process ({\bf nGP}) model of \cite{zhu2013locally}, which relies on a state space model framework for efficient computations, comparable to---but empirically less efficient than---the BTF model \eqref{locLinTrend}. 

We simulated 100 data sets from the model $y_t = y_t^* + \epsilon_t$, where $y_t^*$ is the true function and  $\epsilon_t \stackrel{indep}{\sim}  N(0, \sigma_*^2)$. We use the following true functions $y_t^*$ from \cite{donoho1994ideal}:  \emph{Doppler}, \emph{Bumps}, \emph{Blocks}, and \emph{Heavisine}, implemented in the \texttt{R} package \texttt{wmtsa} \citep{wmtsa}. The noise variance $\sigma_*^2$ is determined by selecting a root-signal-to-noise ratio (RSNR) and computing 
$\sigma_* = \sqrt{\frac{\sum_{t=1}^T (y_t^* - \bar{y}^*)^2}{T-1}} \Big/\mbox{RSNR}$, where $\bar{y}^* = \frac{1}{T}\sum_{t=1}^T y_t^*$. As in \cite{zhu2013locally}, we select RSNR = 7 and use a moderate length time series, $T=128$. 

\begin{figure}[h]
\begin{center}
\includegraphics[width=.5\textwidth]{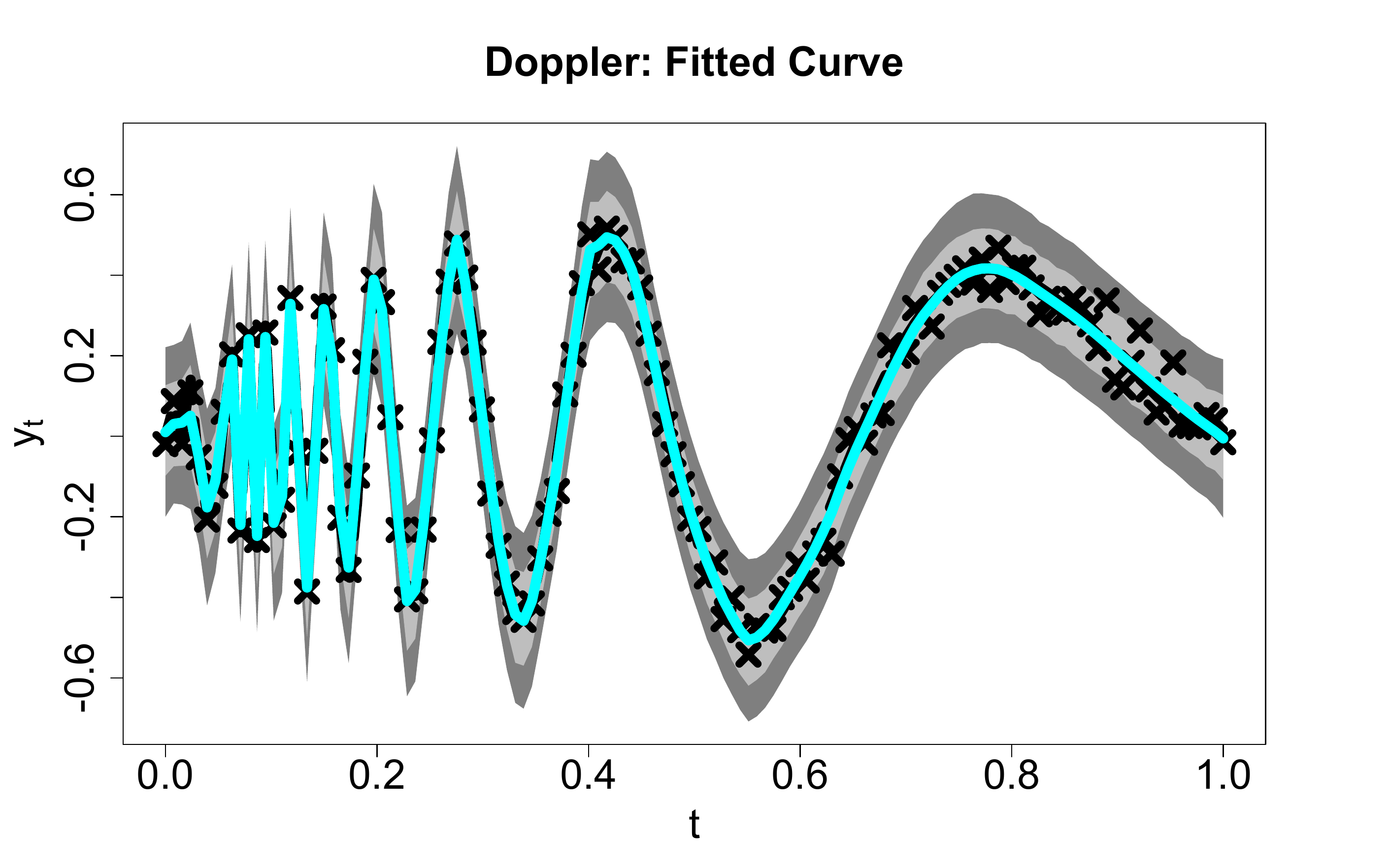}\includegraphics[width=.5\textwidth]{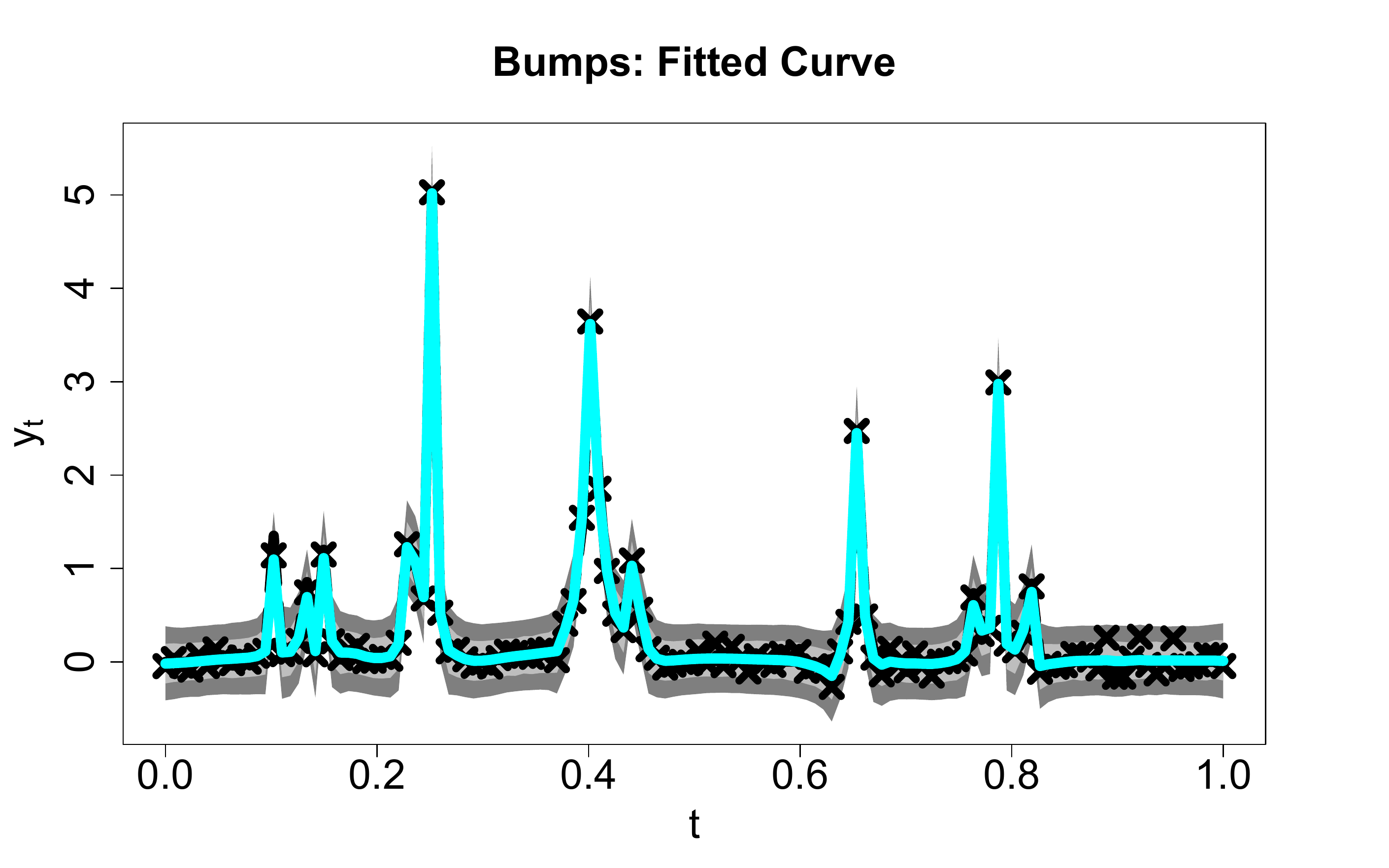}

\includegraphics[width=.5\textwidth]{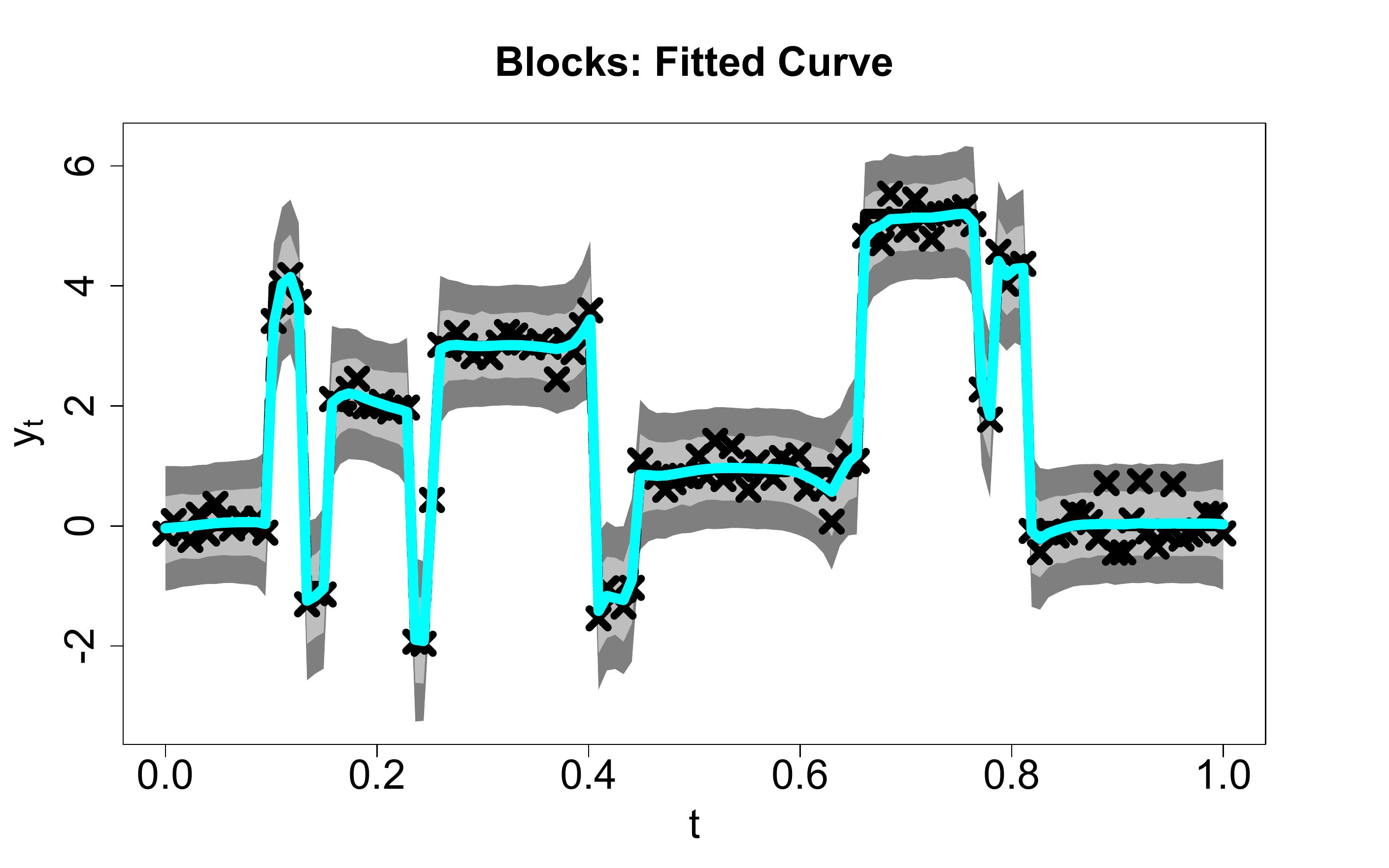}\includegraphics[width=.5\textwidth]{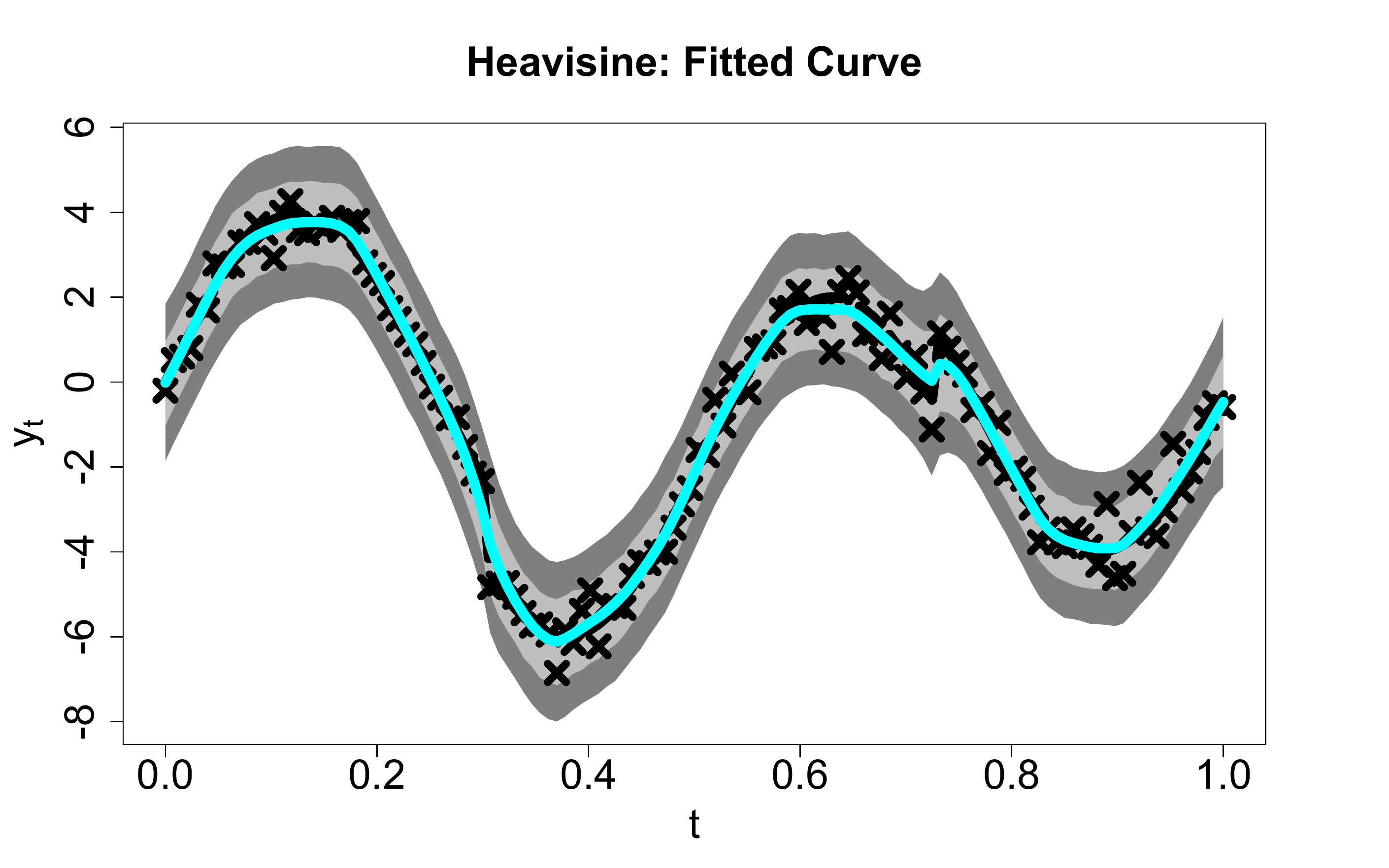}
\caption{Fitted curves for simulated data with $T=128$ and $\text{RSNR}=7$. Each panel includes the simulated observations (x-marks), the posterior expectations of $\beta_t$ (cyan), and the 95\% pointwise HPD credible intervals (light gray) and 95\% simultaneous credible bands (dark gray) for the posterior predictive distribution of $\{y_t\}$ under BTF-DHS model \eqref{locLinTrend} with $D=2$. The proposed estimator, as well as the uncertainty bands, accurately capture both slowly- and rapidly-changing behavior in the underlying functions. 
 \label{fig:fitted_Smoother}}
\end{center}
\end{figure}

In Figure \ref{fig:fitted_Smoother}, we provide an example of each true curve $y_t^*$, together with the proposed BTF-DHS posterior expectations and credible bands. Notably, the Bayesian trend filtering model \eqref{locLinTrend} with $D=2$ and dynamic horseshoe innovations provides an exceptionally accurate fit to each data set. Importantly, the posterior expectations and the posterior credible bands adapt to both slowly- and rapidly-changing behavior in the underlying curves. The implementation is also efficient: the computation time for 10,000 iterations of the Gibbs sampling algorithm, implemented in \texttt{R} (on a MacBook Pro, 2.7 GHz Intel Core i5), is about 45 seconds.

\begin{figure}[h]
\begin{center}
\includegraphics[width=.5\textwidth]{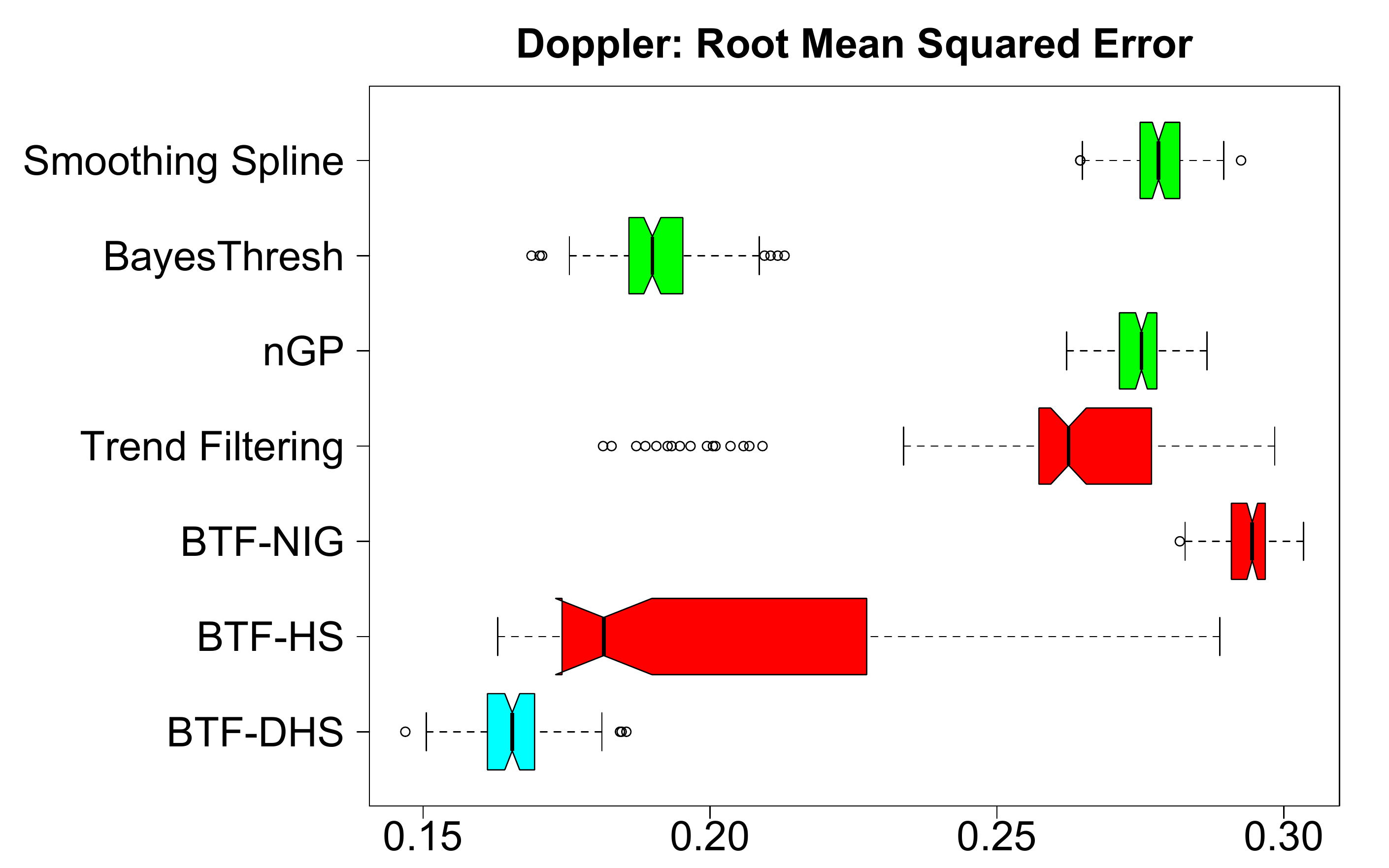}\includegraphics[width=.5\textwidth]{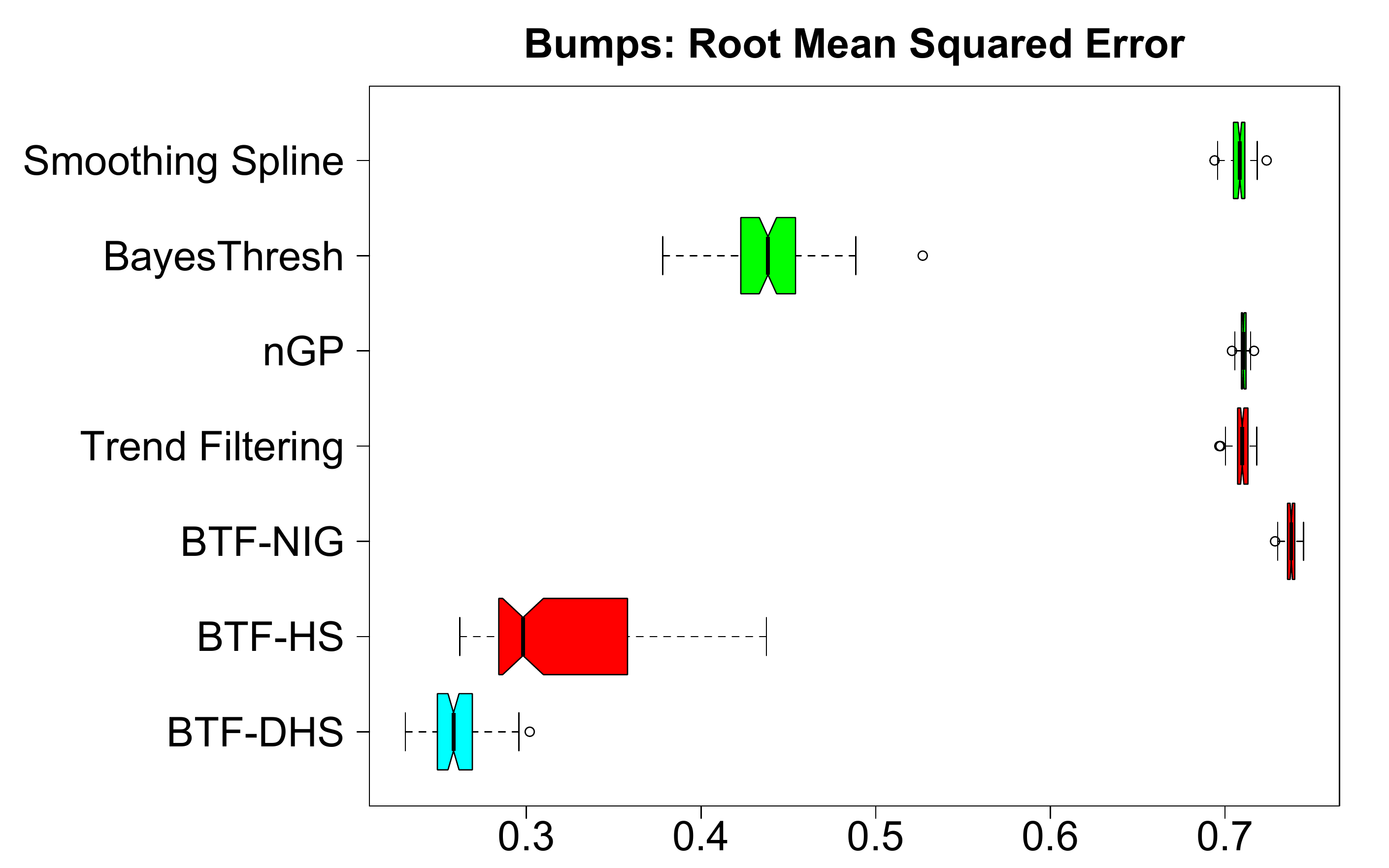}

\includegraphics[width=.5\textwidth]{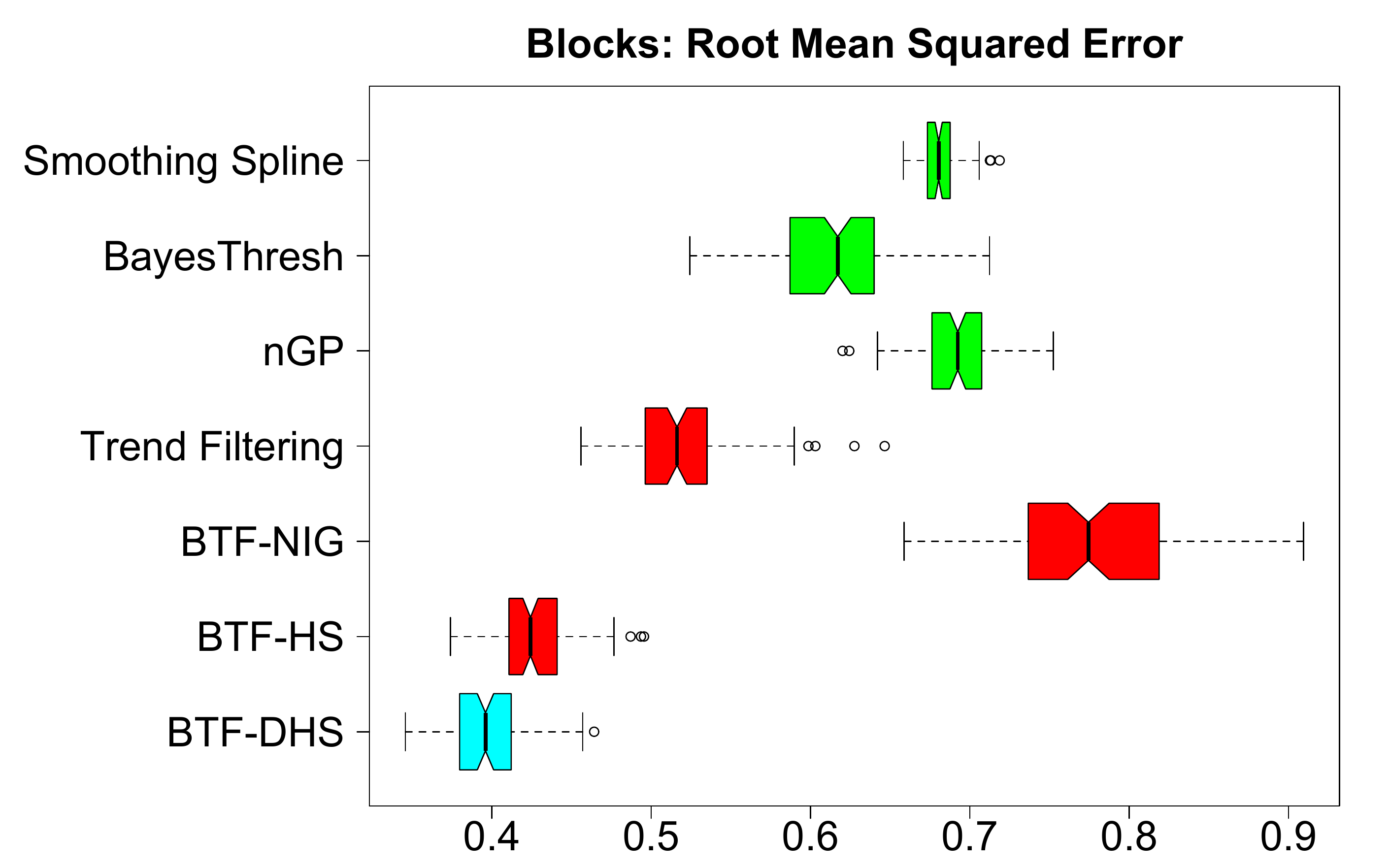}\includegraphics[width=.5\textwidth]{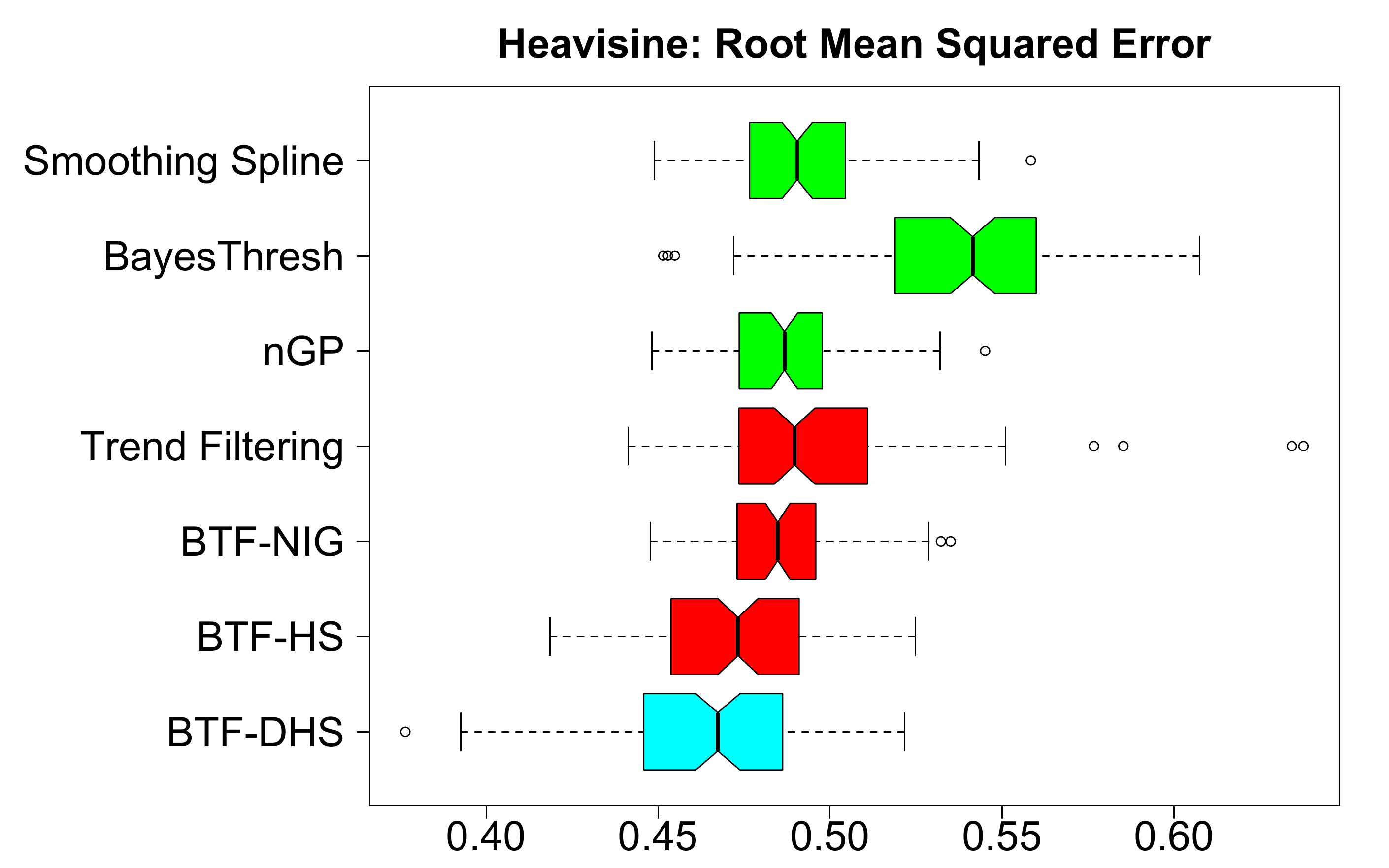}
\caption{Root mean squared errors for simulated data with $T=128$ and $\text{RSNR}=7$. Non-overlapping notches indicate significant differences between medians. The Bayesian trend filtering (BTF) estimators differ in their innovation distributions, which determines the shrinkage behavior of the second order differences ($D=2$): normal-inverse-Gamma (NIG), horseshoe (HS), and dynamic horseshoe (DHS).  %The dynamic horseshoe smoother substantially outperforms all competitors, especially for rapidly-changing curves (Doppler and Bumps). 
\label{fig:RMSE_Smoother}}
\end{center}
\end{figure}

To compare the aforementioned procedures, we compute the root mean squared errors 
$\mbox{RMSE}(\hat{y}) = \sqrt{\frac{1}{T}\sum_{t=1}^T \left(y_t^* - \hat{y}_t\right)^2}
$ for all estimators $\hat{y}$ of the true function, $y^*$. The results are displayed in Figure \ref{fig:RMSE_Smoother}. The proposed BTF-DHS implementation substantially outperforms all competitors, especially for rapidly-changing curves (Doppler and Bumps). The exceptional performance of BTF-DHS is paired with comparably small variability of RMSE, especially relative to non-dynamic horseshoe model (BTF-HS). Interestingly, the magnitude and variability of the RMSEs for BTF-DHS are related to the AR(1) coefficient, $\phi$: the 95\% HPD intervals (corresponding to Figure \ref{fig:fitted_Smoother}) are $(0.77,  0.97)$ (Doppler), $(0.81,  0.97)$ (Bumps), $(0.76,  0.96)$ (Blocks), and $(-0.04,  0.74)$ (Heavisine). For the smoothest function, Heavisine, there is less separation among the estimators. Nonetheless, BTF-DHS performs the best, even though the HPD interval for $\phi$ is wider and contains zero. 

\begin{figure}[h]
\begin{center}
\includegraphics[width=.45\textwidth]{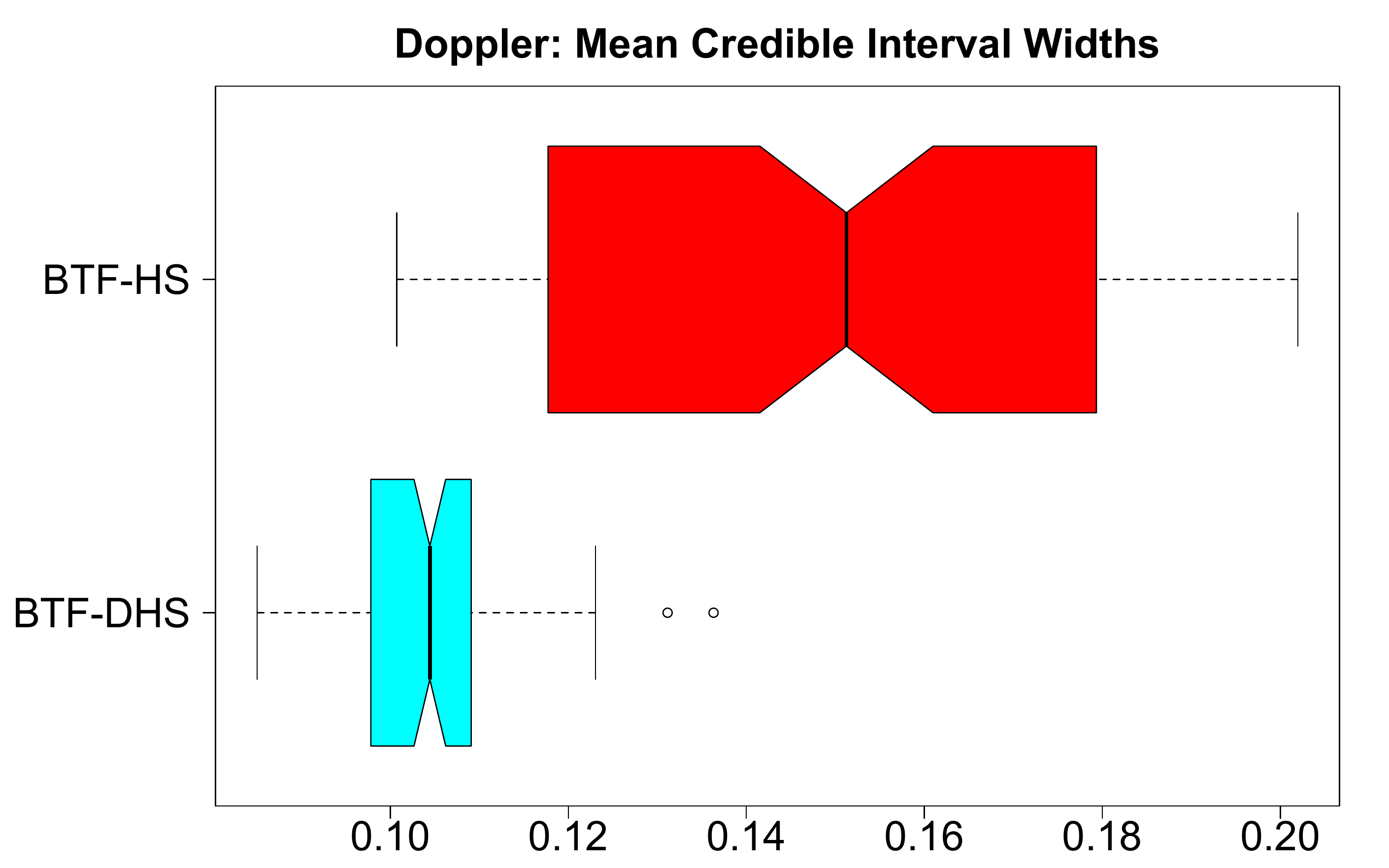}\includegraphics[width=.45\textwidth]{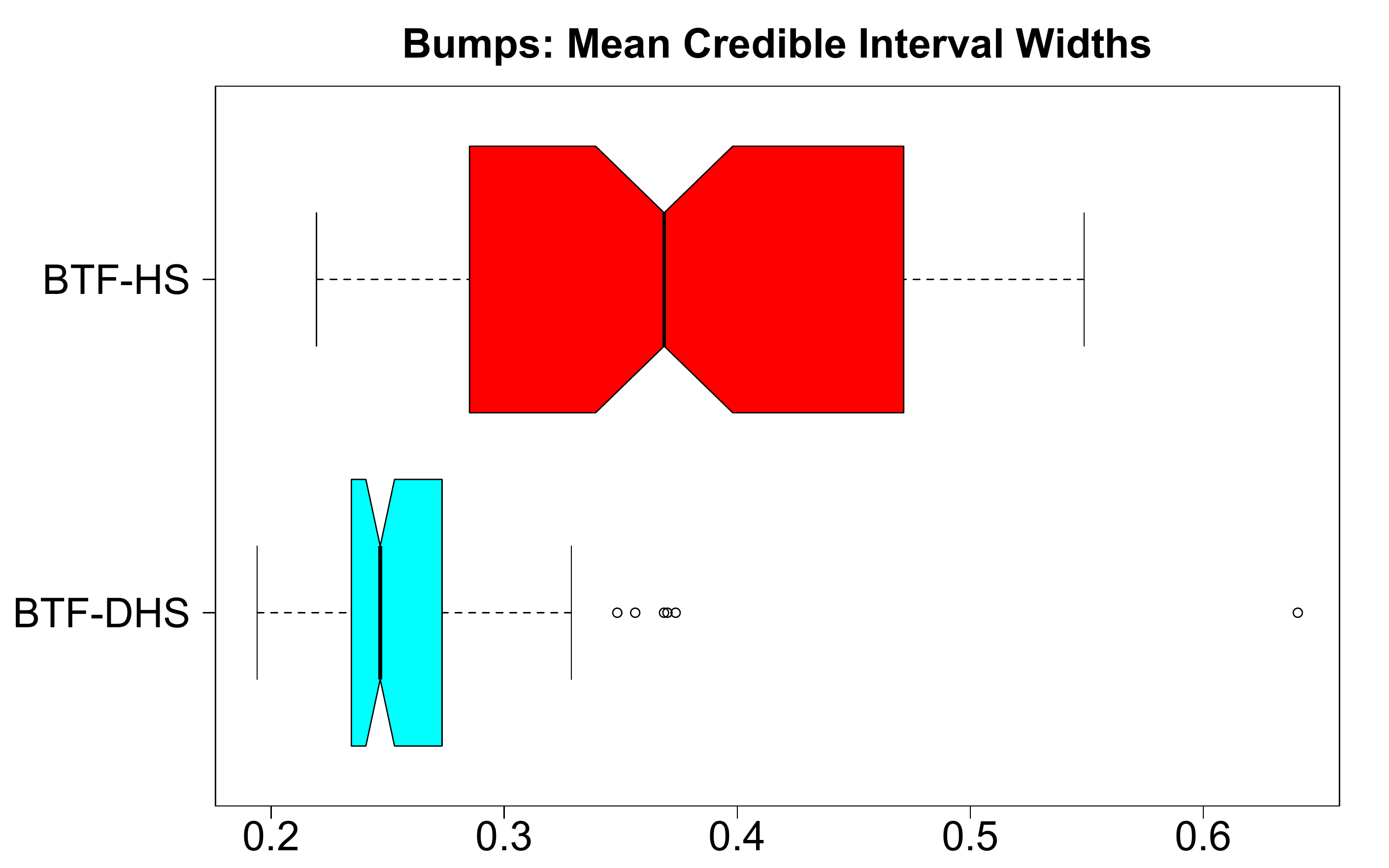}

\includegraphics[width=.45\textwidth]{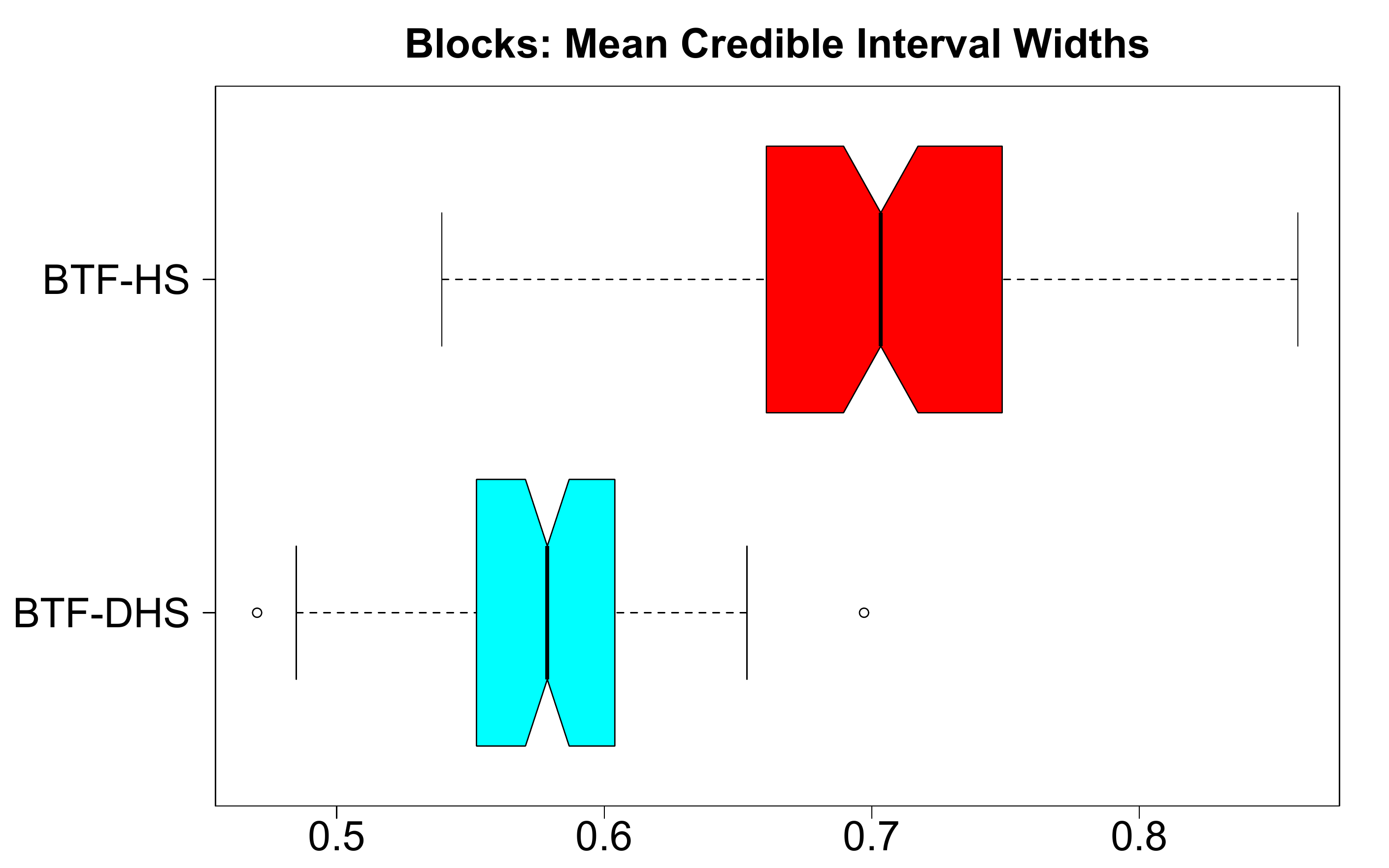}\includegraphics[width=.45\textwidth]{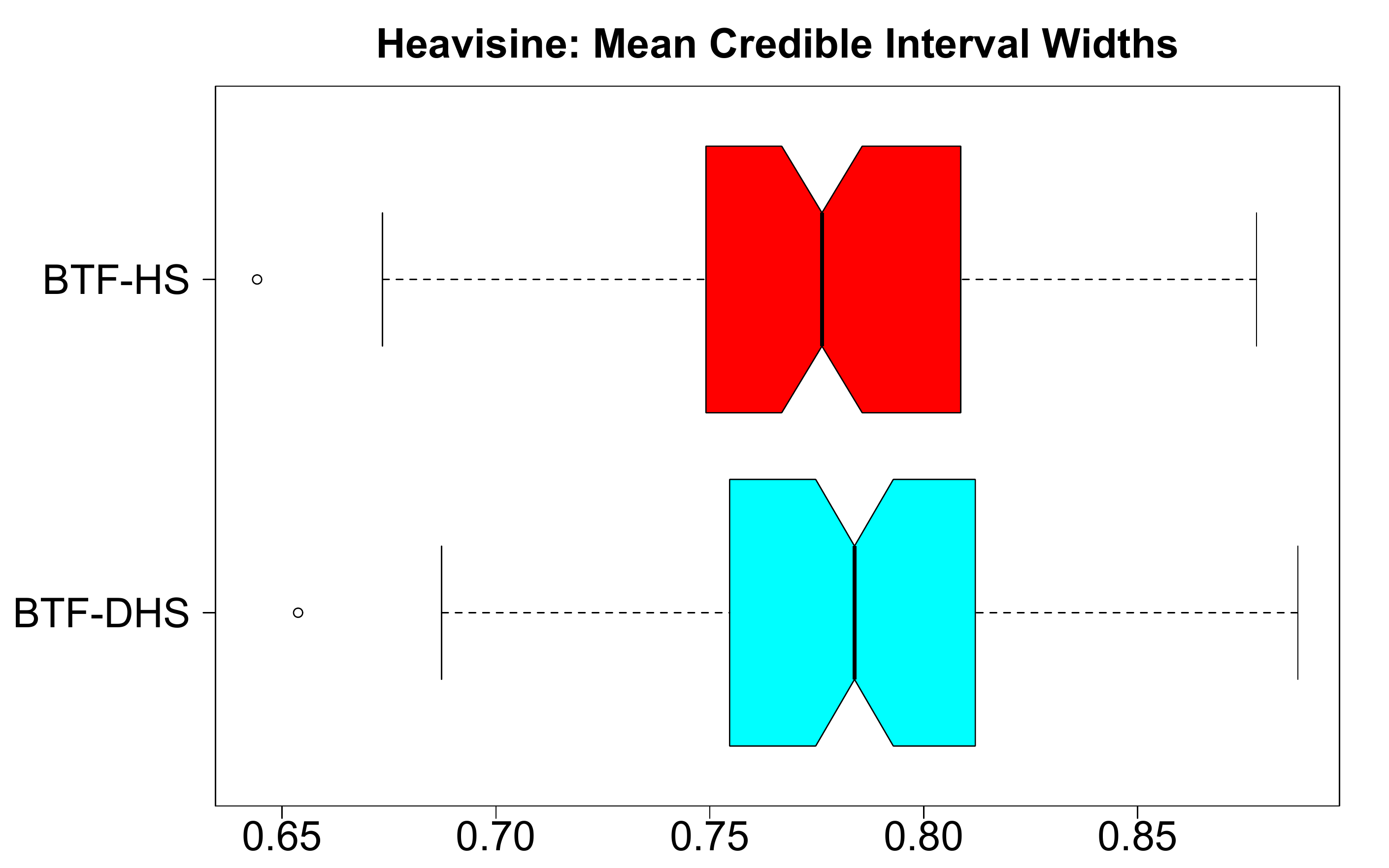}
\caption{Mean credible interval widths for simulated data with $T=128$ and $\text{RSNR}=7$.  Non-overlapping notches indicate significant differences between medians. The Bayesian trend filtering (BTF) estimators differ in their innovation distributions, which determines the shrinkage behavior of the second order differences ($D=2$): normal-inverse-Gamma (NIG), horseshoe (HS), and dynamic horseshoe (DHS).   %The dynamic horseshoe smoother substantially outperforms all competitors, especially for rapidly-changing curves (Doppler and Bumps). 
\label{fig:MCIW_Smoother}}
\end{center}
\end{figure}

We are also interested in uncertainty quantification, and in particular how the dynamic horseshoe prior compares to the horseshoe prior. We compute the mean credible intervals widths 
$
\mbox{MCIW} = \frac{1}{T} \sum_{t=1}^T (\hat \beta_t^{(97.5)} - \hat\beta_t^{(2.5)})
$ 
where $\hat \beta_t^{(97.5)}$ and $\hat \beta_t^{(2.5)}$ are the 97.5\% and 2.5\% quantiles, respectively, of the posterior distribution of $\beta_t$ in \eqref{locLinTrend} for the BTF-DHS and BTF-HS. The results are in Figure \ref{fig:MCIW_Smoother}. The dynamic horseshoe provides massive reductions in MCIW, again in all cases except for Heavisine, for which the methods perform similarly. Therefore, in addition to more accurate point estimation (Figure \ref{fig:RMSE_Smoother}), the BTF-DHS model produces significantly tighter credible intervals---while maintaining the approximately correct nominal (frequentist) coverage.

\subsection{Bayesian Trend Filtering: Application to CPU Usage Data}\label{trendCPU}
To demonstrate the adaptability of the dynamic horseshoe process for model \eqref{locLinTrend}, we consider the CPU usage data in Figure \ref{fig:totalCPUa}. The  data exhibit substantial complexity: an overall smooth intraday trend but with multiple irregularly-spaced jumps, and an increase in volatility from 16:00-18:00. Our goal is to provide an accurate measure of the trend, including jumps, with appropriate uncertainty quantification. For this purpose, we employ the BTF-DHS model \eqref{locLinTrend}, which we extend to include stochastic volatility for the observation error: $y_t \stackrel{indep}{\sim} N(\beta_t, \sigma_t^2)$ with an AR(1) model on $\log \sigma_t^2$ as in \eqref{dhs} with $\eta_t \stackrel{iid}{\sim} N(0, \sigma_\eta^2)$. For the additional sampling step of the stochastic volatility parameters, we use the algorithm of \cite{kastner2014ancillarity} implemented in the \texttt{R} package \texttt{stochvol} \citep{stochvol}.

The resulting model fit is summarized in Figure \ref{fig:totalCPU}. The posterior expectation and posterior credible bands accurately model both irregular jumps and smooth trends, and capture the increase in volatility  from 16:00-18:00 (see Figure \ref{fig:totalCPUc}). By examining regions of nonoverlapping simultaneous posterior credible bands, we may assess change points in the level of the data. In particular, the model fit suggests that the CPU usage followed a slowly increasing trend interrupted by jumps of two distinct magnitudes prior to 16:00, after which the volatility increased and the level decreased until approximately 18:00.

We augment the simulation study of Section \ref{trendSims} with a comparison of out-of-sample estimation and inference of the CPU usage data. We fit each model using 90\% ($T=1296$) of the data selected randomly for training and the remaining 10\% ($T=144$) for testing, which was repeated 100 times. Models were compared using RMSE and MCIW.  % of the 95\% posterior credible intervals.}

%and mean absolute error,  $\mbox{MAE}(\hat{y}) = \frac{1}{T}\sum_{t=1}^T \left|y_t^* - \hat{y}_t\right|$, where $\hat{y}$ is an estimator of the true function, $y^*$. 

Unlike the simulation study in Section \ref{trendSims}, the subsampled data are \emph{not} equally spaced. Taking advantage of the computational efficiency of the proposed BTF methodology, we employ a model-based imputation scheme similar to \cite{elerian2001likelihood}, which is valid for missing observations. For unequally-spaced data $y_{t_i}, i = 1,\ldots,T$,  we expand the operative data set to include missing observations along an equally-spaced grid, $t^* = 1,\ldots,T^*$, such that for each observation point $i$, $y_{t_i} = y_{t^*}$ for some $t^*$. Although  $T^* \ge T$, possibly with $T^* \gg T$, all computations within the sampling algorithm, including the imputation sampling scheme for $\{y_{t^*}: t^* \ne t_i\}$, are linear in the number of (equally-spaced) time points, $T^*$. Therefore, we may apply the same Gibbs sampling algorithm as before, with the additional step of drawing $y_{t^*} \stackrel{indep}{\sim} N(\beta_{t^*}, \sigma_{t^*}^2)$ for each unobserved $t^* \ne t_i$. Implicitly, this procedure assumes that the unobserved points are missing at random, which is satisfied by the aforementioned subsampling scheme.

\begin{figure}[h]
\begin{center}
\includegraphics[width=.45\textwidth]{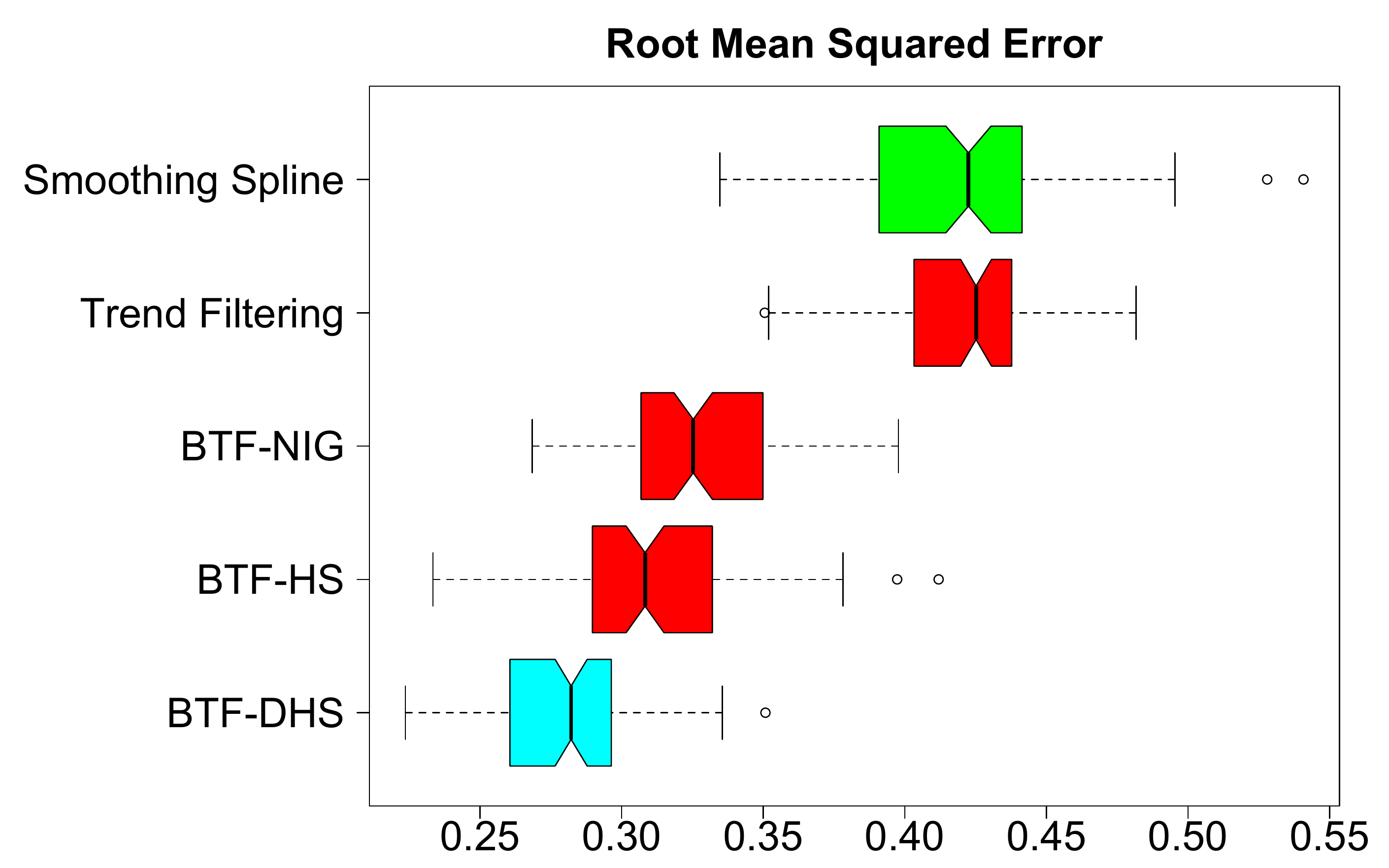}
\includegraphics[width=.45\textwidth]{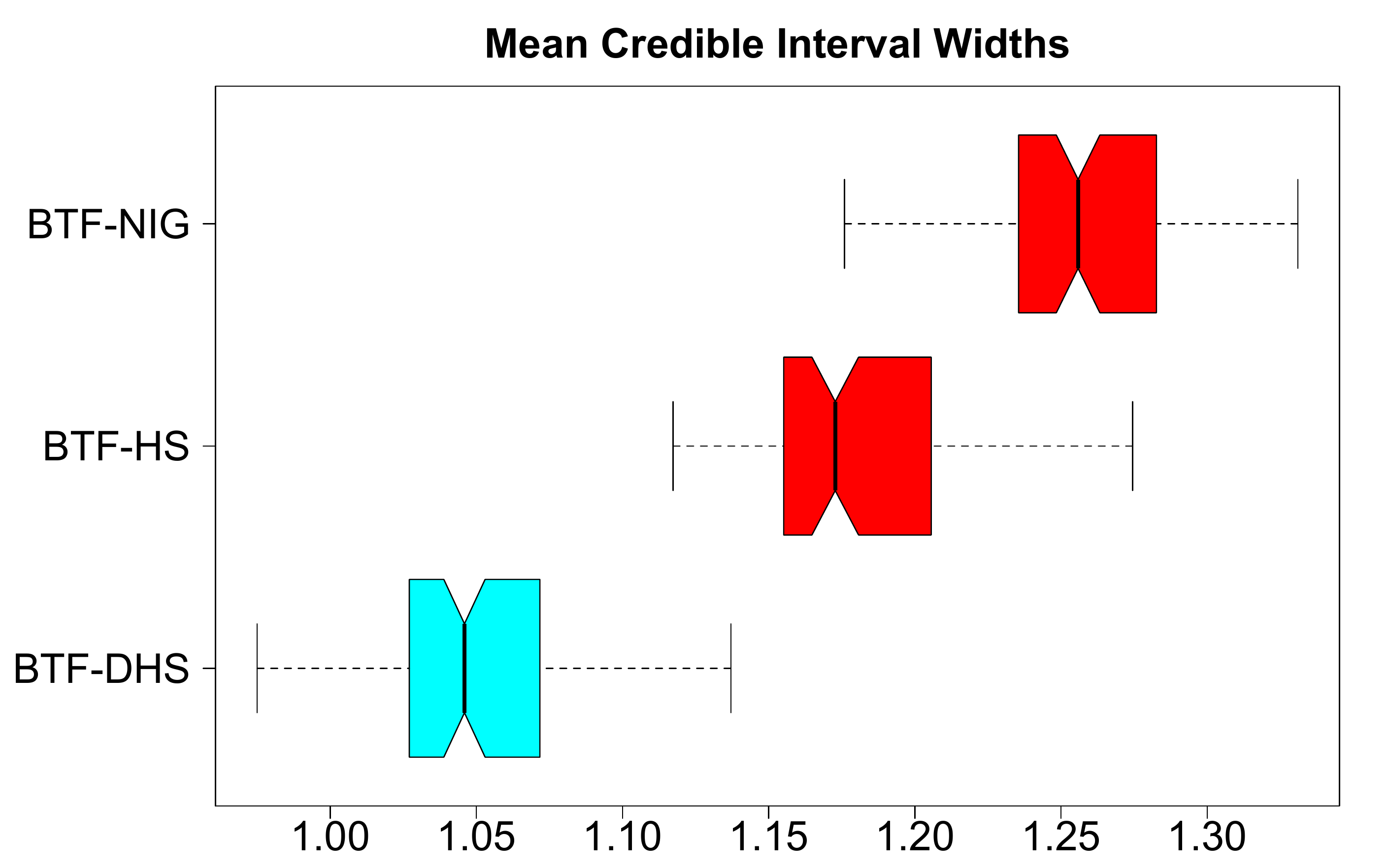}
\caption{\small Root mean squared error ({\bf left}) and mean credible interval widths ({\bf right}) for out-of-sample minute-by-minute CPU usage data. %The models were estimated using 90\% ($T=1296$) of the data selected randomly for training and the remaining 10\% ($T=144$) for testing, which was repeated independently 100 times. 
Non-overlapping notches indicate significant differences between medians. 
The Bayesian trend filtering (BTF) estimators differ in their innovation distributions, which determines the shrinkage behavior of the second order differences ($D=2$): normal-inverse-Gamma (NIG), horseshoe (HS), and dynamic horseshoe (DHS). 
\label{fig:totalCPUboxplot}}
\end{center}
\end{figure}

The results of the out-of-sample estimation study are displayed in Figure \ref{fig:totalCPUboxplot}. The BTF procedures are notably superior to the non-Bayesian trend filtering and smoothing spline estimators, and, as with the simulations of Section \ref{trendSims}, the proposed BTF-DHS model substantially outperforms all competitors. Importantly, the significant reduction in MCIW by BTF-DHS indicates that the posterior credible intervals for the out-of-sample points $y_{t^*}$ are substantially tighter for our method. By reducing uncertainty---while maintaining the approximately correct nominal (frequentist) coverage---the proposed BTF-DHS model provides greater power to detect local features. In addition, the MCMC for the BTF-DHS is fast, despite the imputation procedure: 10,000 iterations runs in about 80 seconds (in \texttt{R} on a MacBook Pro, 2.7 GHz Intel Core i5).

\section{Joint Shrinkage for Time-Varying Parameter Models}\label{tvp}
Dynamic shrinkage processes are appropriate for multivariate time series and functional data models that may benefit from locally adaptive shrinkage properties. As outlined in \cite{dangl2012predictive}, models with \emph{time-varying parameters}  are particularly important in financial and economic applications, due to the inherent structural changes in regulations, monetary policy, market sentiments, and macroeconomic interrelations that occur over time. Consider the following time-varying parameter regression model with multiple dynamic predictors $\bm{x}_t = (x_{1,t},\ldots,x_{p,t})'$:
\begin{equation}\label{globalLocalMult}
\begin{cases}
y_t = \bm{x}_t' \bm{\beta}_t + \epsilon_t, & [\epsilon_t|\sigma_\epsilon] \stackrel{indep}{\sim}  N(0, \sigma_\epsilon^2)\\
\Delta^D \bm{\beta}_{t+1} = \bm{\omega}_t, & [\omega_{j,t} | \tau_0, \{\tau_k\}, \{\lambda_{k,s}\}] \stackrel{indep}{\sim}  N(0, \tau_0^2\tau_j^2 \lambda_{j,t}^2)
 \end{cases}
\end{equation}
where  $\bm{\beta}_t = (\beta_{1,t},\ldots,\beta_{p,t})'$ is the vector of dynamic regression coefficients and $D \in \mathbb{Z}^+$ is the degree of differencing.  Model \eqref{globalLocalMult}
is also a (discretized) \emph{concurrent functional linear model} (e.g., \citealp{silverman2005functional}) and a \emph{varying-coefficient model} \citep{hastie1993varying} in the index $t$, and therefore is broadly applicable. The prior for the innovations $\omega_{j,t}$ incorporates three levels of global-local shrinkage: a global shrinkage parameter $\tau_0$, a predictor-specific shrinkage parameter $\tau_j$, and a predictor- and time-specific local shrinkage parameter $\lambda_{j,t}$. Relative to existing time-varying parameter regression models, our approach incorporates an additional layer of dynamic dependence: not only are the parameters time-varying, but also the \emph{relative influence} of the parameters is time-varying via the shrinkage parameters---which are dynamically dependent themselves.

We also considered a VARIMA alternative to \eqref{globalLocalMult}: $\Delta^D \bm \beta_{t+1} = \bm \Gamma \Delta^D \bm \beta_t + \bm \omega_t$, where $\bm \Gamma$ is a $p \times p$ VAR coefficient matrix. While the VARIMA model allows for lagged cross-correlations between components of $\Delta^D\bm \beta_t$, it does not produce smooth paths for $\beta_{j,t}$, so we do not pursue it further.

%The global shrinkage parameter $\tau_0$ and the local shrinkage parameters $\lambda_{j,t}$ retain the usual interpretations, while the additional parameters $\tau_j$ modulate the shrinkage of $\omega_{j,t}$ across all times $t$, akin to the local shrinkage parameter in the non-dynamic regression setting. %It is a novel contribution of this manuscript to allow the local shrinkage parameters $\lambda_{j,t}$ to be dynamic \emph{and} depend on the history of the shrinkage process, $\lambda_{j,t-1}, \lambda_{j, t-2},\ldots, \lambda_{j,1}$.

To provide jointly localized shrinkage of the dynamic regression coefficients $\{\beta_{j,t}\}$ analogous to the Bayesian trend filtering model of Section \ref{trend}, we extend \eqref{dhs} to allow for multivariate time dependence via a vector autoregression (VAR) on the log-variance:
\begin{equation}\label{SVfullMult}
\begin{cases}
[\omega_{j,t} | \tau_0, \{\tau_k\}, \{\lambda_{k,s}\}] \stackrel{indep}{\sim}  N(0, \tau_0^2\tau_j^2 \lambda_{j,t}^2)\\
h_{j,t} = \log(\tau_0^2\tau_j^2\lambda_{j,t}^2), & j=1,\ldots,p, t= 1,\ldots,T\\
\bm h_{t+1} = \bm \mu + \bm\Phi (\bm h_t -\bm \mu) + \bm \eta_t, & \eta_{j,t} \stackrel{iid}{\sim} Z(\alpha, \beta, 0, 1)
\end{cases}
\end{equation}
where $\bm h_t = (h_{1,t},\ldots, h_{p,t})'$, $\bm \mu = (\mu_1, \ldots, \mu_p)'$, $\bm \eta_t = (\eta_{1,t}, \ldots, \eta_{p,t})'$, and $\bm \Phi$ is the $p \times p$ VAR coefficient matrix. We assume $\bm \Phi = \mbox{diag}\left( \phi_1,\ldots, \phi_p\right)$ for simplicity, but non-diagonal extensions are available.   Contemporaneous dependence may be introduced via a copula model for the log-variance innovations, $\bm \eta_t$ \citep{joe2015dependence}, but may reduce computational and MCMC efficiency.
As in the univariate setting, we use P{\'o}lya-Gamma mixtures (independently) for the log-variance evolution errors, 
$[\eta_{j,t} | \xi_{j,t}] \stackrel{indep}{\sim} N( \xi_{j,t}^{-1}[\alpha-\beta]/2, \xi_{j,t}^{-1})$ with $ \xi_{j,t} \stackrel{iid}{\sim} \mbox{PG}(\alpha + \beta, 0)$ and $\alpha = \beta = 1/2$. We augment model \eqref{SVfullMult} with half-Cauchy priors for the predictor-specific and global parameters, $\tau_j  \stackrel{indep}{\sim} C^+(0, 1)$ and  $\tau_0 \sim C^+(0,\sigma_\epsilon/\sqrt{Tp})$, in which we scale by the observation error variance and the number of innovations $\{\omega_{j,t}\}$  \citep{piironen2016hyperprior}. These priors may be equivalently represented on the log-scale using the P{\'o}lya-Gamma parameter expansion  $[\mu_j|\mu, \xi_{\mu_j}] \sim N(\mu, \xi_{\mu_j}^{-1})$ and $[\mu_0 | \sigma_\epsilon, \xi_{\mu_0}] \sim N(\log \sigma_\epsilon^2 - \log T, \xi_{\mu_0}^{-1})$ with $\xi_{\mu_j},\xi_{\mu_0} \stackrel{iid}{\sim} \mbox{PG}(1, 0)$ and the identification $\mu_j = \log(\tau_0^2\tau_j^2)$ and $\mu_0 = \log(\tau_0^2)$.

\subsection{Time-Varying Parameter Models: Simulations} \label{multSims}
%The time-varying parameter regression model \eqref{globalLocalMult} is a popular approach for inference and forecasting, particularly in economics and finance. A variety of options exist for the observation error $\epsilon_t$ and the evolution error $\bm \omega_t$, such as stochastic volatility, heavy-tailed distributions, and shrinkage priors. 
We conducted a simulation study to evaluate competing variations of the time-varying parameter regression model \eqref{globalLocalMult}, in particular relative to the proposed dynamic shrinkage process ({\bf DHS}) in \eqref{SVfullMult}. Similar to the simulations of Section \ref{trendSims}, we focus on the distribution of the innovations, $\omega_{j,t}$, and again include the normal-inverse-Gamma ({\bf NIG}) and the (static) horseshoe ({\bf HS}) as competitors, in each case selecting $D=1$.  We also include  \cite{belmonte2014hierarchical}, which uses the Bayesian Lasso as a prior on the innovations ({\bf BL}). Lastly, we include \cite{kalli2014time}, which offers an alternative approach for dynamic shrinkage ({\bf KG}).
Among models with non-dynamic regression coefficients, we include a lasso regression \citep{tibshirani1996regression} and an ordinary linear regression.   These non-dynamic methods were non-competitive and are excluded from the figures.

We simulated 100 data sets of length $T=200$ from the model $y_t = \bm{x}_t'\bm{\beta}_t^* + \epsilon_t$, where the $p=20$ predictors are $x_{1,t} = 1$ and  $x_{j,t} \stackrel{iid}{\sim} N(0,1)$  for $j > 2$, and $\epsilon_t \stackrel{iid}{\sim}  N(0, \sigma_*^2)$. We also consider autocorrelated predictors $x_{j,t}$ in the supplement with similar results. The true regression coefficients $\bm \beta_t^* = (\beta_{1,t}^*,\ldots,\beta_{p,t}^*)'$ are the following: $\beta_{1,t}^* = 2$ is constant; $\beta_{2,t}^*$ is piecewise constant with $\beta_{2,t}^* = 0$ everywhere except $\beta_{2,t}^* = 2$ for $t = 41,\ldots,80$ and $\beta_{2,t}^* = -2$ for $t = 121,\ldots,160$; $\beta_{3,t}^* = \frac{1}{\sqrt{100}} \sum_{s=1}^t Z_s$ with $Z_s \stackrel{iid}{\sim}N(0,1)$ is a scaled random walk for $t \le 100$ and $\beta_{3,t}^* = 0$ for $t > 100$; and $\beta_{j,t}^* = 0$ for $j =4,\ldots, p=20$. The predictor set contains a variety of functions: a constant nonzero function, a locally constant function, a slowly-varying function that thresholds to zero for $t > 100$, and  17 true zeros. The noise variance $\sigma_*^2$ is determined by selecting a root-signal-to-noise ratio (RSNR) and computing $\sigma_* = \sqrt{\frac{\sum_{t=1}^T (y_t^* - \bar{y}^*)^2}{T-1}} \Big/\mbox{RSNR}$, where  $ y_t^* =  \bm{x}_t'\bm{\beta}_t^*$ and $\bar{y}^* = \frac{1}{T}\sum_{t=1}^T y_t^*$. We select $\mbox{RSNR} = 3$.% which is a challenging setting. 

%In Figure \ref{fig:RMSE_Reg_Fit}, we show the true regression functions $\beta_{j,t}^*$, together with the proposed BTF-DHS posterior expectations and credible bands for $\beta_{j,t}$. Despite the challenge presented by the Bumps function, the proposed model \eqref{globalLocalMult} with innovation distribution \eqref{SVfullMult} adequately identifies the constant and zero curves, captures the important features of the Bumps function, and accurately estimates the smoother Heavisine function. 

\begin{figure}[h]
\begin{center}
\includegraphics[width=.5\textwidth]{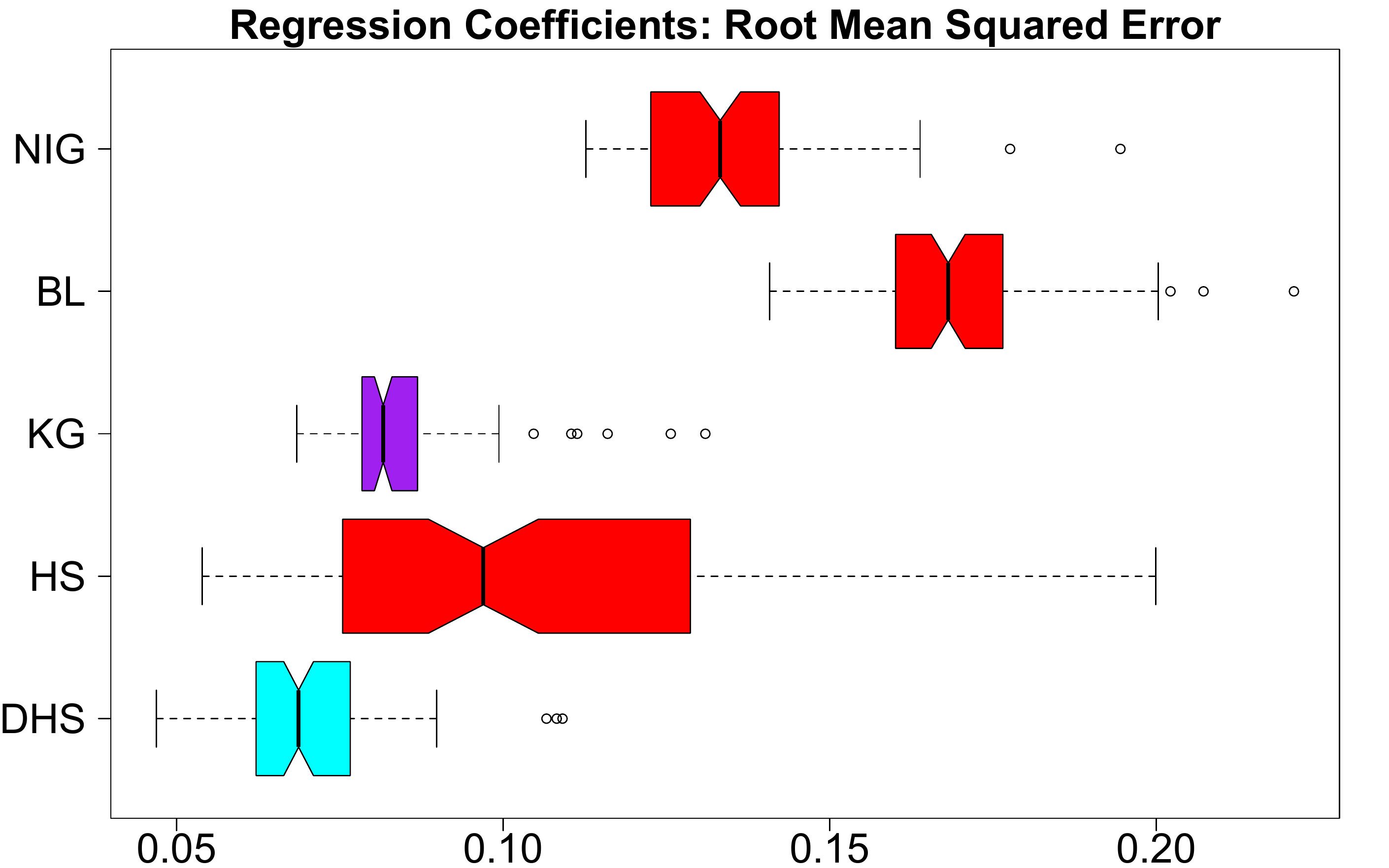}\includegraphics[width=.5\textwidth]{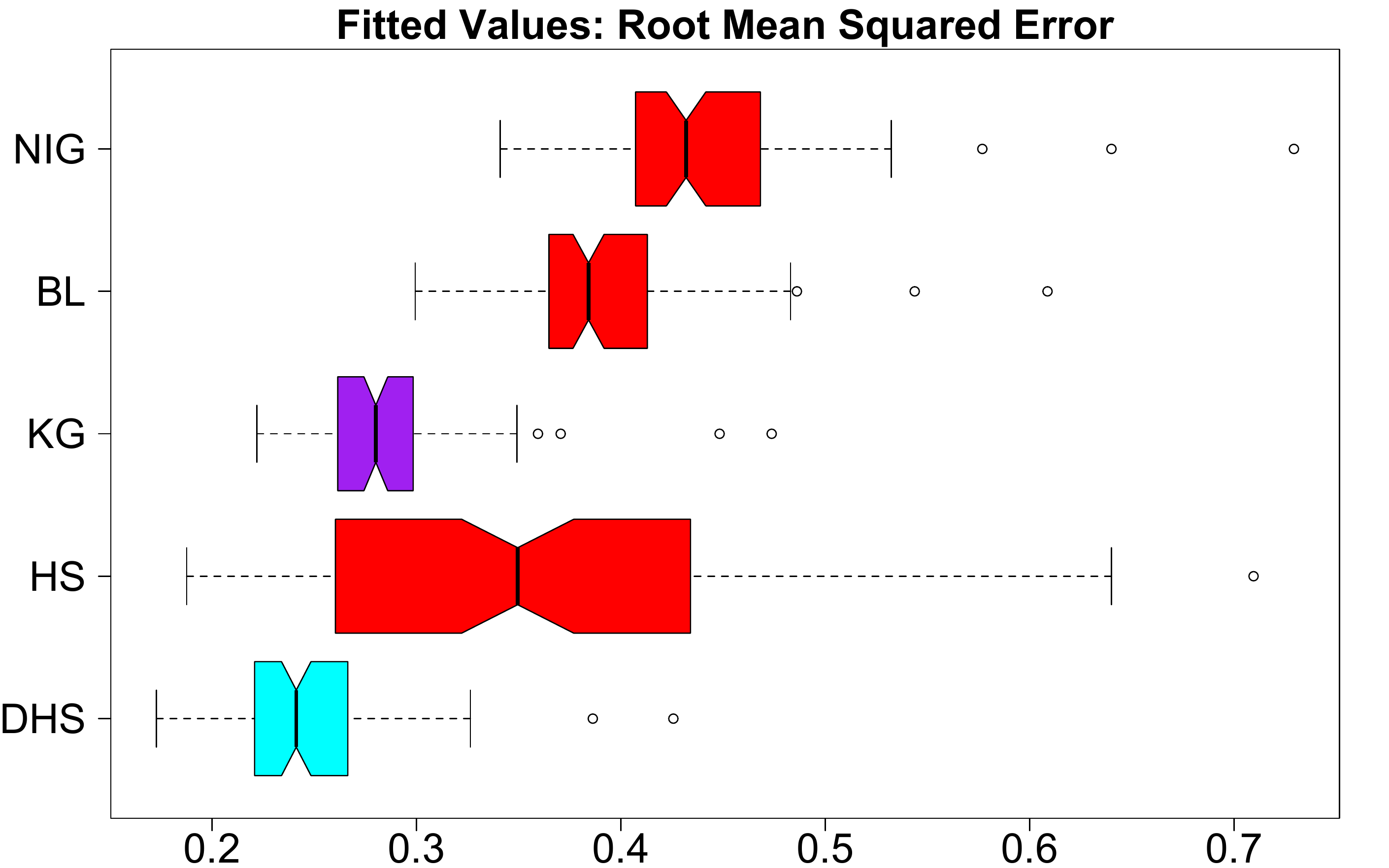}
\caption{Root mean squared errors for the regression coefficients, $\beta_{j,t}^*$ ({\bf left}) and the true curves, $y_t^* =\bm{x}_t'\bm{\beta}_t^*$ ({\bf right}) for simulated data. Non-overlapping notches indicate significant differences between medians.
\label{fig:RMSE_Reg}}
\end{center}
\end{figure}

We evaluate competing methods using RMSEs for both $y_t^*$ and $\bm{\beta}_t^*$ defined by $\mbox{RMSE}(\hat{y}) = \sqrt{\frac{1}{T}\sum_{t=1}^T \left(y_t^* - \hat{y}_t\right)^2}$ and $\mbox{RMSE}(\bm{ \hat{\beta}}) = \sqrt{\frac{1}{Tp}\sum_{t=1}^T \sum_{j=1}^p \left(\beta_{j,t}^* - \hat{\beta}_{j,t}\right)^2}$ for all estimators  $ \bm{\hat \beta_t} $ of the true regression functions, $ \bm\beta_t^*$ with $\hat{y}_t = \bm x_t' \bm{\hat \beta_t}$. The results are displayed in Figure \ref{fig:RMSE_Reg}. The proposed BTF-DHS model substantially outperforms the competitors in both recovery of the true regression functions, $\beta_{j,t}^*$ and estimation of the true curves, $y_t^*$. %Notably, the dynamic (BTF) procedures offer massive gains over the models with static regression coefficients. 
 Our closest competitor is \cite{kalli2014time}, which also uses dynamic shrinkage, yet is less accurate in estimating the regression coefficients $\beta_{j,t}^*$ and the fitted values $y_t^*$. In addition, our MCMC algorithm is vastly more efficient: for 10,000 MCMC iterations, the \cite{kalli2014time} sampler ran for 3 hours and 40 minutes (using Matlab code from Professor Griffin's website), while our proposed algorithm completed in 6 minutes (on a MacBook Pro, 2.7 GHz Intel Core i5).

% Heavisine: HPD interval for $\phi$ is  (-0.006, 0.758)

\subsection{Time-Varying Parameter Models: The Fama-French Asset Pricing Model}\label{fama}
Asset pricing models commonly feature highly structured factor models to parsimoniously model the co-movement of stock returns. Such fundamental factor models identify common risk factors among assets, which may be treated as exogenous predictors in a time series regression. Popular approaches include the one-factor Capital Asset Pricing Model (CAPM, \citealp{sharpe1964capital}) and the three-factor Fama-French model (FF-3, \citealp{fama1993common}). Recently, the five-factor Fama-French model (FF-5, \citealp{fama2015five}) was proposed as an extension of FF-3 to incorporate additional common risk factors. However, outstanding questions remain regarding which, and how many, factors are necessary. Importantly, an attempt to address these questions must consider the dynamic component: the relevance of individual factors may change over time, particularly for different assets.

We apply model \eqref{globalLocalMult} to extend these fundamental factor models to the dynamic setting, in which the factor loadings are permitted to vary---perhaps rapidly---over time. 
%The model is also a generalization of \cite{kim2016capital}, who develop a time-varying parameter model with stochastic volatility for the one-factor CAPM, and demonstrate the importance of dynamic factor loadings. 
For further generality, we append the momentum factor of \cite{carhart1997persistence} to FF-5 to produce a fundamental factor model with six factors and dynamic factor loadings. Importantly, the shrinkage towards sparsity induced by the dynamic horseshoe process allows the model to effectively select out unimportant factors, which also may change over time.  As in Section \ref{trendCPU}, we modify model \eqref{globalLocalMult} to include stochastic volatility for the observation error, $[\epsilon_t |\{\sigma_s\}] \stackrel{indep}{\sim} N(0, \sigma_t^2)$.

To study various market sectors, we use weekly industry portfolio data from the website of Kenneth R. French, which provide the value-weighted return of stocks in the given industry. We focus on manufacturing (Manuf) and healthcare (Hlth). For a given industry portfolio, the response variable is the returns in excess of the risk free rate, $y_t = R_{t} - R_{F,t}$, with predictors $\bm x_t = (1, R_{M,t} - R_{F,t}, \mathit{SMB}_t, \mathit{HML}_t, \mathit{RMW}_t, \mathit{CMA}_t, \mathit{MOM}_t)'$, defined as follows: the \emph{market risk factor}, $R_{M,t} - R_{F,t}$ is the return on the market portfolio $R_{M,t}$ in excess of the risk free rate $R_{F,t}$;  the \emph{size factor}, $\mathit{SMB}_t$ (small minus big) is the difference in returns between portfolios of small and large market value stocks; the \emph{value factor}, $\mathit{HML}_t$ (high minus low) is the difference in returns between portfolios of high and low book-to-market value stocks; the \emph{profitability factor}, $\mathit{RMW}_t$ is the difference in returns between portfolios of robust and weak profitability stocks; the \emph{investment factor}, $\mathit{CMA}_t$ is the difference in returns between portfolios of stocks of low and high investment firms; and the \emph{momentum factor}, $\mathit{MOM}_t$ is the difference in returns between portfolios of stocks with high and low prior returns. These data are publicly available on Kenneth R. French's website, which provides additional details on the portfolios. %Important special cases include $\beta_{3,t} = \cdots = \beta_{7,t} = 0$ (CAPM), $\beta_{5,t} = \beta_{6,t}  = \beta_{7,t} = 0$ (FF-3), and $\beta_{7,t} = 0$ (FF-5). Note that the factors are time-dependent, but are common for all assets under consideration. 
We standardize all predictors and the response to have unit variance.

We conduct inference on the time-varying regression coefficients $\beta_{j,t}$ using simultaneous credible bands. Unlike pointwise credible intervals, simultaneous credible bands control for multiple testing, and may be computed as in \cite{ruppert2003semiparametric}. Letting $B_{j,t}(\alpha)$ denote the $(1-\alpha)$\% simultaneous credible band for predictor $j$ at time $t$, we compute \emph{Simultaneous Band Scores} (SimBaS; \citealp{meyer2015bayesian}), $P_{j,t} = \min\{\alpha: 0 \ne B_{j,t}(\alpha)\}$.  The SimBaS $P_{j,t}$ indicate the minimum level for which the simultaneous bands do not include zero, while controlling for multiple testing, and therefore may be used to detect which predictors $j$ are important at time $t$. Globally, we compute \emph{global Bayesian p-values} (GBPVs; \citealp{meyer2015bayesian}), $P_j = \min_t\{P_{j,t}\}$ for each predictor $j$, which indicate whether or not a predictor is important at \emph{any} time $t$. SimBaS and GBPVs have proven useful in functional regression models, but also are suitable for time-varying parameter regression models to identify important predictors while controlling for multiple testing.

\begin{figure}[h]
\begin{center}
\includegraphics[width=1\textwidth]{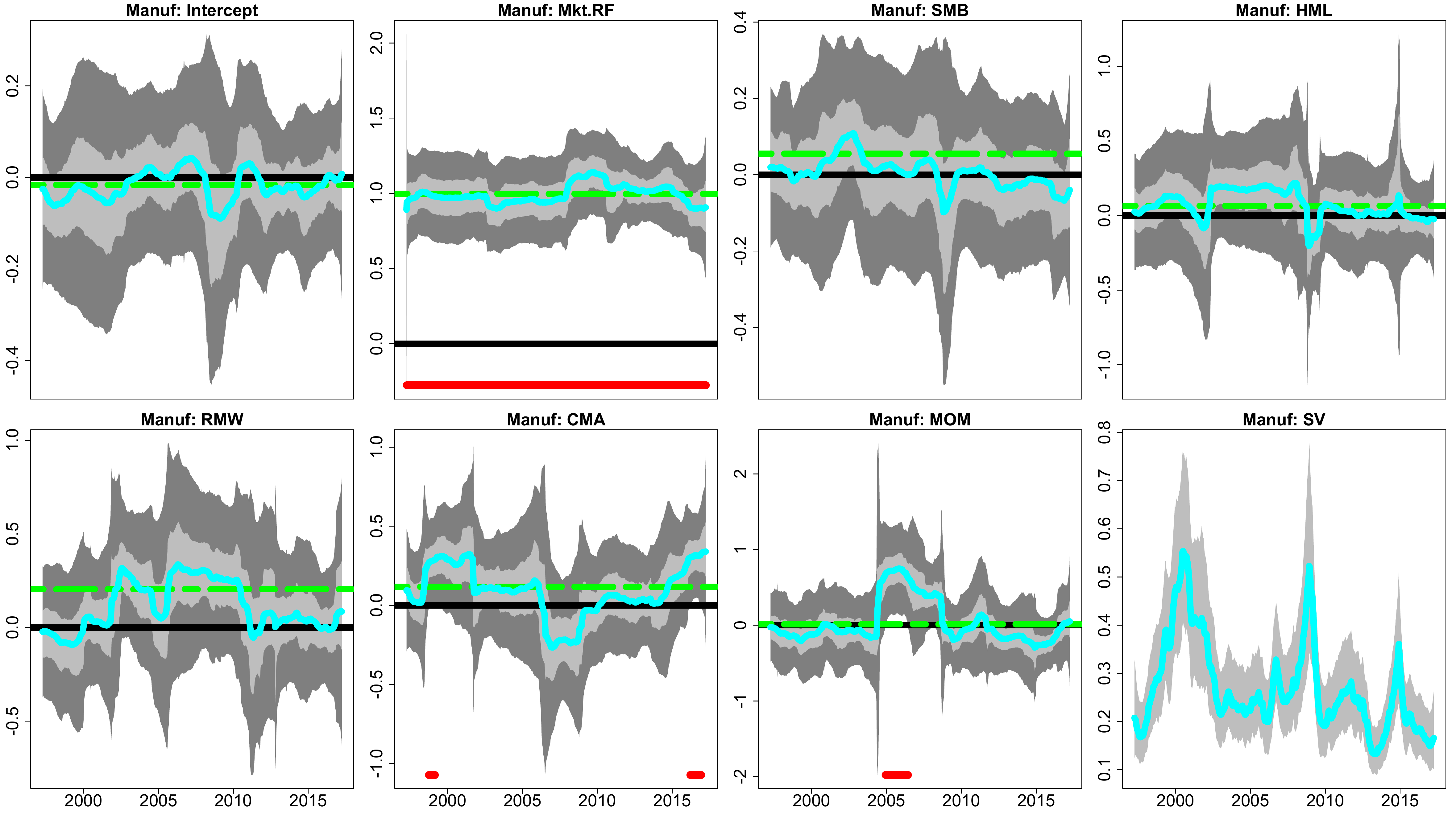}
\caption{Posterior expectations (cyan), 95\% pointwise HPD credible intervals (light gray)  and 95\% simultaneous credible bands (dark gray) for $\beta_{j,t}$ and $\sigma_t$ (bottom right) under the BTF-DHS model given by \eqref{globalLocalMult} and \eqref{SVfullMult} for value-weighted manufacturing industry returns. The solid black line is zero, the dashed green line is the ordinary linear regression estimate, and the solid red line indicates periods for which the 95\% simultaneous credible bands do not contain zero, or, equivalently, $P_{j,t}$ (SimBaS) is less than 0.05.
\label{fig:RMSE_Reg_Fit_Manuf}}
\end{center}
\end{figure}

\begin{figure}[h]
\begin{center}
\includegraphics[width=1\textwidth]{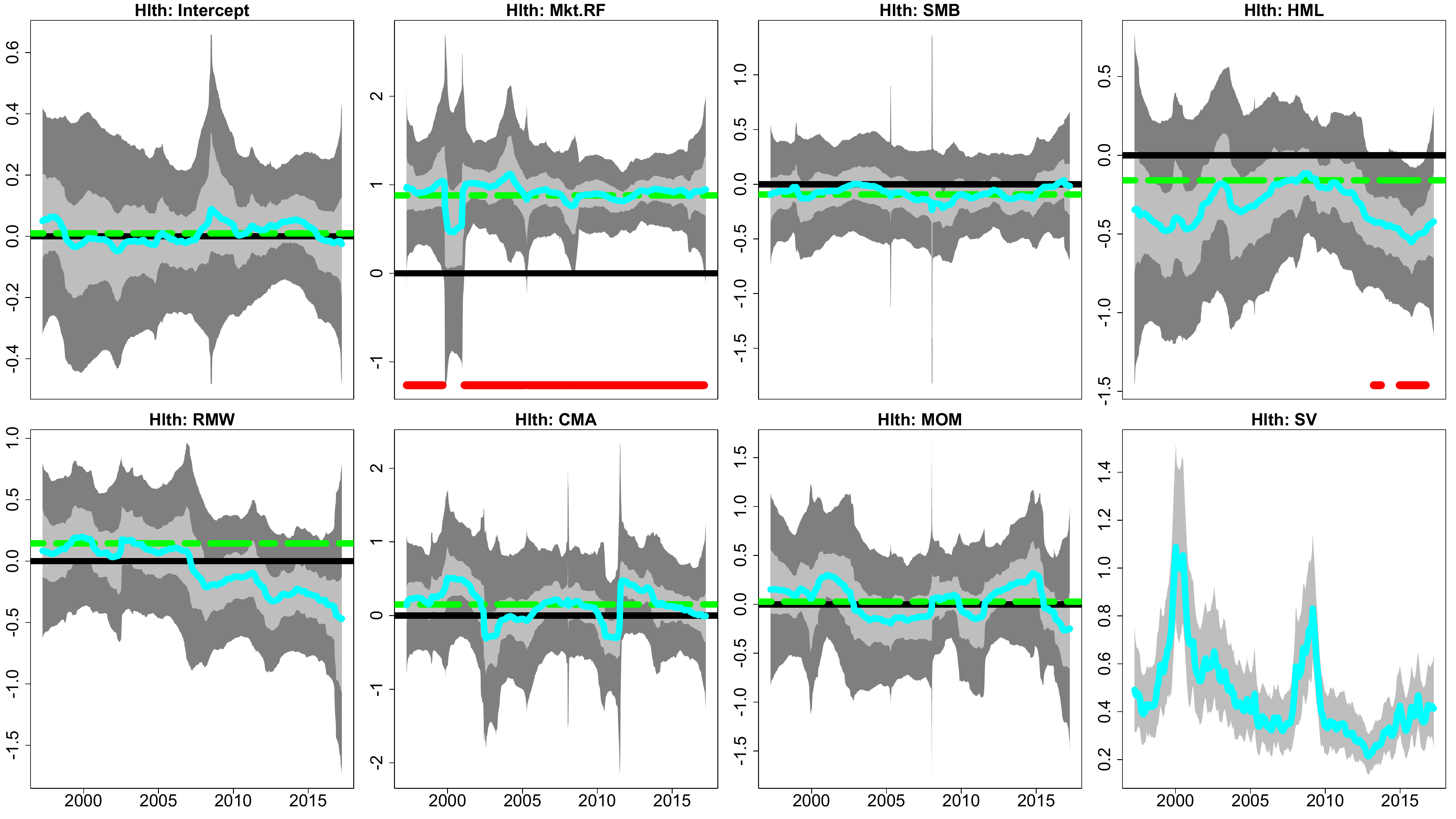}
\caption{Posterior expectations (cyan), 95\% pointwise HPD credible intervals (light gray)  and 95\% simultaneous credible bands (dark gray) for $\beta_{j,t}$ and $\sigma_t$ (bottom right) under the BTF-DHS model given by \eqref{globalLocalMult} and \eqref{SVfullMult} for value-weighted healthcare industry returns. The solid black line is zero, the dashed green line is the ordinary linear regression estimate, and the solid red line indicates periods for which the 95\% simultaneous credible bands do not contain zero, or, equivalently, $P_{j,t}$ (SimBaS) is less than 0.05.
\label{fig:RMSE_Reg_Fit_Hlth}}
\end{center}
\end{figure}

In Figures \ref{fig:RMSE_Reg_Fit_Manuf} and \ref{fig:RMSE_Reg_Fit_Hlth}, we plot the posterior expectation and credible bands for the time-varying regression coefficients and observation error stochastic volatility for the weekly manufacturing and healthcare industry data sets, respectively, from 4/1/2007 - 4/1/2017 ($T = 522$). 
%The 95\% simultaneous credible bands (dark gray) indicate which coefficients are significantly different from zero, and if so, at which times. 
For the manufacturing industry, the important factors identified by the GBPVs at the 5\% level are the market risk ($R_{M,t} - R_{F,t}$, $\mbox{GBPV} = 0.000$), investment ($\mathit{CMA}_t$, $\mbox{GBPV} = 0.024$), and momentum ($\mathit{MOM}_t$, $\mbox{GBPV} = 0.019$). However, the SimBaS 
$P_{j,t}$ for $\mathit{CMA}_t$ and $\mathit{MOM}_t$ are below 5\% only for brief periods (red lines), which suggests that these important effects are intermittent. For the healthcare industry, the GBPVs identify market risk ($\mbox{GBPV} = 0.001$) and value ($\mathit{HML}_t$, $\mbox{GBPV} = 0.023$) as the only important factors. Notably, the only common factor flagged by GBPVs in both the manufacturing and healthcare industries under model \eqref{globalLocalMult} over this time period is the market risk. This result suggests that the aggressive shrinkage behavior of the dynamic shrinkage process is important in this setting, since several factors may be effectively irrelevant for some or all time points.

%By comparison, an ordinary linear regression does \emph{not} find $\mathit{MOM}_t$ to be significant at the 5\% level, since the non-dynamic model ignores the fluctuations from 2008-2012, but does identify the market risk, profitability ($\mathit{RMW}_t$), and investment as significant factors (see the supplement for details). 

%, and profitability. By comparison, the ordinary linear regression identifies these factors as well as size ($\mathit{SMB}_t$) as significant at the 5\% level (see the supplement for details). Notably, the only common factor significant in both the manufacturing and healthcare industries under model \eqref{globalLocalMult} over this time period is the market risk. This result suggests that the aggressive shrinkage behavior of the dynamic shrinkage process is important in this setting, since several factors may be effectively irrelevant for some or all time points.

% Some conclusions:
	% Definitely need dynamic factors
	% SMB is not that important
	% Surprisingly, the CAPM factor may not be dynamic (after conditioning on other (dynamic) factors)

% Some notes:
	% Really a joint model across assets; independence assumption allows us to run in parallel
	% Mention the goals of asset pricing:
		% Estimating costs of capital
		% Computing optimal asset allocations
		% Measuring performance evaluations
	% Typically, the analysis would be done on many stocks (100s), but we can run in parallel (indep)
	% As motivated by \cite{carvalho2011dynamic}, fundamental factor models produce parsimonious
		%dynamic covariance models justified by economy theory.  
	%\cite{carvalho2011dynamic} incorporate latent factors into the dynamic FF-3 model.
	%Key point: by assuming a fundamental factor structure, i.e., a \emph{known} correlation structure 		% based on exogenous variables, we can focus on developing a more flexible TSs model

\section{MCMC Sampling Algorithm and Computational Details}\label{algorithm}
We design a Gibbs sampling algorithm for the dynamic shrinkage process. The sampling algorithm is both computationally and MCMC efficient, and builds upon two main components: (i) a  log-variance sampling algorithm \citep{kastner2014ancillarity} augmented with a P{\'o}lya-Gamma sampler \citep{polson2013bayesian}; and (ii) a Cholesky Factor Algorithm (CFA, \citealp{rue2001fast}) for sampling the state variables in the dynamic linear model. %Alternative sampling algorithms exist for more general DLMs, such as the simulation smoothing algorithm of \cite{durbin2002simple}. However, as demonstrated by \cite{mccausland2011simulation} and explored in \cite{chan2009efficient} and \cite{chan2013moving}, the CFA sampler is often more efficient. 
Importantly, both components employ algorithms that are linear in the number of time points, which produces a highly efficient sampling algorithm. 

The general sampling algorithm is as follows: (i) sample the dynamic shrinkage components (the log-volatilities $\{h_t\}$, the P{\'o}lya-Gamma mixing parameters $\{\xi_t\}$, 
the unconditional mean of log-variance $\mu$, the AR(1) coefficient of log-variance $\phi$, and the discrete mixture component indicators $\{s_t\}$); (ii) sample the state variables $\{\bm \beta_t\}$; and (iii) sample the observation error variance $\sigma_\epsilon^2$. We provide details of the dynamic shrinkage process sampling algorithm in Section \ref{sampleDSP} and include the details for sampling steps (ii) and (iii) in the supplement.

\subsection{Efficient Sampling for the Dynamic Shrinkage Process}\label{sampleDSP}
Consider the (univariate) dynamic shrinkage process in \eqref{dhs} with the P{\'o}lya-Gamma parameter expansion of Theorem \ref{PGthm}. We provide implementation details for the dynamic horseshoe process with $\alpha = \beta = 1/2$, but extensions to other cases are straightforward. The sampling framework of \cite{kastner2014ancillarity} represents the likelihood for $h_t$ on the log-scale, and approximates the ensuing $\log \chi_1^2$ distribution for the errors via a known discrete mixture of Gaussian distributions. In particular, let $\tilde y_t = \log(\omega_t^2 +c)$, where $c$ is a small offset to avoid numerical issues. Conditional on the mixture component indicators $s_t$, the likelihood is  $\tilde y_t \stackrel{indep}{\sim} N(h_t + m_{s_t}, v_{s_t})$ where $m_i$ and $v_i, i =1,\ldots,10$ are the pre-specified mean and variance components of the 10-component Gaussian mixture provided in \cite{omori2007stochastic}.  Under model \eqref{dhs}, the evolution equation is $h_{t+1} = \mu + \phi(h_t- \mu)  + \eta_t$ with initialization $h_1 = \mu + \eta_0$ and innovations $ [\eta_t | \xi_t] \stackrel{indep}{\sim} N( 0, \xi_t^{-1})$ for $[\xi_t] \stackrel{iid}{\sim}\mbox{PG}(1,0)$. Note that model \eqref{dsp} provides a more general setting, which similarly may be combined with the Gaussian likelihood for $\tilde y_t$ above.

To sample $\bm h = (h_1,\ldots,h_T)$ jointly, we directly compute the posterior distribution of $\bm h$ and exploit the tridiagonal structure of the resulting posterior precision matrix. In particular, we equivalently have 
$ \bm{\tilde y} \sim N(\bm m + \bm{\tilde h} + \tilde{\bm \mu}, \bm \Sigma_v) $ and $\bm D_\phi \bm{\tilde h} \sim N(\bm 0, \bm \Sigma_\xi)$, where $\bm m = (m_{s_1},\ldots, m_{s_T})'$, $\bm{\tilde h} = (h_1 - \mu, \ldots, h_T - \mu)'$, $\bm{\tilde \mu} = (\mu, (1-\phi)\mu,\ldots, (1-\phi)\mu)'$, $\bm\Sigma_v = \mbox{diag}\left(\{v_{s_t}\}_{t=1}^T\right)$, $\bm\Sigma_\xi = \mbox{diag}\left(\{\xi_{t}^{-1}\}_{t=1}^T\right)$, and $\bm D_\phi$ is a lower triangular matrix with ones on the diagonal, $-\phi$ on the first off-diagonal, and zeros elsewhere. We sample from the posterior distribution of $\bm h$  by sampling from the posterior distribution of $\bm{\tilde h}$ and setting $\bm h = \bm{\tilde h} + \mu\bm{1}$ for $\bm{1}$ a $T$-dimensional vector of ones. The required posterior distribution is $\bm{\tilde h} \sim N\big(\bm{Q}_{\tilde h}^{-1}\bm{\ell}_{\tilde h}, \bm{Q}_{\tilde h}^{-1}\big)$, where $\bm{Q}_{\tilde h} = \bm \Sigma_v^{-1} + \bm D_\phi' \bm \Sigma_\xi^{-1} \bm D_\phi$ is a tridiagonal symmetric matrix with diagonal elements $\bm d_0(\bm{Q}_{\tilde h})$ and first off-diagonal elements $\bm d_1(\bm{Q}_{\tilde h})$ defined as
\begin{align*}
&\bm d_0(\bm{Q}_{\tilde h})= \Big[(v_{s_1}^{-1} + \xi_1 + \phi^2 \xi_2), (v_{s_2}^{-1} + \xi_2 + \phi^2 \xi_3), \ldots, (v_{s_{T-1}}^{-1} + \xi_{T-1} + \phi^2 \xi_T), (v_{s_T}^{-1} + \xi_T) \Big],\\
&\bm d_1(\bm{Q}_{\tilde h}) = \big[(-\phi \xi_2), (-\phi \xi_3),\ldots, (-\phi \xi_{T-1})\big], \mbox{ and} \\
&\bm{\ell}_{\tilde h} = \bm \Sigma_v^{-1}\big(\bm{\tilde y} - \bm m - \bm{\tilde \mu}\big) = \Big[\frac{\tilde y_1 - m_{s_1} - \mu}{v_{s_1}},  \frac{\tilde y_2 - m_{s_2} - (1-\phi)\mu}{v_{s_2}}, \ldots,\frac{\tilde y_T - m_{s_T} - (1-\phi)\mu}{v_{s_T}}\Big]'.
\end{align*}
%$\bm{\ell}_{\tilde h} = \bm \Sigma_v^{-1}\big(\bm{\tilde y} - \bm m - \bm{\tilde \mu}\big) = \big[\big(\tilde y_1 - m_{s_1} - \mu\big)/v_{s_1}, \big(\tilde y_2 - m_{s_2} - (1-\phi)\mu\big)/v_{s_2}, \ldots,\big(\tilde y_T - m_{s_T} - (1-\phi)\mu\big)/v_{s_T}\big]'.$  
Drawing from this posterior distribution is straightforward and efficient, using band back-substitution described in  \cite{kastner2014ancillarity}: (i) compute the Cholesky decomposition $\bm{Q}_{\tilde h} = \bm L \bm L'$, where $\bm L$ is lower triangle; (ii) solve $\bm L \bm a = \bm{\ell}_{\tilde h}$ for $\bm a$; and (iii) solve $\bm L' \bm{\tilde h} = \bm a + \bm e$ for $\bm{\tilde h}$, where $\bm e \sim N(\bm 0, \bm I_T)$.

Conditional on the log-volatilities $\{h_t\}$, we sample the AR(1) evolution parameters: the log-innovation precisions $\{\xi_t\}$, the autoregressive coefficient $\phi$, and the unconditional mean $\mu$. The precisions are distributed $[\xi_t | \eta_t] \sim \mbox{PG}(1, \eta_t)$ for $\eta_t = h_{t+1} - \mu - \phi(h_t - \mu)$, which we sample using the \texttt{rpg()} function in the \texttt{R} package \texttt{BayesLogit} \citep{polson2013bayesian}. The P{\'o}lya-Gamma sampler is efficient: using only exponential and inverse-Gaussian draws, \cite{polson2013bayesian} construct an accept-reject sampler for which the probability of acceptance is uniformly bounded below at 0.99919, which does not require any tuning. Next, we assume the prior $[(\phi + 1)/2] \sim \mbox{Beta}(a_\phi, b_\phi)$, which restricts $|\phi| < 1$ for stationarity, and sample from the full conditional distribution of $\phi$ using the slice sampler of \cite{neal2003slice}. We select $a_\phi = 10$ and $b_\phi =2$, which places most of the mass for the density of $\phi$ in $(0,1)$ with a prior mean of 2/3 and a prior mode of 4/5 to reflect the likely presence of persistent volatility clustering. The prior for the global scale parameter is $\tau \sim C^+(0, \sigma_\epsilon/\sqrt T)$, which implies $\mu = \log(\tau^2)$ is $[\mu |\sigma_\epsilon, \xi_\mu] \sim N(\log (\sigma_\epsilon^2/T), \xi_\mu^{-1})$ with $\xi_\mu \sim \mbox{PG}(1,0)$. Including the initialization $h_1 \sim N(\mu,  \xi_0^{-1})$ with $\xi_0 \sim \mbox{PG}(1,0)$, the posterior distribution for $\mu$ is $\mu \sim N(Q_\mu^{-1} \ell_\mu, Q_\mu^{-1} )$ with $Q_\mu = \xi_\mu + \xi_0 + (1-\phi)^2 \sum_{t=1}^{T-1} \xi_t$ and $\ell_\mu = \xi_\mu \log (\sigma_\epsilon^2/T) + \xi_0 h_1 + (1-\phi)\sum_{t=1}^{T-1} \xi_t (h_{t+1} - \phi h_t)$.  Sampling $\xi_\mu$ and $\xi_0$ follows the P{\'o}lya-Gamma sampling scheme above.

Finally, we sample the discrete mixture component indicators $s_t$. The discrete mixture probabilities are straightforward to compute: the prior mixture probabilities are the mixing proportions given by \cite{omori2007stochastic} and the likelihood is $\tilde y_t \stackrel{indep}{\sim} N(h_t + m_{s_t}, v_{s_t})$; see \cite{kastner2014ancillarity} for details.

Note that the use of the discrete mixture approximation for log-variance models as a component within a larger MCMC sampling algorithm has been used widely in the literature 
\citep{clark2011real,d2013macroeconomic,belmonte2014hierarchical,carriero2015realtime,kastner2017efficient}.
%carriero2016common

\section{Discussion and Future Work}\label{conclusions}
Dynamic shrinkage processes provide a computationally convenient and widely applicable mechanism for incorporating adaptive shrinkage and regularization into existing models. By extending a broad class of global-local shrinkage priors to the dynamic setting, the resulting processes inherit the desirable shrinkage behavior, but with greater time-localization. The success of dynamic shrinkage processes suggests that other priors may benefit from log-scale or other appropriate representations, with or without additional dependence modeling.

%These results show that the Bayesian trend filtering model \eqref{locLinTrend} with dynamic horseshoe innovations substantially improves upon existing curve-fitting procedures, and due to both its computational efficiency and the availability of posterior inference, may provide a useful procedure for a wide variety of applications. 

As demonstrated in Sections \ref{trend} and \ref{tvp}, dynamic shrinkage processes are particularly appropriate for dynamic linear models, including trend filtering and time-varying parameter regression. In both settings, the dynamic linear models with dynamic horseshoe innovations outperform all competitors in simulated data, and produce reasonable and interpretable results for real data applications. Dynamic shrinkage processes may be useful in other dynamic linear models, such as incorporating seasonality or change points with appropriately-defined (dynamic) shrinkage. Given the exceptional curve-fitting capabilities of the Bayesian trend filtering model \eqref{locLinTrend} with dynamic horseshoe innovations (BTF-DHS), a natural extension would be to incorporate the BTF-DHS into more general additive, functional, or longitudinal data models in order to capture irregular or local curve features. 

An important extension of the dynamic fundamental factor model of Section \ref{fama} is to incorporate a large number of assets, possibly with residual correlation among stock returns beyond the common factors of FF-5. Building upon \cite{carvalho2011dynamic}, a reasonable approach may be to combine a set of known factors, such as the Fama-French factors, with a set of unknown factors to be estimated from the data, where \emph{both} sets of factor loadings are endowed with dynamic shrinkage processes to provide greater adaptability yet sufficient capability for shrinkage of irrelevant factors. 

Another promising area for applications of the proposed methodology is compressive sensing and signal processing, which commonly rely on approximations for estimation and prediction (e.g., \citealp{ziniel2013dynamic,wang2016variational}). The linear time complexity of our MCMC algorithm for Bayesian trend filtering with dynamic shrinkage may offer the computational scalability to provide full Bayesian inference, and perhaps improved prediction accompanied by adequate uncertainty quantification, which is notably absent from the papers cited above.

%\singlespacing

\bibliographystyle{apalike}
\bibliography{BFDLMbib}

%\section*{Appendix}

\appendix

\section{Proofs}\label{appendix:proofs}

\begin{proof}(Proposition \ref{prop:distequiv})
Proposition \ref{prop:distequiv} follows from Proposition \ref{prop:distequiv2} with $\mu_z = 0$. 
\end{proof}
\begin{proof}(Proposition \ref{prop:distequiv2})
Let $\eta \sim Z(\alpha, \beta, \mu_z, 1)$ with density \eqref{zdist}, i.e.,
\begin{equation*}
[z] = \big[\sigma B(\alpha, \beta)\big]^{-1}  \big\{ \exp\big[(z-\mu_z)/\sigma_z\big]\big\}^\alpha
 \big\{ 1 + \exp\big[(z-\mu_z)/\sigma_z\big]\big\}^{-(\alpha+\beta)}.
\end{equation*}
The density of $\lambda^2 = \exp(\eta)$ is 
\begin{align*}
\big[\lambda^2\big] &\propto \big(\lambda^2\big)^{-1}\big\{ \exp\big[\log(\lambda^2)-\mu_z\big]\big\}^\alpha \big\{ 1 + \exp\big[\log(\lambda^2)-\mu_z\big]\big\}^{-(\alpha+\beta)}\\
&\propto \big(\lambda^2\big)^{\alpha - 1} \big[1 + \lambda^2/\exp(\mu_z)\big]^{-(\alpha + \beta)}
\end{align*}
and therefore the density of $\kappa = 1/(1 + \lambda^2)$ is 
\begin{align*}
[\kappa] &\propto  \kappa^{-2}\big[\kappa^{-1} - 1\big]^{\alpha - 1} \big[1 + (\kappa^{-1} - 1)/\exp(\mu_z)\big]^{-(\alpha + \beta)} \\
&\propto \kappa^{-2 - (\alpha - 1)} (1-\kappa)^{\alpha - 1}\left\{\kappa^{-1} \big[\kappa \exp(\mu_z) + (1-\kappa)\big]\right\}^{-(\alpha + \beta)}
\\
&\propto(1-\kappa)^{\alpha - 1} \kappa^{\beta - 1} \big[\kappa \exp(\mu_z) + (1-\kappa)\big]^{-(\alpha + \beta)}
\end{align*}
i.e., $\kappa \sim \mbox{TPB}(\beta, \alpha, \exp(\mu_z))$.
\end{proof}

\begin{proof}(Theorem  \ref{theorem:condTPB})
Under model \eqref{dhs}, i.e., 
\begin{equation*}
h_{t+1} =  \mu + \phi (h_t - \mu) + \eta_t,  \quad \eta_t \stackrel{iid}{\sim} Z(\alpha, \beta, 0, 1),
\end{equation*}
we have $[h_{t+1} | h_t, \phi, \mu] \sim Z(\alpha, \beta, \mu + \phi(h_t - \mu),1)$. Using Proposition \ref{prop:distequiv2}, the conditional distribution for $\kappa_{t+1}$ is $[\kappa_{t+1} | h_t, \phi, \mu] \sim \mbox{TPB}(\beta, \alpha, \exp(\mu + \phi(h_t - \mu)))$. By substituting $\tau = \exp(\mu)$ and $\lambda_t = \exp(h_t - \mu)$, we equivalently have 
$[\kappa_{t+1} | \lambda_t, \phi, \tau] \sim \mbox{TPB}(\beta, \alpha, \tau^2\lambda_t^{2\phi})$. Noting  $\tau^2\lambda_t^{2\phi} =  \tau^{2(1-\phi)} \left[\frac{1 - \kappa_t}{\kappa_t}\right]^\phi$ completes the proof.
\end{proof}

\begin{proof}{(Theorem \ref{theorem:conc})}
Let $\gamma_t = \big[(1 - \kappa_t)/\kappa_t\big]^\phi$ and note that $\kappa \mapsto \kappa^{-1/2}$ and  $\kappa \mapsto \big[1 + (\gamma_t - 1)\kappa\big]^{-1}$ are decreasing in $\kappa$ for $ \gamma_t > 1$. It follows that, for $\gamma_t > 1$, 
\begin{align*}
\mathbb{P}\big(\kappa_{t+1} > \varepsilon  \big| \{\kappa_s\}_{s \le t}, \phi \big) &= \int_\varepsilon^1 \pi^{-1} \gamma_t^{1/2} \kappa_{t+1}^{-1/2 }(1-\kappa_{t+1})^{-1/2}\left[1 + (\gamma_t - 1)\kappa_{t+1}\right]^{-1} \, d\kappa_{t+1} \\
%& \le \pi^{-1} \gamma_t^{1/2} \varepsilon^{-1/2}  \int_\varepsilon^1(1-\kappa_{t+1})^{-1/2}\left[1 + (\gamma_t - 1)\kappa_{t+1}\right]^{-1} \, d\kappa_{t+1} \\
& \le \pi^{-1} \gamma_t^{1/2} \varepsilon^{-1/2} \left[1 + (\gamma_t - 1)\varepsilon\right]^{-1}  \int_\varepsilon^1(1-\kappa_{t+1})^{-1/2} \, d\kappa_{t+1} \\
&\le 2\pi^{-1}\varepsilon^{-1/2} (1 - \varepsilon)^{1/2} \frac{\gamma_t^{1/2}}{1 + (\gamma_t - 1)\varepsilon}
\end{align*}
converges to zero as $\kappa_t \rightarrow 0$, since $\kappa_t \rightarrow 0$ implies $\gamma_t \rightarrow \infty$. %The result follows immediately.
\end{proof}

\begin{proof}(Theorem  \ref{theorem:DattaGhosh})
Marginalizing over $\omega_t$, the likelihood is $[y_{t+1} | \{\kappa_s\}] \stackrel{indep}{\sim} N(0, \kappa_{t+1}^{-1})$. From Theorem \ref{theorem:condTPB}, the posterior distribution of $\kappa_{t+1}$ may be computed as
\begin{align*}
[\kappa_{t+1} | y_{t+1}, \{\kappa_s\}_{s \le t}, \phi, \tau] &\propto \left\{\kappa_{t+1}^{\beta- 1} (1-\kappa_{t+1})^{\alpha - 1} \big[1 + (\gamma_t - 1)\kappa_{t+1}\big]^{-(\alpha + \beta)}\right\} 
\\&\quad \times \left\{ \kappa_{t+1}^{1/2}\exp\left(-y_{t+1}^2\kappa_{t+1}/2\right)\right\} \\
&\propto (1-\kappa_{t+1})^{-1/2} \big[1 + (\gamma_t - 1)\kappa_{t+1}\big]^{-1} \exp\left(-y_{t+1}^2\kappa_{t+1}/2\right) 
\end{align*}
for $\alpha = \beta = 1/2$, where $\gamma_t = \tau^{2(1-\phi)} \big[(1 - \kappa_t)/\kappa_t\big]^\phi$. Defining $p_1(\kappa) = (1 - \kappa)^{-1/2}$, $p_2(\kappa | \gamma_t) = \big[1 + (\gamma_t - 1)\kappa\big]^{-1}$, and $p_3(\kappa | y_{t+1}) = \exp\left(-y_{t+1}^2\kappa/2\right) $ for $\kappa \in (0,1)$, observe that $p_1(\cdot)$ is increasing in $\kappa$, $p_2(\kappa  | \gamma_t) \le \big[p_1(\kappa)\big]^2$ for all $\gamma_t \ge 0$, and $p_3(\cdot)$ is decreasing in $\kappa$. Similar to  \cite{datta2013asymptotic}, the following inequalities hold for $\varepsilon \in (0,1)$ with $\varepsilon' = 1-\varepsilon$:
\begin{align*}
\mathbb{P}\big(\kappa_{t+1} < \varepsilon'  \big| y_{t+1}, \{\kappa_s\}_{s \le t}, \phi, \tau\big) & 
\le \frac{\mathbb{P}\big(\kappa_{t+1} < \varepsilon'  \big| y_{t+1}, \{\kappa_s\}_{s \le t}, \phi, \tau\big)}{\mathbb{P}\big(\kappa_{t+1} > \varepsilon'  \big| y_{t+1}, \{\kappa_s\}_{s \le t}, \phi, \tau\big)} \\
&\le \frac{\int_0^{\varepsilon'} (1 - \kappa_{t+1})^{-3/2} \exp\left(-y_{t+1}^2 \kappa_{t+1}/2\right) \, d\kappa_{t+1}}{\int_{\varepsilon'}^1 \big[1 + (\gamma_t - 1)\kappa_{t+1}\big]^{-3/2} \exp\left(-y_{t+1}^2 \kappa_{t+1}/2\right) \, d\kappa_{t+1}} \\
&\le \frac{\int_0^{\varepsilon'} (1 - \kappa_{t+1})^{-3/2} \, d\kappa_{t+1}}{\exp\left(-y_{t+1}^2/2\right)  \int_{\varepsilon'}^1 \big[1 + (\gamma_t - 1)\kappa_{t+1}\big]^{-3/2} \, d\kappa_{t+1}} \\
&\le \frac{ 2 \big[(1 - \varepsilon')^{-1/2} - 1\big]}{\exp\left(-y_{t+1}^2/2\right) 
2(\gamma_t - 1)^{-1}\big\{\big[1 + (\gamma_t - 1)\varepsilon'\big]^{-1/2} - \gamma_t^{-1/2}\big\}} \\
& \le \big[(1 - \varepsilon')^{-1/2} - 1\big]\exp\left(y_{t+1}^2/2\right) 
\gamma_t^{1/2} 
\\ & \quad \times \left\{
\frac{1 - \gamma_t}{1 - \gamma_t^{1/2}/[1 + (\gamma_t - 1)\varepsilon']^{1/2}}
\right\}.
\end{align*}
Noting the final term in curly braces converges to 1 as $\gamma_t \rightarrow 0$, we obtain $\mathbb{P}\big(\kappa_{t+1} < 1 - \varepsilon  \big| y_{t+1}, \{\kappa_s\}_{s \le t}, \phi, \tau\big)  \rightarrow 0$ as $\gamma_t \rightarrow 0$. The result for (a) follows immediately. 

For $\varepsilon \in (0,1)$ and $\gamma_t < 1$, and observing that $p_2(\kappa|\gamma_t)$ is increasing in $\kappa$ for $\gamma_t < 1$, then for any $\delta \in (0,1)$,
\begin{align*}
\mathbb{P}\big(\kappa_{t+1} > \varepsilon  \big| y_{t+1}, \{\kappa_s\}_{s \le t}, \phi, \tau\big) 
%&= \frac{\int_{\varepsilon}^1 (1-\kappa_{t+1})^{-1/2} \big[1 + (\gamma_t - 1)\kappa_{t+1}\big]^{-1} \exp\left(-y_{t+1}^2\kappa_{t+1}/2\right)  \, d\kappa_{t+1} }{\int_{0}^1 (1-\kappa_{t+1})^{-1/2} \big[1 + (\gamma_t - 1)\kappa_{t+1}\big]^{-1} \exp\left(-y_{t+1}^2\kappa_{t+1}/2\right)   \, d\kappa_{t+1} } \\
& \le \frac{ \gamma_t^{-1} \exp\left(-y_{t+1}^2\varepsilon/2\right) \int_{\varepsilon}^1 (1-\kappa_{t+1})^{-1/2}  \, d\kappa_{t+1} }{\int_0^{\delta \epsilon'} \exp\left(-y_{t+1}^2\kappa_{t+1}/2\right)  \, d\kappa_{t+1} }, \\%\quad \mbox{for any } \delta \in (0,1) \\
& \le \frac{ \gamma_t^{-1} \exp\left(-y_{t+1}^2\varepsilon/2\right)  2(1 - \varepsilon)^{1/2}}{\exp\left(-y_{t+1}^2\delta\varepsilon/2\right) \delta\varepsilon} \\
&= \exp\left(-y_{t+1}^2 \varepsilon[1-\delta]/2\right) \gamma_t^{-1} 2(1 - \varepsilon)^{1/2}(\delta\varepsilon)^{-1}
\end{align*}
which converges to zero as $|y_{t+1}| \rightarrow \infty$, proving (b).
\end{proof}

\begin{proof}{(Theorem \ref{PGthm})}
The density of $\eta \sim Z(\alpha, \beta, 0 ,1)$ may be written
\begin{align*}
[\eta] &=  \frac{1}{B(\alpha, \beta)}\frac{[\exp(\eta)]^\alpha}{[1+ \exp(\eta)]^{\alpha+\beta}} \\
&=  \frac{1}{B(\alpha, \beta)}
2^{-(\alpha+\beta)} \exp\{\eta[\alpha - (\alpha + \beta)/2]\} \int_0^\infty \exp(-\eta^2\xi/2)p_{\alpha+\beta}(\xi)\,d\xi
\end{align*}
using Theorem 1 of \cite{polson2013bayesian}, where $p_b(\xi)$ is the density of the random variable $\xi \sim \mbox{PG}(b, 0), b >0$. It follows that
$$
[\eta] \propto \int_0^\infty \exp\Big\{-\frac{1}{2} \big[\eta^2 \xi - \eta ( \alpha - \beta)  \big]\Big\} p_{\alpha+\beta}(\xi)\,d\xi \propto \int_0^\infty f_N\big(\eta; \xi^{-1}[\alpha-\beta]/2, \xi^{-1}\big)p_{\alpha+\beta}(\xi)\,d\xi
$$
where $f_N(\eta; \mu_N, \sigma_N^2)$ is the density of the random variable $\eta \sim N(\mu_N, \sigma_N^2)$. 

The conditional distribution $[\xi | \eta]\sim \mbox{PG}(\alpha + \beta, \eta)$ is a direct result of \cite{polson2013bayesian}. 
%[Note that they use the theorem to construct scale-mixtures for logistic distribution only, but it would work more generally for $Z$-distributions]
\end{proof}

\section{MCMC Sampling Algorithm and Computational Details}\label{appendix:algorithm}
We design a Gibbs sampling algorithm for the dynamic shrinkage process. The sampling algorithm is both computationally and MCMC efficient, and builds upon two main components: (1) a stochastic volatility sampling algorithm \citep{kastner2014ancillarity} augmented with a P{\'o}lya-Gamma sampler \citep{polson2013bayesian}; and (2) a Cholesky Factor Algorithm (CFA, \citealp{rue2001fast}) for sampling the state variables in the dynamic linear model. Alternative sampling algorithms exist for more general DLMs, such as the simulation smoothing algorithm of \cite{durbin2002simple}. However, as demonstrated by \cite{mccausland2011simulation} and explored in \cite{chan2009efficient} and \cite{chan2013moving}, the CFA sampler is often more efficient. Importantly, both components employ algorithms that are linear in the number of time points, which produces a highly efficient sampling algorithm. 

The general sampling algorithm is as follows, with the details provided in the subsequent sections: 
\begin{enumerate}
\item Sample the dynamic shrinkage components (Section \ref{sampleDSP})
\begin{enumerate}
\item Log-volatilities, $\{h_t\}$
\item P{\'o}lya-Gamma mixing parameters, $\{\xi_t\}$ 
\item Unconditional mean of log-volatility, $\mu$
\item AR(1) coefficient of log-volatility, $\phi$
\item Discrete mixture component indicators, $\{s_t\}$ 
\end{enumerate}
\item Sample the state variables, $\{\bm \beta_t\}$ (Section \ref{appendix:sampleStates})
\item Sample the observation error variance, $\sigma_\epsilon^2$.
\end{enumerate}
For the observation error variance, we follow \cite{carvalho2010horseshoe} and assume the Jeffreys' prior $[\sigma_\epsilon^2] \propto 1/\sigma_\epsilon^2$. The full conditional distribution is $[\sigma_\epsilon |  \{y_t\}_{t=1}^T,  \{\beta_t\}_{t=1}^T, \tau^2]  \propto \sigma_\epsilon^{-1} \times \sigma_\epsilon^{-T}  \exp\big\{-\frac{1}{2\sigma_\epsilon^2} \sum_{t=1}^T (y_t - \beta_t)^2  \big\} \times \frac{\sqrt{T}}{\sigma_\epsilon (1+ T\tau^2/\sigma_\epsilon^2)}$, where the last term arises from $\tau \sim C^+(0, \sigma_\epsilon/\sqrt{T})$. We sample from this distribution using the slice sampler of \cite{neal2003slice}.

If we instead use a stochastic volatility model for the observation error variance as in Sections  \ref{trendCPU} and \ref{fama}, we replace this step with a stochastic volatility sampling algorithm (e.g., \citealp{kastner2014ancillarity}), which requires additional sampling steps for the corresponding log-volatility and the unconditional mean, AR(1) coefficient, and evolution error variance of log-volatility. An efficient implementation of such a sampler is available in the \texttt{R} package \texttt{stochvol} \citep{stochvol}. In this setting, we do not scale $\tau$ by the standard deviation, and instead assume $\tau \sim C^+(0, 1/\sqrt{T})$.

\begin{figure}[h]
\begin{center}
\includegraphics[width=1\textwidth]{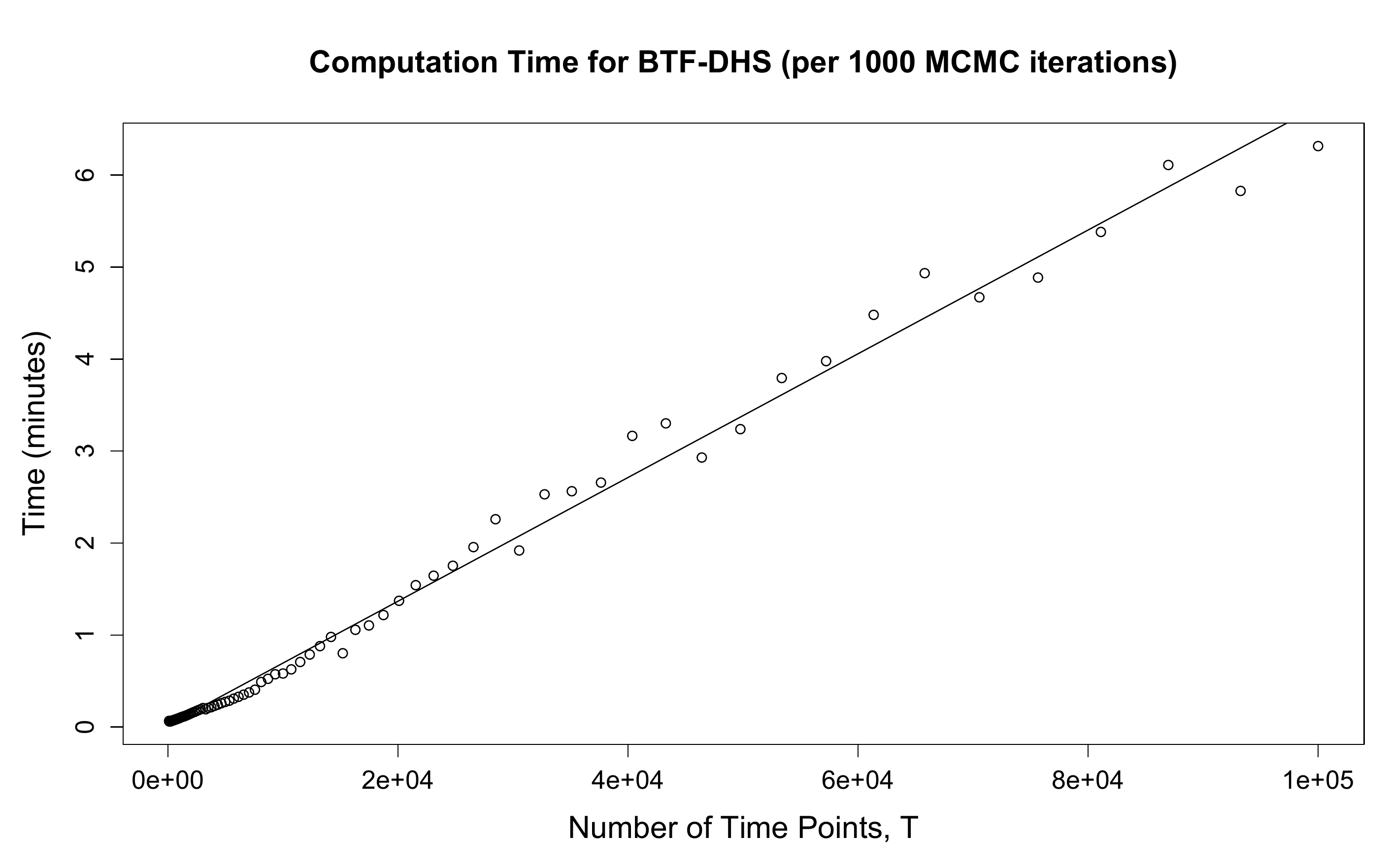}
\caption{Computation time per $1000$ MCMC iterations for the Bayesian trend filtering model with dynamic horseshoe innovations (BTF-DHS). \label{appendix:fig:linear}}
\end{center}
\end{figure}

In Figure \ref{appendix:fig:linear}, we provide empirical evidence for the linear time $\mathcal{O}(T)$ computations of the Bayesian trend filtering model with dynamic horseshoe innovations. The runtime per 1000 MCMC iterations is less than 6 minutes  (on a MacBook Pro, 2.7 GHz Intel Core i5)
for samples sizes up to $T = 10^5$, so the Gibbs sampling algorithm is scalable. 

\subsection{Efficient Sampling for the Dynamic Shrinkage Process}\label{appendix:sampleDSP}
Consider the (univariate) dynamic shrinkage process in \eqref{dhs} with the P{\'o}lya-Gamma parameter expansion of Theorem \ref{PGthm}. We provide implementation details for the dynamic horseshoe prior with $\alpha = \beta = 1/2$, but extensions to other cases are straightforward. The SV sampling framework of \cite{kastner2014ancillarity} represents the likelihood for $h_t$ on the log-scale, and approximates the ensuing $\log \chi_1^2$ distribution for the errors via a known discrete mixture of Gaussian distributions. In particular, let $\tilde y_t = \log(\omega_t^2 +c)$, where $c$ is a small offset to avoid numerical issues. Conditional on the mixture component indicators $s_t$, the likelihood is  $\tilde y_t \stackrel{indep}{\sim} N(h_t + m_{s_t}, v_{s_t})$ where $m_i$ and $v_i, i =1,\ldots,10$ are the pre-specified mean and variance components of the 10-component Gaussian mixture provided in \cite{omori2007stochastic}. The evolution equation is $h_{t+1} = \mu + \phi(h_t- \mu)  + \eta_t$ with initialization $h_1 = \mu + \eta_0$ and innovations $ [\eta_t | \xi_t] \stackrel{indep}{\sim} N( 0, \xi_t^{-1})$ for $[\xi_t] \stackrel{iid}{\sim}\mbox{PG}(1,0)$.

To sample $\bm h = (h_1,\ldots,h_T)$ jointly, we directly compute the posterior distribution of $\bm h$ and exploit the tridiagonal structure of the resulting posterior precision matrix. In particular, we equivalently have 
$ \bm{\tilde y} \sim N(\bm m + \bm{\tilde h} + \tilde{\bm \mu}, \bm \Sigma_v) $ and $\bm D_\phi \bm{\tilde h} \sim N(\bm 0, \bm \Sigma_\xi)$, where $\bm m = (m_{s_1},\ldots, m_{s_T})'$, $\bm{\tilde h} = (h_1 - \mu, \ldots, h_T - \mu)'$, $\bm{\tilde \mu} = (\mu, (1-\phi)\mu,\ldots, (1-\phi)\mu)'$, $\bm\Sigma_v = \mbox{diag}\left(\{v_{s_t}\}_{t=1}^T\right)$, $\bm\Sigma_\xi = \mbox{diag}\left(\{\xi_{t}^{-1}\}_{t=1}^T\right)$, and $\bm D_\phi$ is a lower triangular matrix with ones on the diagonal, $-\phi$ on the first off-diagonal, and zeros elsewhere. We sample from the posterior distribution of $\bm h$  by sampling from the posterior distribution of $\bm{\tilde h}$ and setting $\bm h = \bm{\tilde h} + \mu\bm{1}$ for $\bm{1}$ a $T$-dimensional vector of ones. The required posterior distribution is $\bm{\tilde h} \sim N\big(\bm{Q}_{\tilde h}^{-1}\bm{\ell}_{\tilde h}, \bm{Q}_{\tilde h}^{-1}\big)$, where $\bm{Q}_{\tilde h} = \bm \Sigma_v^{-1} + \bm D_\phi' \bm \Sigma_\xi^{-1} \bm D_\phi$ is a tridiagonal symmetric matrix with diagonal elements $\bm d_0(\bm{Q}_{\tilde h})$ and first off-diagonal elements $\bm d_1(\bm{Q}_{\tilde h})$ defined as
\begin{align*}
&\bm d_0(\bm{Q}_{\tilde h})= \Big[(v_{s_1}^{-1} + \xi_1 + \phi^2 \xi_2), (v_{s_2}^{-1} + \xi_2 + \phi^2 \xi_3), \ldots, (v_{s_{T-1}}^{-1} + \xi_{T-1} + \phi^2 \xi_T), (v_{s_T}^{-1} + \xi_T) \Big],\\
&\bm d_1(\bm{Q}_{\tilde h}) = \big[(-\phi \xi_2), (-\phi \xi_3),\ldots, (-\phi \xi_{T-1})\big], \mbox{ and} \\
&\bm{\ell}_{\tilde h} = \bm \Sigma_v^{-1}\big(\bm{\tilde y} - \bm m - \bm{\tilde \mu}\big) \\
& \quad = \Big[\frac{\tilde y_1 - m_{s_1} - \mu}{v_{s_1}},  \frac{\tilde y_2 - m_{s_2} - (1-\phi)\mu}{v_{s_2}}, \ldots,\frac{\tilde y_T - m_{s_T} - (1-\phi)\mu}{v_{s_T}}\Big]'.
\end{align*}
%$\bm{\ell}_{\tilde h} = \bm \Sigma_v^{-1}\big(\bm{\tilde y} - \bm m - \bm{\tilde \mu}\big) = \big[\big(\tilde y_1 - m_{s_1} - \mu\big)/v_{s_1}, \big(\tilde y_2 - m_{s_2} - (1-\phi)\mu\big)/v_{s_2}, \ldots,\big(\tilde y_T - m_{s_T} - (1-\phi)\mu\big)/v_{s_T}\big]'.$  
Drawing from this posterior distribution is straightforward and efficient, using band back-substitution described in  \cite{kastner2014ancillarity}: (1) compute the Cholesky decomposition $\bm{Q}_{\tilde h} = \bm L \bm L'$, where $\bm L$ is lower triangle; (2) solve $\bm L \bm a = \bm{\ell}_{\tilde h}$ for $\bm a$; and (3) solve $\bm L' \bm{\tilde h} = \bm a + \bm e$ for $\bm{\tilde h}$, where $\bm e \sim N(\bm 0, \bm I_T)$.

Conditional on the log-volatilities $\{h_t\}$, we sample the AR(1) evolution parameters: the log-innovation precisions $\{\xi_t\}$, the autoregressive coefficient $\phi$, and the unconditional mean $\mu$. The precisions are distributed $[\xi_t | \eta_t] \sim \mbox{PG}(1, \eta_t)$ for $\eta_t = h_{t+1} - \mu - \phi(h_t - \mu)$, which we sample using the \texttt{rpg()} function in the \texttt{R} package \texttt{BayesLogit} \citep{polson2013bayesian}. The P{\'o}lya-Gamma sampler is efficient: using only exponential and inverse-Gaussian draws, \cite{polson2013bayesian} construct an accept-reject sampler for which the probability of acceptance is uniformly bounded below at 0.99919, which does not require any tuning. Next, we assume the prior $[(\phi + 1)/2] \sim \mbox{Beta}(a_\phi, b_\phi)$, which restricts $|\phi| < 1$ for stationarity, and sample from the full conditional distribution of $\phi$ using the slice sampler of \cite{neal2003slice}. We select $a_\phi = 10$ and $b_\phi =2$, which places most of the mass for the density of $\phi$ in $(0,1)$ with a prior mean of 2/3 and a prior mode of 4/5 to reflect the likely presence of persistent volatility clustering. The prior for the global scale parameter is $\tau \sim C^+(0, \sigma_\epsilon/\sqrt T)$, which implies $\mu = \log(\tau^2)$ is $[\mu |\sigma_\epsilon, \xi_\mu] \sim N(\log (\sigma_\epsilon^2/T), \xi_\mu^{-1})$ with $\xi_\mu \sim \mbox{PG}(1,0)$. Including the initialization $h_1 \sim N(\mu,  \xi_0^{-1})$ with $\xi_0 \sim \mbox{PG}(1,0)$, the posterior distribution for $\mu$ is $\mu \sim N(Q_\mu^{-1} \ell_\mu, Q_\mu^{-1} )$ with $Q_\mu = \xi_\mu + \xi_0 + (1-\phi)^2 \sum_{t=1}^{T-1} \xi_t$ and $\ell_\mu = \xi_\mu \log (\sigma_\epsilon^2/T) + \xi_0 h_1 + (1-\phi)\sum_{t=1}^{T-1} \xi_t (h_{t+1} - \phi h_t)$.  Sampling $\xi_\mu$ and $\xi_0$ follows the P{\'o}lya-Gamma sampling scheme above.

Finally, we sample the discrete mixture component indicators $s_t$. The discrete mixture probabilities are straightforward to compute: the prior mixture probabilities are the mixing proportions given by \cite{omori2007stochastic} and the likelihood is $\tilde y_t \stackrel{indep}{\sim} N(h_t + m_{s_t}, v_{s_t})$; see \cite{kastner2014ancillarity} for details. 

In the multivariate setting $p > 1$ of \eqref{SVfullMult} with $\bm \Phi = \mbox{diag}\left( \phi_1,\ldots, \phi_p\right)$, we may modify the log-volatility sampler of $\{h_{j,t}\}$ by redefining relevant quantities using the ordering $\bm h = (h_{1,1},\ldots, h_{1,T}, h_{2,1},\ldots, h_{p,T})'$. In particular, the posterior precision matrix is again tridiagonal, but with diagonal elements $d_0(\bm{Q}_{\tilde h}) = \big[d_{0, 1}(\bm{Q}_{\tilde h}),\ldots, d_{0,p}(\bm{Q}_{\tilde h})\big]$ and first off-diagonal elements $\bm d_1(\bm{Q}_{\tilde h})  =  \big[d_{1, 1}(\bm{Q}_{\tilde h}), 0, d_{1, 2}(\bm{Q}_{\tilde h}), 0,\ldots, 0, d_{1,p}(\bm{Q}_{\tilde h})\big]$, where $d_{0, j}(\bm{Q}_{\tilde h})$ and $d_{1, j}(\bm{Q}_{\tilde h})$ are the diagonal elements and first off-diagonal elements, respectively, for predictor $j$ as computed in the univariate case above. Similarly, the linear term $\bm{\ell}_{\tilde h} = \big[\bm{\ell}_{\tilde h, 1}', \ldots, \bm{\ell}_{\tilde h, p}'\big]'$ where $\bm{\ell}_{\tilde h, j}$ is the linear term  for predictor $j$ as computed in the univariate case. The parameters $\xi_{j,t}$, $\phi_j$, and $s_{j,t}$ may be sampled independently as in the univariate case, while samplers for $\{\mu_j\}$ and $\mu_0$ proceed as in a standard hierarchical Gaussian model. For the more general case of non-diagonal $\bm \Phi$, we may use a simulation smoothing algorithm (e.g., \citealp{durbin2002simple}) for the log-volatilities $\{h_{j,t}\}$, while the sampler for $\bm \Phi$ will depend on the chosen prior.

\subsection{Efficient Sampling for the State Variables}\label{appendix:sampleStates}
In the univariate setting of \eqref{locLinTrend}, the sampler for $\bm \beta = (\beta_1,\ldots,\beta_T)$ is similar to the log-volatility sample in Section \ref{sampleDSP}. We provide the details for $D=2$; the $D=1$ case is similar to Section \ref{sampleDSP} with $\phi=1$, $\mu= 0$, and $m_{s_t} = 0$. Model \eqref{locLinTrend} may be written $\bm y \sim N(\bm \beta, \bm \Sigma_\epsilon)$ and $\bm D_2 \bm \beta \sim N(\bm 0, \bm \Sigma_\omega)$, where $\bm y = (y_1,\ldots,y_T)'$, $\bm \Sigma_\epsilon = \mbox{diag}\left(\{\sigma_t^2\}_{t=1}^T\right)$, $\bm \Sigma_\omega = \mbox{diag}\left(\{\sigma_{\omega_t}^2\}_{t=1}^T\right)$ for $\sigma_{\omega_t}^2 = \tau^2 \lambda_t^2$, and $\bm D_2$ is a lower triangular matrix with ones on the diagonal, $\left(0, -2,\ldots,-2\right)$ on the first off-diagonal, ones on the second off-diagonal, and zeros elsewhere. Note that we allow the observation error variance $\sigma_t^2$ to be time-dependent for full generality, as in Section \ref{fama}. The posterior for $\bm \beta$ is $\bm \beta \sim N\left(\bm{Q}_{\bm\beta}^{-1}\bm{\ell}_{\bm\beta}, \bm{Q}_{\bm\beta}^{-1}\right)$, where $\bm{Q}_{\bm\beta} = \bm \Sigma_\epsilon^{-1} + \bm D_2'\bm\Sigma_\omega^{-1} \bm D_2$ is a pentadiagonal symmetric matrix with diagonal elements  
$\bm d_0(\bm{Q}_{\bm\beta})$, first off-diagonal elements $\bm d_1(\bm{Q}_{\bm\beta})$, and  second-off diagonal elements $\bm d_2(\bm{Q}_{\bm\beta})$ defined as 
\begin{align*}
\bm d_0(\bm{Q}_{\bm\beta}) &= \Big[ \big(\sigma_1^{-2} + \sigma_{\omega_1}^{-2} + \sigma_{\omega_3}^{-2}\big), \big(\sigma_2^{-2} + \sigma_{\omega_2}^{-2} + 4\sigma_{\omega_3}^{-2} + \sigma_{\omega_4}^{-2}\big), \ldots, \\ 
& \quad \big(\sigma_t^{-2} + \sigma_{\omega_t}^{-2} + 4\sigma_{\omega_{t+1}}^{-2} + \sigma_{\omega_{t+2}}^{-2}\big), \ldots, \\
& \quad \big(\sigma_{T-2}^{-2} + \sigma_{\omega_{T-2}}^{-2} + 4\sigma_{\omega_{T-1}}^{-2} + \sigma_{\omega_{T}}^{-2}\big),  \big(\sigma_{T-1}^{-2} + \sigma_{\omega_{T-1}}^{-2} + 4\sigma_{\omega_{T}}^{-2}\big), \big(\sigma_{T}^{-2} + \sigma_{\omega_{T}}^{-2}\big)\Big], \\
\bm d_1(\bm{Q}_{\bm\beta}) &= [-2\sigma_{\omega_3}^{-2}, -2\big(\sigma_{\omega_3}^{-2}  +\sigma_{\omega_4}^{-2} \big), \ldots, -2\big(\sigma_{\omega_t}^{-2}  +\sigma_{\omega_{t+1}}^{-2} \big), \ldots,  -2\big(\sigma_{\omega_{T-1}}^{-2}  +\sigma_{\omega_{T}}^{-2} \big), -2\sigma_{\omega_T}^{-2}],\\
\bm d_2(\bm{Q}_{\bm\beta}) &= \big[\sigma_{\omega_3}^{-2},\ldots, \sigma_{\omega_t}^{-2}, \ldots,\sigma_{\omega_T}^{-2}\big],
\end{align*}
 and $\bm{\ell}_{\bm\beta} = \bm \Sigma_\epsilon^{-1}\bm y = \big[y_1/\sigma_1^2,\ldots, y_t/\sigma_t^2,\ldots, y_T/\sigma_T^2 \big]'$. Drawing from the posterior distribution is straightforward and efficient using the back-band substitution algorithm described in Section \ref{sampleDSP}.

In the multivariate setting of \eqref{globalLocalMult}, we similarly derive the posterior distribution for $\bm \beta = (\bm \beta_1',\ldots,\bm\beta_T')' = (\beta_{1,1},\beta_{2,1},\ldots,\beta_{p,1}, \beta_{1,2},\ldots,\beta_{p,T})'$. Let $\bm X = \mbox{blockdiag}\left(\{\bm{x}_t'\}_{t=1}^T\right)$ denote the $T \times Tp$ block-diagonal matrix of predictors and $\bm \Sigma_\omega = \mbox{diag}\left(\{\sigma_{\omega_{j,t}}^2\}_{j,t}\right)$ for $\sigma_{\omega_{j,t}}^2 = \tau_0^2\tau_j^2\lambda_{j,t}^2$. The posterior distribution is $\bm \beta \sim N\left(\bm{Q}_{\bm\beta}^{-1}\bm{\ell}_{\bm\beta}, \bm{Q}_{\bm\beta}^{-1}\right)$, where 
$$
\bm{Q}_{\bm\beta} = \bm{X}' \bm \Sigma_\epsilon^{-1}\bm{X} + \bm \left(\bm D_2' \otimes \bm I_p\right)\bm\Sigma_\omega^{-1} \left(\bm D_2\otimes \bm I_p\right)
$$ 
and 
$$
\bm{\ell}_{\bm\beta}  = \bm{X}' \bm \Sigma_\epsilon^{-1} \bm{y} = \left[\bm{x}_1'y_1/\sigma_1^2, \ldots, \bm{x}_t'y_t/\sigma_t^2, \ldots, \bm{x}_T'y_T/\sigma_T^2\right]'.
$$
Note that $\bm{Q}_{\bm\beta}$ may be constructed directly as above, but is now $2p$-banded. Alternatively, the regression coefficients $\{\beta_{j,t}\}$ may be sampled jointly using the simulation smoothing algorithm of \cite{durbin2002simple}.

\section{Additional Simulation Results} \label{appendix:sims}
We augment the simulation study of Section \ref{tvp} by considering autocorrelated predictors. Following the simulation design from Section \ref{tvp}, we simulated 100 data sets of length $T=200$ from the model $y_t = \bm{x}_t'\bm{\beta}_t^* + \epsilon_t$ with $\epsilon_t \stackrel{iid}{\sim}  N(0, \sigma_*^2)$. The $p=20$ predictors are $x_{1,t} = 1$ and, for $j=2,\ldots,p$, the time series $\{x_{j,t}\}_{t=1}^T$ is simulated from an AR(1) process with an autoregressive coefficient of 0.8, Gaussian innovations, unconditional mean zero, and unconditional standard deviation one. The true regression coefficients $\bm \beta_t^* = (\beta_{1,t}^*,\ldots,\beta_{p,t}^*)'$ are the following: $\beta_{1,t}^* = 2$ is constant; $\beta_{2,t}^*$ is piecewise constant with $\beta_{2,t}^* = 0$ everywhere except $\beta_{2,t}^* = 2$ for $t = 41,\ldots,80$ and $\beta_{2,t}^* = -2$ for $t = 121,\ldots,160$; $\beta_{3,t}^* = \frac{1}{\sqrt{100}} \sum_{s=1}^t Z_s$ with $Z_s \stackrel{iid}{\sim}N(0,1)$ is a scaled random walk for $t \le 100$ and $\beta_{3,t}^* = 0$ for $t > 100$; and $\beta_{j,t}^* = 0$ for $j =4,\ldots, p=20$. The predictor set contains a variety of functions: a constant nonzero function, a locally constant function, a slowly-varying function that thresholds to zero for $t > 100$, and  17 true zeros. The noise variance $\sigma_*^2$ is determined by selecting a root-signal-to-noise ratio (RSNR) and computing $\sigma_* = \sqrt{\frac{\sum_{t=1}^T (y_t^* - \bar{y}^*)^2}{T-1}} \Big/\mbox{RSNR}$, where  $ y_t^* =  \bm{x}_t'\bm{\beta}_t^*$ and $\bar{y}^* = \frac{1}{T}\sum_{t=1}^T y_t^*$. We select $\mbox{RSNR} = 3$.% which is a challenging setting. 

\begin{figure}[h]
\begin{center}
\includegraphics[width=.5\textwidth]{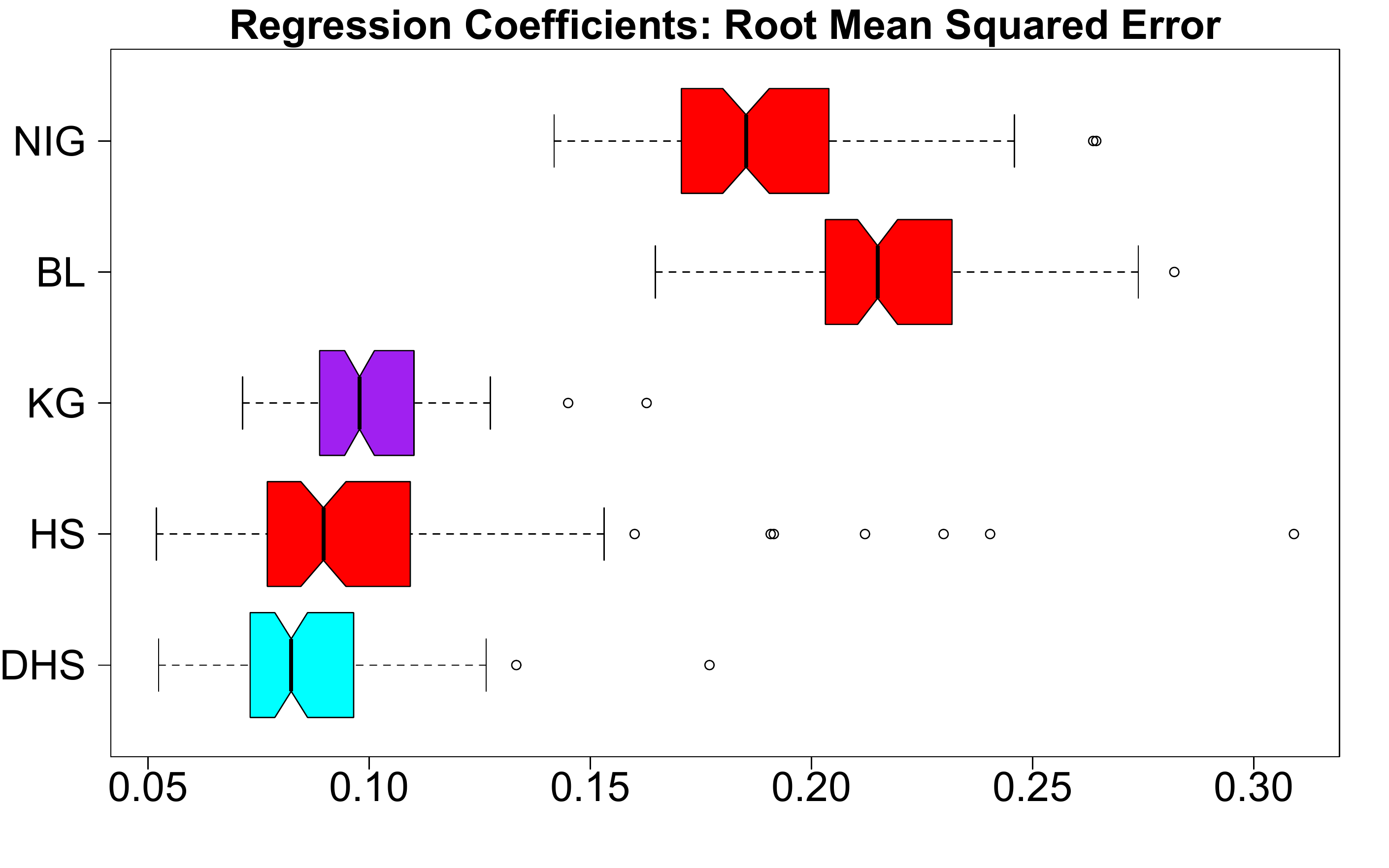}\includegraphics[width=.5\textwidth]{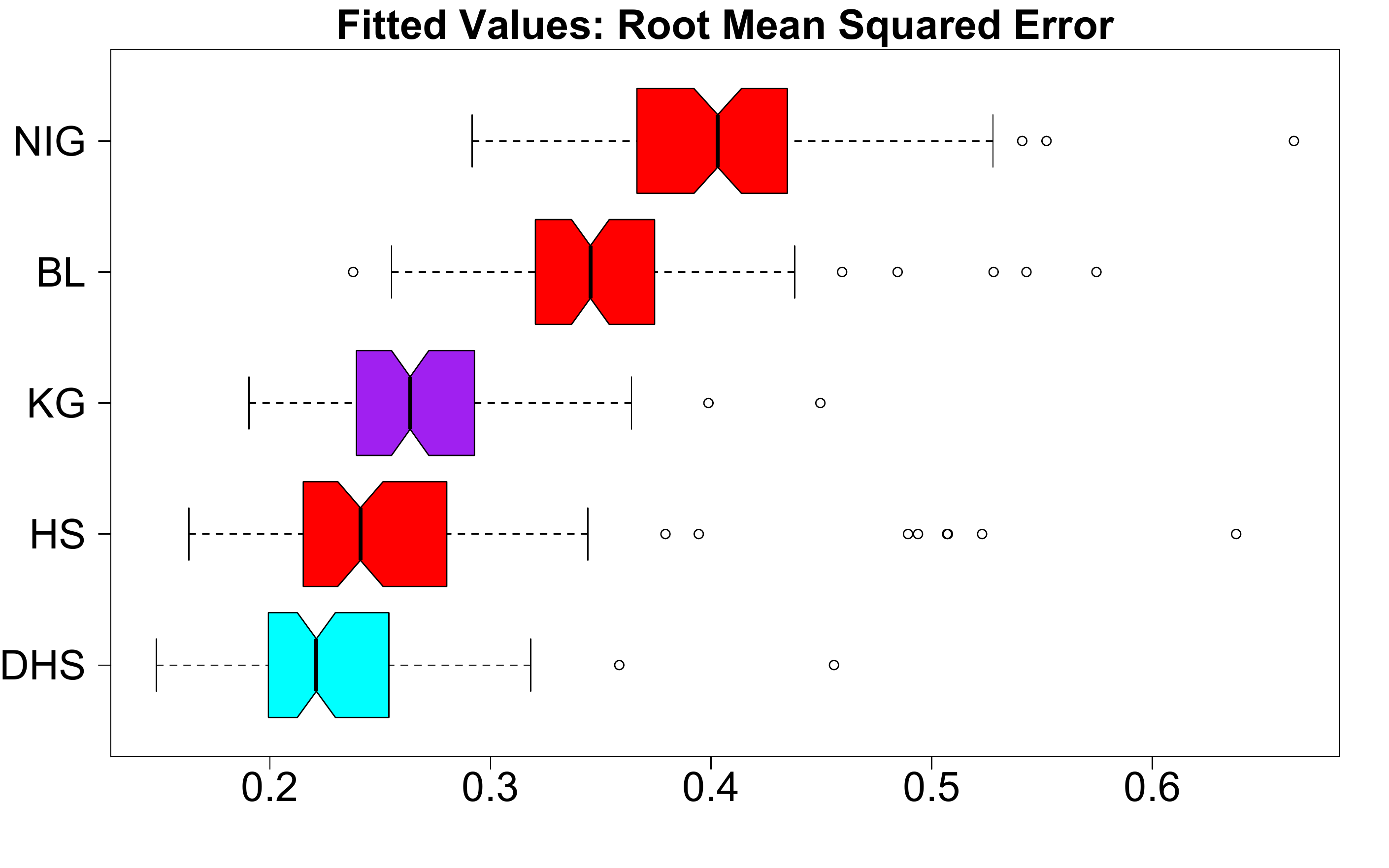}
\caption{Root mean squared errors for the regression coefficients, $\beta_{j,t}^*$ ({\bf left}) and the true curves, $y_t^* =\bm{x}_t'\bm{\beta}_t^*$ ({\bf right}) for simulated data. Non-overlapping notches indicate significant differences between medians.
\label{fig:RMSE_Reg_ar}}
\end{center}
\end{figure}

We evaluate competing methods using RMSEs for both $y_t^*$ and $\bm{\beta}_t^*$ defined by $\mbox{RMSE}(\hat{y}) = \sqrt{\frac{1}{T}\sum_{t=1}^T \left(y_t^* - \hat{y}_t\right)^2}$ and $\mbox{RMSE}(\bm{ \hat{\beta}}) = \sqrt{\frac{1}{Tp}\sum_{t=1}^T \sum_{j=1}^p \left(\beta_{j,t}^* - \hat{\beta}_{j,t}\right)^2}$ for all estimators  $ \bm{\hat \beta_t} $ of the true regression functions, $ \bm\beta_t^*$ with $\hat{y}_t = \bm x_t' \bm{\hat \beta_t}$. The results are displayed in Figure \ref{fig:RMSE_Reg_ar}. The proposed BTF-DHS model outperforms the competitors in both recovery of the true regression functions, $\beta_{j,t}^*$ and estimation of the true curves, $y_t^*$.

\end{document}